\documentclass[journal=jacsat,manuscript=article]{achemso}
\usepackage{graphicx}
\usepackage{rotating}
\usepackage{array}
\usepackage{multirow}
\usepackage{epstopdf}
\usepackage{amsmath}
\usepackage{amsfonts}
\usepackage{setspace}
\usepackage{braket}
\usepackage{xcolor}
\usepackage{adjustbox}
\usepackage{makecell}
\usepackage{subcaption}
\usepackage{nameref} 
\usepackage{hyperref}
\usepackage{changes}
\usepackage{xr-hyper}
\title{Self-consistent vertex corrected $GW$ with static and dynamic screening using tensor hypercontraction: assessment of molecular charged excitations}

\author{Munkhorgil Wang}
\affiliation{Department of Chemistry, University of Michigan, Ann Arbor, Michigan 48109, USA}
\author{Ming Wen}
\affiliation{Department of Chemistry, University of Michigan, Ann Arbor,  Michigan 48109, USA}
\author{Pavel Pokhilko}
\affiliation{Department of Chemistry, University of Michigan, Ann  Arbor,  Michigan 48109, USA}
\alsoaffiliation{Current address: 
Institute of Theoretical Chemistry, Faculty of Chemistry, University of Vienna, Währinger Straße 17, 1090 Vienna, Austria}
\author{Chia-Nan Yeh}
\affiliation{Center for Computational Quantum Physics, Flatiron Institute, New York, New York 10010, USA}
\author{Miguel A. Morales}
\affiliation{Center for Computational Quantum Physics, Flatiron Institute, New York, New York 10010, USA}
\author{Dominika Zgid}
\affiliation{Department  of  Chemistry,  University  of  Michigan,  Ann  Arbor,  Michigan  48109, USA}
\alsoaffiliation{Department of Physics, University of Michigan, Ann Arbor, Michigan 48109, USA }
\alsoaffiliation{Faculty of Physics, University of Warsaw, 02-093 Warsaw, Poland }
\email{zgid@umich.edu}

\begin{document}
\maketitle

\begin{abstract}
We investigate self-consistent vertex corrections to the $GW$ self-energy for ionization potentials (IPs) and electron affinities (EAs). We benchmark IPs against $\Delta$CCSD(T) references in the $G_0W_0\Gamma$29 and GW100 sets and compare GW100 EAs with EOM-CCSD references. Because many anions are metastable, these addition energies are model quantities and should not be interpreted as experimental EAs. 
Tensor hypercontraction (THC) of the Coulomb integrals enables efficient self-consistent $GW\Gamma_{\Sigma}$ implementations, where vertex corrections are included only in the self-energy.
We establish a hierarchy of vertex-corrected self-energies relative to sc$GW$, ordered from least to most negative as SOX $>$ SOSEX $>$ G3W2 $>$ 2SOSEX $>$ sc$GW$. Equivalently, the absolute magnitude increases along this sequence. This trend follows an effective-screening picture, in which increasing screening progressively reduces exchange contributions. Static and dynamic variants show consistent differences due to the frequency dependence of the screened interaction. Across all methods, vertex corrections act as an approximately frequency-uniform self-energy renormalization rather than altering its spectral structure.
In terms of accuracy, sc$GW\Gamma_{\Sigma}$ does not uniformly improve IPs or the EA model quantities over sc$GW$. For IPs, SOX and SOSEX usually degrade performance, whereas 2SOSEX and G3W2 remain close to sc$GW$, with only marginal improvements for selected systems at higher cost. Although the tested variants reduce EA MAEs, this should not be interpreted as a general improvement for physical anions because many nominal EA states are metastable.
These results indicate that generic vertex insertions are insufficient to outperform sc$GW$; systematic improvements require designed diagrammatic approximations combined with efficient tensor factorization.

\end{abstract}

\section{Introduction}
Many-body perturbation theory (MBPT) formulated in terms of Green's functions provides a natural framework for describing excitation spectra.~\cite{Fetter:Walecka:2012,Stefanucci:vanLeeuwen:book:2013}
Within this formalism, Hedin's equations~\cite{Hedin65} establish a closed set of relations connecting the Green's function $G$, self-energy $\Sigma$, polarization $P$, screened Coulomb interaction $W$, and vertex function $\Gamma$, and serve as the foundation for diagrammatic approximations such as $GW$.

Despite its formal simplicity, the fully self-consistent $GW$ scheme including vertex corrections in both the self-energy and polarization (sc$GW\Gamma$) has long remained computationally prohibitive. Even for the homogeneous electron gas, an implementation including the full vertex structure has only recently been reported.~\cite{Kutepov17_Vertex}

In the $GW$ implementations executed on the real-frequency axis, a primary difficulty arises from achieving self-consistency, where the number of poles in the Green's function increases during iterations.~\cite{Holm98,Rinke_GW_review}
As a consequence, most practical applications rely on non-self-consistent or partially self-consistent approaches, including $G_0W_0$~\cite{G0W0_Pickett84,G0W0_Hybertsen86,Strinati_PhysRevB,Strinati_PhysRevLett.45.290}, quasiparticle self-consistent $GW$~\cite{QPGW_Schilfgaarde,Visscher:qpGW:scaling:2021,QSGW_Kotani07}, and eigenvalue self-consistent $GW$~\cite{Rangel::evGW_benchmark::2016,Wilhelm:evGW:2016}.
These approximations introduce starting-point dependence, resulting in the $G_0W_0$@reference scheme, whose accuracy is severely reference dependent, affecting energy differences and ionization potentials~\cite{Modrzejewski:RPA:intermol:2020,Forster:SOSEX:2022,Chibotaru:BS-G0W0:2020,caruso_benchmark_2016,Maggio:GWVertexCorrected:2017}, and may violate conservation laws, leading to ambiguities in the definition of the Green's function.~\cite{Baym61,Rinke_GW_review,Harsha:GW_unbiased:2024}

Formulations on the imaginary-frequency axis avoid the proliferation of poles and yield numerically smooth Green's functions. However, they introduce additional challenges, including the need for analytical continuation, the absence of a natural separation between occupied and virtual states, and increased computational cost.~\cite{Stan06,Koval14,Rinke_GW_review}
Recent developments in frequency grids~\cite{Kananenka:grids:2016,Kananenka16,Yoshimi:IR:2017,Iskakov_Chebychev_2018,dong2020legendrespectral}, tensor factorization techniques~\cite{Yeh:THC-RPA:2023,Yeh:THC-GW:2024,Iskakov20,Yeh:GPU:GW:2022,Yeh:X2C:GW:2022,Green:2025}, and advanced convergence algorithms~\cite{Pokhilko:algs:2022,PhysRevB.100.085112,Pokhilko:homotopy:2025} have enabled routine fully self-consistent $GW$ calculations without vertex corrections.
Such calculations satisfy conservation laws and provide access to total energies and energy differences, enabling accurate predictions of magnetic interactions, Néel temperatures, intermolecular interactions, and charge-transfer phenomena in both molecules and solids.~\cite{Pokhilko:tpdm:2021,Pokhilko:local_correlators:2021,Pokhilko:BS-GW:solids:2022,Pokhilko:Neel_T:2022,Pokhilko:THC-GWSOX:2024,Pokhilko:homotopy:2025}

Within Hedin's formalism, vertex corrections modify both the self-energy $\Sigma$ and the polarization $P$ through approximations to the vertex function $\Gamma$. Fully self-consistent implementations based on bold diagrammatic expansions remain scarce due to their substantial computational cost, with only a limited number of studies reported.~\cite{Kutepov:scGW:CrI3:2021,Kutepov:scG3W2:2022,Foster:G3W2stat:2022,vanLeeuwen:PSD:electron_gas:2014}
In practice, vertex corrections are most often treated in a non-self-consistent or stochastic manner~\cite{vlcekStochasticVertexCorrections2019,Mejuto-Zaera::AreQPimpo::2021,Weng:EmbeddingVertex:2023}, which raises open questions regarding their quantitative accuracy, their dependence on the underlying mean-field reference, and the extent to which cancellations between vertex contributions in $\Sigma$ and $P$ may occur.~\cite{Shishkin07,wen_comparing_2024,shirleySelfconsistentGWHigherorder1996a,vanLeeuwen:PSD:electron_gas:2014,brunevalGWMiracleManyBody2021,Foster:G3W2dyn:2024,lewisVertexCorrectionsPolarizability2019}
To address these issues, we eliminate starting-point dependence by implementing both sc$GW$ and the self-energy-only sc$GW\Gamma_{\Sigma}$ approximations fully self-consistently. Previous work has shown that the advantages of $G_0W_0\Gamma$ can be offset by its sensitivity to the mean-field starting point reference, with fully self-consistent $GW$ often yielding comparable or improved results.~\cite{wen_comparing_2024,lewisVertexCorrectionsPolarizability2019,Foster:G3W2dyn:2024}
In the present work, we therefore assess the intrinsic impact of selected self-energy corrections while retaining RPA screening.

We consider a hierarchy of sc$GW\Gamma_{\Sigma}$ approximations, including second-order exchange (SOX)~\cite{Szabo_ostlund}, second-order screened exchange (SOSEX)~\cite{Rinke:SOSEX:2015,Forster:SOSEX:2022}, complete second-order screened exchange (2SOSEX)~\cite{Bruneval_2SOSEX_25}, and the $G3W2$ approximation~\cite{Foster:G3W2dyn:2024,Kutepov:scG3W2:2022}, with both static and dynamic variants.

A central computational bottleneck in these approaches is the evaluation of exchange-like diagrams in a self-consistent setting. While density fitting and related techniques reduce the cost of Coulomb terms~\cite{Whitten:integrals:73,Dunlap:DF:1979,RIpaper1,RI_auxbasis,Beebe:Cholesky:77,Koch:Cholesky:2003,Aquilante:Cholesky:2007,Koch:CholMethodSpec:2008,Aquilante:Cholesky:2009,Aquilante:Cholesky2:2009}, they do not eliminate the effective four-index scaling associated with exchange contributions.
To address this limitation, we employ tensor hypercontraction (THC) with interpolative separable density fitting (ISDF)~\cite{Martinez:LS-THC:2012,Lee:THC:2020,Matthews:THC:2020,Yang:ISDF:THC:2023,Lu:ISDF:2015,Lu:ISDF_Bloch:2016,LinLin:ISDF:2017,Yang:ISDF:THC:2023,Yeh:THC-RPA:2023}, enabling efficient evaluation of exchange-like self-energies within a finite-temperature sc$GW$ framework.~\cite{Yeh:THC-GW:2024,Yeh:GPU:GW:2022,Yeh:X2C:GW:2022,Green:2025}
This framework provides a unified and computationally efficient backbone for all sc$GW\Gamma_{\Sigma}$ variants considered in this work.~\cite{Pokhilko:THC-GWSOX:2024,Pokhilko:THC-G3W2:2025}

Using these approaches, we systematically benchmark sc$GW$ and sc$GW\Gamma_{\Sigma}$ for the charged excitation energies, including IPs and EAs of small molecules, providing rich benchmark data set and a controlled assessment of vertex effects.

The remainder of the paper is organized as follows: In Sec. II we introduce the theoretical framework; Sec. III describes the computational protocol; Sec. IV presents results and discussion; and Sec. V summarizes our conclusions.

\section{Theory} 
\subsection{Finite-temperature Green's functions}

We start by defining the contour-ordered Green's function, referencing Luttinger-Ward's work~\cite{Luttinger60}, which depends on two contour times $z$ and $z'$ (which later will become real time or imaginary time). Eventually, these will be taken either along the real‐time Keldysh contour or the imaginary‐time Matsubara contour. Denoted by the annihilation~$a_p(z)$ and creation~$a^\dagger_q(z')$ operators in the Heisenberg picture evolving along the contour under the full grand-canonical Hamiltonian:
\begin{equation}
    G_{pq}(z, z')=\epsilon(z, z')\langle\mathcal{P}\mathbf{(}{a_p(z)a^\dagger_q(z')\mathbf{)}}\rangle
    \label{eq:contour_GF}
\end{equation}
where $\mathcal{P}$ is the contour-ordering operator. The sign function $\epsilon(z, z')=\pm1$ enforces fermionic anti-commutation when exchanging the order of operators on the contour.

We work in the grand-canonical ensemble at inverse temperature $\beta=\frac{1}{k_BT}$ and chemical potential $\mu$, using the grand-canonical Hamiltonian $\hat{H}-\mu \hat{N}$. 

Following the Luttinger-Ward formulation, we begin from the contour-ordered one-body Green's function $G_{pq}(z, z')$. Restricting the contour to the imaginary-time Matsubara branch $z\rightarrow\tau\in [0,\beta]$, the Heisenberg operators are defined as
\begin{align}
    a_p(\tau)&=e^{\tau(\hat{H}-\mu \hat{N})}pe^{-\tau(\hat{H}-\mu \hat{N})} \\
    a^\dagger_q(\tau)&=e^{\tau(\hat{H}-\mu \hat{N})}q^\dagger e^{-\tau(\hat{H}-\mu \hat{N})}
    \label{eq:Heisenberg_operator}
\end{align}
The contour ordering operator $\mathcal{P}$ reduces to an ordinary imaginary-time ordering operator $\mathcal{T_{\tau}}$, and $\epsilon(\tau, \tau')$ imposes the anti-periodicity 
$G(\tau+\beta)=-G(\tau)$. Thus, the Matsubara Green's function is re-expressed as
\begin{equation}
    G_{pq}(\tau,\tau')=-\langle\mathcal{T}a_p(\tau)a_q^\dagger(\tau')\rangle
\label{eq:imaginary_contour_GF}
\end{equation}

Thermal expectation values are taken with the equilibrium density operator
\begin{equation}
    \rho=\frac{1}{\mathcal{Z}}e^{-\beta(\hat{H}-\mu \hat{N})}
    \label{eq:rho}    
\end{equation}
where $\mathcal{Z}$ is the partition function defined as
\begin{equation}
    \mathcal{Z}=\mathrm{Tr}[e^{-\beta(\hat{H}-\mu \hat{N})}]
    \label{eq:density_parition}
\end{equation}
Since the expectation value of any operator $\mathcal{O}$ can be evaluated from $\langle\mathcal{O}\rangle=\mathrm{Tr}[\rho\mathcal{O}]$, we can insert  Eq.~\eqref{eq:rho} into Eq.~\eqref{eq:imaginary_contour_GF} gives  
\begin{equation}
\begin{split}
    G_{pq}(\tau,\tau')&=-\mathrm{Tr}[\rho\mathcal{T}a_p(\tau)a_q^\dagger(\tau')]\\
    &=    -\frac{1}{\mathcal{Z}}\mathrm{Tr}[e^{-\beta(\hat{H}-\mu \hat{N})}\mathcal{T}a_p(\tau)a_q^\dagger(\tau')]
\end{split}
\label{eq:grand-canonical_GF}
\end{equation}

Since $\rho\propto e^{-\beta(\hat{H}-\mu \hat{N})}$ is a function of $\hat{H}-\mu \hat{N}$, it therefore commutes with $[\hat{H}-\mu \hat{N}, \rho] = 0$. Moreover, $\hat{H}-\mu \hat{N}$  is time independent, so the theory is invariant under a simultaneous shift of both imaginary times. 
Therefore, the Matsubara Green's function depends only on the imaginary-time difference   
\begin{equation}
    G_{pq}(\tau, \tau')=G_{pq}(\tau-\tau', 0)
    \label{eq:stationary_GF}
\end{equation}
By convention, we set the earlier time to zero $\tau'=0$, resulting in the following simplifications $\Delta\tau=\tau\ $ and consequently
\begin{equation}
  G_{pq}(\tau)\equiv G_{pq}(\tau,0)
    \label{eq:zero_GF}
\end{equation}

For $0<\tau<\beta$, time ordering is automatic (since $\tau>0)$, so
\begin{equation}
    G_{pq}(\tau)= G_{pq}(\tau,0)=-\frac{1}{\mathcal{Z}}\mathrm{Tr}[e^{-(\beta-\tau)(\hat{H}-\mu \hat{N})}a_pe^{-\tau(\hat{H}-\mu \hat{N})}a_q^\dagger]
    \label{eq:imaginary_GF}
\end{equation}
Eq.~\eqref{eq:imaginary_GF} is the standard Matsubara Green's function expressed in the interacting picture, where  $\hat{H}_0=\hat{H}-\mu \hat{N}$. One then performs a Fourier transform between the imaginary time $\tau$ and fermionic Matsubara frequencies $\omega_n=\frac{(2n+1)\pi}{\beta}$ to obtain
\begin{equation}
    G_{pq}(i\omega_n)=\int_0^{\beta}d\tau G_{pq}(\tau)e^{i\omega_n\tau}
    \label{eq:ft_GF}
\end{equation}

Given the Matsubara Green's function definition
\begin{equation}
    G_{pq}(\tau)=-\langle\mathcal{T}a_p(\tau)a_q^\dagger(0)\rangle
\end{equation}
the one-body (correlated) density matrix is obtained from the equal-time limit of the imaginary-time propagator, where $\tau=0^-$. This results in 
$G_{pq}(\tau=0^-)=-\langle\mathcal{T}a_p(\tau= 0^-)a_q^\dagger(0)\rangle=-\langle -(a_q^\dagger(0) a_p(\tau=0^-))\rangle = \gamma_{pq}$. Using the anti-periodicity $G(\tau+\beta)=-G(\tau)$ and the fact that $G_{pq}(\tau=0^-)=\gamma_{pq}$
one finds the standard identity
\begin{equation}
    \gamma_{pq} = \langle a_q^\dagger a_p\rangle = G_{pq}(\tau=0^-) = -G_{pq}(\tau=\beta^-).
\label{eq:gamma_from_G}
\end{equation}

In the Green's function formalism, to express all the many-body interactions between electrons that arise beyond a chosen mean-field reference, we define a quantity called self-energy $\Sigma(\omega)$. This quantity is defined via the Dyson equation 
\begin{equation}
    \Sigma(i\omega)=G^{-1}_0(i\omega)-G^{-1}(i\omega)
\end{equation}
where $G_0(i\omega)$ is defined as a chosen reference Green's function (usually a non-interacting or mean-field Green's function), while $G(i\omega)$ is defined as the interacting Green's function that illustrates all the many-body effects present in a system of interest at a given level of theory.
Diagrammatically, as expressed in Ref.~\citenum{Luttinger60}, the self-energy is a sum of all one-particle irreducible diagrams. Since the evaluation of all such diagrams is impossible for even simple systems, finding approximations to such an expansion is necessary. 

\subsection{Hedin's equations and $GW$ approximation}
\begin{figure}[H]
    \centering
    \includegraphics[width=\linewidth]{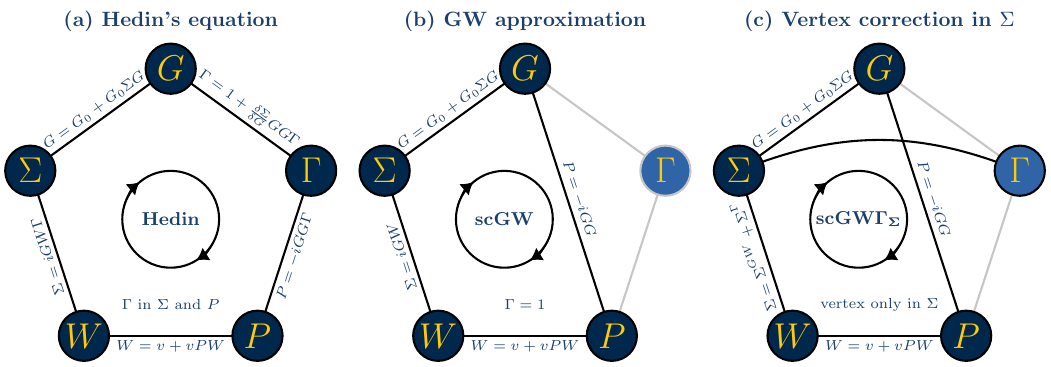}
    \caption{Schematic comparison of three self-consistent frameworks. (a) The full Hedin's equations, in which the vertex $\Gamma$ enters both the self-energy $\Sigma$ and irreducible polarizability $P$. (b) The sc$GW$ approximation, in which $\Gamma=1$, yielding the $GW$ self-energy and bubble polarizability. (c) The self-energy-only vertex-corrected approximation employed in this work, denoted sc$GW\Gamma_{\Sigma}$, in which selected exchange-type vertex corrections are included only in $\Sigma$, while $P$ retains the bubble form and generates RPA screening.}
    \label{fig:Hedin}
\end{figure}

Within the Hedin's framework~\cite{Hedin65}, the many-electron problem is recast into a closed, self-consistent cycle in which the screened Coulomb interaction $W$, the polarizability $P$, the self-energy $\Sigma$, the Green's function $G$, and the vertex function $\Gamma$ are mutually dependent. 
Figure~\ref{fig:Hedin} distinguishes the three levels of approximation relevant here. Panel~(a) shows the full Hedin's equations, denoted sc$GW\Gamma$, where the same vertex structure enters both $\Sigma$ and $P$. Panel~(b) shows the sc$GW$ approximation, where $\Gamma=1$ and the polarization is the RPA bubble form. Panel~(c) shows the approximation employed in this work, denoted sc$GW\Gamma_{\Sigma}$, where selected vertex corrections are inserted only into the self-energy, while the polarization remains at the sc$GW$ bubble level. Although $P$ and $W$ are updated through the interacting $G$ during self-consistency, no explicit polarization vertex is included in sc$GW\Gamma_{\Sigma}$. This construction has been extensively reviewed in the literature.~\cite{GW_Aryasetiawan98,Onida02,Rinke_GW_review}

Here, we adopt a compact index notation to denote the state of real space-time $\mathbf{r}_1,t_1$ and spin $\sigma_1$ as $1\equiv(\mathbf{r}_1,\sigma_1,t_1)$. The bare Coulomb interaction $v$ is the instantaneous  repulsion between two charges that vanishes as $1/|\mathbf{r}-\mathbf{r'}|$.  The screened interaction $W$ is written as 
\begin{equation}
    W(12) = v(12) + W(13)P(34)v(42)
    \label{eq:W_def}
\end{equation}
which states that screening dresses the bare Coulomb kernel by repeated particle-hole excitations encoded in polarizability $P$, where $W$ is a time/frequency dependent response function.~\cite{GW_Aryasetiawan98,Mahan00} 

The irreducible polarizability is defined as
\begin{equation}
    P(34) = iG(45)G(64)\Gamma(56;3)
    \label{eq:P_def}
\end{equation}
$P$ quantifies the density response to a change in the total potential $V$ and thus sets the dielectric properties of the system. The product of two Green;s functions describes the creation and annihilation of a particle-hole pair, while the vertex factor $\Gamma$ injects exchange-correlation corrections beyond an independent-particle picture. 

The vertex function is introduced as
\begin{equation}
\begin{split}
\label{eq:vertex}
    \Gamma(12;3) {}&\equiv \delta(12)\delta(13) + \frac{\delta \Sigma(12)}{\delta V(3)} \\
    &=\delta(13) \delta(23)+\frac{\delta \Sigma (12)}{\delta G(45)} G(46) G(75) \Gamma(67;3)
\end{split}
\end{equation}
and it collects all exchange-correlation corrections associated with the insertion of an external perturbation $V$.
This relation follows from the Dyson equation together with the chain rule, and shows that the vertex depends on the functional derivative $\delta\Sigma/\delta G$. When $\Gamma(12;3)=\delta(12)\delta(13)$, vertex corrections are neglected and the response reduces to the bare particle-hole bubble. An exact evaluation of $\Gamma$ is computationally demanding because it requires a four-point functional derivative and the self-consistent solution of a Bethe-Salpeter-like integral equation.~\cite{Bethe:Salpeter:1951,Onida02}

Within the $GW$ approximation, the vertex $\Gamma$ is truncated at the zeroth order, resulting in the vertex expression
\begin{equation}\label{eq:vertex=1}
    \Gamma(12;3) \approx \delta(12)\delta(13)
\end{equation}
that leads to neglecting vertex-mediated local-field and ladder corrections while retaining dynamic screening via $W$. These approximations are the base of practical working expressions and define the widely used $GW$ self-energy expression~\cite{G0W0_Pickett84,G0W0_Hybertsen86,GW_Aryasetiawan98}
\begin{equation}
\begin{split}
\label{eq: self-energy}
\Sigma(12) {}&= iW(13)G(14)\Gamma(42;3)\\
& \approx iG(12)W(12)
\end{split}
\end{equation}

When Eq. \eqref{eq:vertex=1} is invoked, the polarizability becomes the independent-particle bubble:
\begin{equation}
\begin{split}
\label{eq: polarizability}
P(12) {}&= -iG(23)G(42)\Gamma(34;1)\\
& \approx -iG(12)G(21)
\end{split}
\end{equation}
In periodic solids, this reduces to the Adler-Wiser form and yields the random-phase approximation (RPA) of the dielectric function,~\cite{Ren:RI-RPA:2012} when no self-consistency is invoked. In the self-consistent implementations ~\cite{scGW_Andrey09,Yeh:GPU:GW:2022,Kutepov:scGW:CrI3:2021}, the use of $\Phi$-derivable approximations ensures conservation laws.~\cite{Baym61, BaymKadanoff62}

\subsection{Self-energy-only vertex corrections: sc$GW\Gamma_{\Sigma}$} \label{sec:vertex}

\begin{figure}[H]
    \centering
    \includegraphics[width=\linewidth]{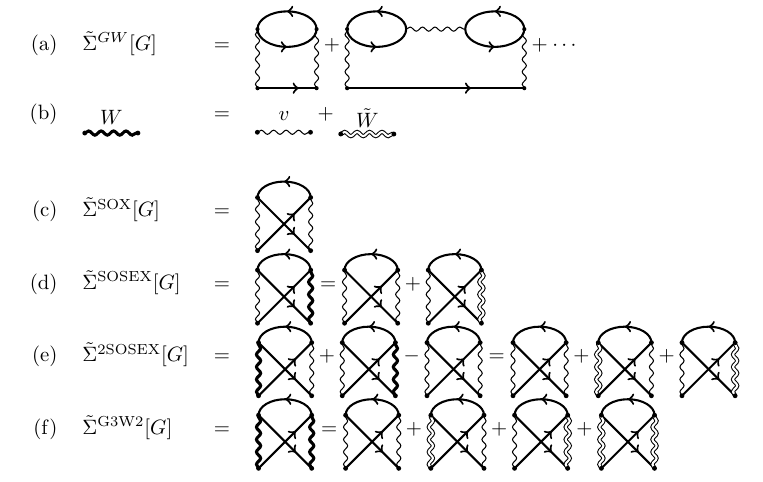}
    \caption{
    Skeleton Feynman diagrams for the self-energy approximations considered in this work. Panel (a) shows the $GW$ self-energy, and panel (b) shows the interaction decomposition introduced in Eq.~\eqref{eq:pol_W}. Panels (c)--(f) depict the vertex self-energies, $\tilde{\Sigma}^{\text{SOX}(U(1),U(2))}$, of SOX, SOSEX, 2SOSEX, and $G3W2$, respectively. For each vertex-corrected approximation, both the compact form in terms of the full screened interaction $W$ and the expanded form obtained by substituting Eq.~\eqref{eq:pol_W}---with terms written explicitly in $v$ and $\tilde{W}$---are shown. Bold wavy lines denote the full screened interaction $W$, while double-wavy lines denote the screening contribution $\tilde{W}$.
    }
    \label{fig:Feynman_diagrams}
\end{figure}

In this work, the vertex corrections are included only at the level of the self-energy. The polarization $P$ remains at the $GW$ approximation level used in Eq.~\eqref{eq:vertex=1}. Thus, no polarization vertex is applied in the construction of screened interaction $W$, and all beyond-$GW$ corrections enter exclusively through the vertex-corrected self-energy. However, since our $GW$ procedure is performed self-consistently, $P$ and $W$ are updated in the presence of vertex-corrected self-energy.

This approximation should be distinguished from a fully consistent vertex-corrected solution of Hedin's equations. In the exact theory, the same vertex structure that modifies the self-energy also enters the irreducible polarizability, and the two channels can partially cancel in one-particle excitation energies. In the present work, the screened interaction is generated from the RPA bubble form of $P$, although this bubble is evaluated self-consistently with the interacting Green's function obtained from the vertex-corrected self-energy. Thus, screening responds indirectly through the updated $G$, but no explicit polarization vertex is included.

This limitation is closely related to the diagrammatic analysis given in the Appendix of Ref.~\cite{Pokhilko:THC-G3W2:2025} There, we showed that the screened self-energy-only SOSEX, 2SOSEX, and $G3W2$ approximations are not $\Phi$-derivable when the corresponding vertex insertions into the polarizability are omitted. A fully dynamic self-energy-only $G3W2$ formulation and benchmark was also presented by Bruneval and F\"{o}rster.~\cite{Foster:G3W2dyn:2024} Diagrammatically, the same higher-order functional generates contributions to both the self-energy and the polarization; retaining only the self-energy branch therefore breaks the balanced cancellation expected in a fully consistent construction. Consequently, the trends reported below should be interpreted as the effect of selected exchange-type self-energy corrections on top of self-consistent RPA screening, rather than as the result of a complete vertex correction in Hedin's equations. This imbalance may contribute to the limited and nonuniform improvement over sc$GW$ observed for IPs and for the EA model quantities.

To make this separation explicit, we decompose the self-energy into a static and a dynamical contribution,
\begin{equation}
\Sigma[G](i\omega_n)=\Sigma_{\infty}[G]+\tilde{\Sigma}[G](i\omega_n)
\label{eq:full_se}
\end{equation}
where $\Sigma_{\infty}[G]$ is frequency independent and $\tilde{\Sigma}[G](i\omega_n)$ carries the Matsubara-frequency dependence. Transforming the imaginary-time Dyson equation to Matsubara frequencies and working in a possibly non-orthogonal atomic orbital (AO) basis yields the matrix Dyson relation  
\begin{align} 
    [G(i\omega_n)]^{-1}&=(i\omega_n+\mu)S-H_0-\Sigma(i\omega_n)\\
    &\equiv [G_0(i\omega_n)]^{-1}-\Sigma(i\omega_n)
\end{align}  
with  
\begin{equation}  
[G_0(i\omega_n)]^{-1}=(i\omega_n+\mu)S-H_0
\end{equation}
Here, $S$ denotes the overlap matrix in the non-orthogonal atomic-orbital basis.

To introduce screening in the exchange diagrams, we separate the screened interaction $W$ into an instantaneous bare part $v$ and a purely polarizable $\tilde{W}$ that is frequency-dependent. Consistent with Eq.~\eqref{eq:W_def}, the screened interaction admits the standard polarization expansion in the frequency space, where $\Omega_m=\frac{2m\pi}{\beta}$ denotes a bosonic Matsubara frequency,
\begin{align}
    W(\Omega_m)&=v+vP(\Omega_m)v+vP(\Omega_m)vP(\Omega_m)v+\cdots \\
    &\equiv v+\tilde{W}(\Omega_m)
\label{eq:pol_W}
\end{align}
and $\tilde{W}(\Omega_m)\equiv W(\Omega_m)-v$ collects all terms containing at least one polarization insertion (i.e., all dynamical screening beyond the bare interaction).

Figure~\ref{fig:Feynman_diagrams} summarizes the hierarchy of self-energy approximations studied in this work using skeleton Feynman diagrams. All vertex-corrected schemes are organized around the same second-order exchange (SOX) topology, which constitutes the minimal vertex insertion into the self-energy and restores the second-order Pauli exchange missing in conventional $GW$. More elaborate vertex corrections reuse the same SOX backbone while replacing one or both interaction lines according to Eq.~\eqref{eq:pol_W}, leading to the SOSEX/2SOSEX and $G3W2$ approximations depicted in the figure.

The corresponding general SOX contribution to the self-energy is
\begin{equation}
\begin{split}
    \tilde{\Sigma}^{\text{SOX}(U(1),U(2))}_{ij}(\tau_{ij})=-\sum_{klmnpq}U(1)_{(ip\vert km)}(\tau_{ip,km})\\\times U(2)_{(jl\vert qn)}(\tau_{jl,qn})G_{pq}(\tau_{pq})G_{kl}(\tau_{kl})G_{mn}(\tau_{mn})
\end{split}
    \label{eq:sox}
\end{equation}
where $U(1)$ and $U(2)$ can be the bare Coulomb interaction $v$ or screened Coulomb interaction $W$ and the imaginary-time differences are defined as $\tau_{xy}=\tau_x-\tau_y$. The most efficient implementation of Eq.~\eqref{eq:sox} depends on the structure of the interaction kernels, whether $U$ is bare or screened, and whether $W$ is treated statically or dynamically, which directly affects the optimal contraction order and computational scaling. Implementation details and algorithmic optimizations are discussed in Ref.~\citenum{Pokhilko:THC-G3W2:2025}.

We evaluate the vertex-corrected self-energy by augmenting the dynamical, frequency-dependent part of the $GW$ self-energy with an additive SOX-type vertex contribution in agreement with Figure.~\ref{fig:Feynman_diagrams}. Specifically, we define
\begin{equation}
    \tilde{\Sigma}^{GW\Gamma_{\Sigma}}[G](i\omega_n)
    = \tilde{\Sigma}^{GW}[G](i\omega_n) +\tilde{\Sigma}^{\text{SOX}(U(1),U(2))}[G](i\omega_n)
    \label{eq:gwg}
\end{equation}
where $\tilde{\Sigma}^{GW}[G](i\omega_n)$ is the $GW$ contribution and $\tilde{\Sigma}^{\text{SOX}(U(1),U(2))}[G](i\omega_n)$ denotes the chosen vertex correction built from the Green's function $G$ and the interaction kernels $U(1),U(2)$.

Importantly, the vertex correction is introduced only at the level of the dynamical self-energy $\tilde{\Sigma}[G](i\omega_n)$. Nevertheless, because the calculation is fully self-consistent, both the static $\Sigma_{\mathrm{\infty}}[G]$ and dynamic $\tilde{\Sigma}[G](i\omega_n)$ parts of self-energy are recomputed from the updated Green's function in each iteration.

When no vertex corrections are included, the $GW$ self-energy omits the second-order Pauli exchange. The SOX~\cite{Szabo_ostlund} approximation restores this contribution by choosing $U(1)=U(2)=v$ in Eq.~\eqref{eq:sox}, yielding a second-order, Fock-like exchange built entirely from the bare interaction. This SOX diagram provides a heavily truncated yet non-trivial approximation to the vertex function $\Gamma$.~\cite{MaromPRB2012, brunevalGWMiracleManyBody2021}

Second-order screened exchange (SOSEX)~\cite{Rinke:SOSEX:2015} retains the SOX topology but introduces screening by replacing one interaction line by $W$. Using the decomposition $W=v+\tilde{W}$, we define
\begin{equation}
    \tilde{\Sigma}^\text{SOSEX}_{ij}=\tilde{\Sigma}^{\text{SOX}(v,W)}_{ij} = \tilde{\Sigma}^{\text{SOX}(v,v)}_{ij} + \tilde{\Sigma}^{\text{SOX}(v,\tilde{W})}_{ij}
    \label{eq:sosex}
\end{equation}
When the full frequency dependence of the screened interaction is retained $\tilde{W} \to \tilde{W}(i\Omega_m)$ the resulting approximation includes the time- or frequency-dependent character of electronic screening. We refer to this formulation as dynamic-SOSEX. In contrast, a commonly used simplification is to replace the screened interaction by its zero-frequency value $\tilde{W}\rightarrow \tilde{W}(i\Omega_m=0)$ thereby neglecting its dependence on the bosonic Matsubara frequency $\Omega_m$. This yields static-SOSEX, in which screening is treated as instantaneous. Thus, the distinction between the two variants is that dynamic-SOSEX retains the full dynamical response of the medium, whereas static-SOSEX keeps only the static, zero-frequency limit of that response. SOSEX is sometimes interpreted as an antisymmetrization correction to the Coulomb-hole contribution in $GW$.

The 2SOSEX approximation further extends this construction by adding the complementary time ordering in which the screened interaction appears on the other interaction line
\begin{equation}
    \tilde{\Sigma}^\text{2SOSEX}_{ij}=\tilde{\Sigma}^{\text{SOX}(v,v)}_{ij}+\tilde{\Sigma}^{\text{SOX}(v,\tilde{W})}_{ij}+\tilde{\Sigma}^{\text{SOX}(\tilde{W},v)}_{ij}
    \label{eq:2sosex}
\end{equation}
which, again, can be formulated either in a dynamic form using the dynamical $\tilde{W}(i\Omega_m)$ or in a static form using only $\tilde{W}(i\Omega_m=0)$, as illustrated in Figure.~\ref{fig:Feynman_diagrams}.

Finally, the $G3W2$ family~\cite{Foster:G3W2stat:2022} includes the full second order $W$, thereby collecting all exchange diagrams composed from three Green's functions and two screened interactions as
\begin{equation}
    \begin{split}
    &\tilde{\Sigma}^{G3W2}_{ij}=\tilde{\Sigma}^{\text{SOX}(W,W)}_{ij}\\&=\tilde{\Sigma}^{\text{SOX}(v,v)}_{ij}+\tilde{\Sigma}^{\text{SOX}(v,\tilde{W})}_{ij}+\tilde{\Sigma}^{\text{SOX}(\tilde{W},v)}_{ij}+\tilde{\Sigma}^{\text{SOX}(\tilde{W},\tilde{W})}_{ij}
    \end{split}
\label{eq:g3w2}
\end{equation}
This construction is closely tied to the vertex in Eq.~\eqref{eq:vertex}, which upon linearization can be written as
\begin{equation}
    \Gamma \approx 1+\frac{\delta\Sigma^{GW}}{\delta G}
\end{equation}
In a fully dynamical formulation, both screened interactions depend on the bosonic Matsubara frequency, so the evaluation of Eq.~\eqref{eq:g3w2} involves a double convolution over frequency variables. In a direct quadrature implementation, this leads to a computational cost scaling as $\mathcal{O}(n_{\mathrm{AO}}^5n_{\mathrm{\Omega}}^2)$, making dynamic-$G3W2$ considerably more demanding than its static counterpart. In the static approximation, both interaction lines are replaced by their zero-frequency limits, $\tilde{W}(i\Omega_m=0)$, which removes the dynamical frequency dependence and greatly simplifies the numerical treatment. Therefore, in this work we restrict $G3W2$ to the static limit.~\cite{Foster:G3W2stat:2022}

\subsection{Decomposition of two-electron Coulomb interactions}
The two-electron repulsion integral (ERIs) is defined as
\begin{equation}
\begin{split}
    v_{pqrs}&=(pq|rs) \\&= \iint d\mathbf{r_1} d\mathbf{r_2} \phi^{\dagger}_{p}(\mathbf{r_1})\phi_{q}(\mathbf{r_1})\frac{1}{|\mathbf{r_1}-\mathbf{r_2}|}\phi^{\dagger}_{r}(\mathbf{r_2})\phi_{s}(\mathbf{r_2})
\end{split}
\end{equation}
Here, $\phi_p(\mathbf r)$ denotes an AO basis function. Throughout this section, $\mathbf r$ refers to spatial coordinates. The spin dependence can be incorporated by treating the orbital labels $\{p,q,r,s\}$ as composite (space+spin) indices. This is consistent with our implementation, which allows for the introduction of the spin dependence in the THC representation when needed. Decomposition schemes of two-electron integrals are essential for reducing both memory and computational costs. The most common approaches include Cholesky decomposition \cite{Beebe:Cholesky:77, Koch:Cholesky:2003, Aquilante:Cholesky:2007, Koch:CholMethodSpec:2008, Aquilante:Cholesky:2009, Aquilante:Cholesky2:2009} and density fitting (DF) or resolution-of-identity (RI) \cite{Whitten:integrals:73, Dunlap:DF:1979, Eichkorn:RIbasis:95}.

In our original sc$GW$ implementation,~\cite{Yeh:GPU:GW:2022, Yeh:X2C:GW:2022, Green:2025} we employ standard three-index decompositions such as Cholesky decomposition~\cite{Beebe:Cholesky:77,Koch:Cholesky:2003,Aquilante:Cholesky:2007, Koch:CholMethodSpec:2008,Aquilante:Cholesky:2009,Aquilante:Cholesky2:2009} and DF/RI,~\cite{Whitten:integrals:73,Dunlap:DF:1979,Eichkorn:RIbasis:95} which approximate the ERIs as
\begin{equation}\label{eq:eri_df}
    (pq|rs) \approx \sum_{\mathcal{Q}}V_{pq}(\mathcal{Q}) V_{rs}(\mathcal{Q})
\end{equation}

By factorizing the ERIs into products of three-index intermediates, the DF approximation reduces storage from $\mathcal{O}(n_{\mathrm{AO}}^4)$ to $\mathcal{O}(n_{\mathcal{Q}} n_{\mathrm{AO}}^2)$, where $n_{\mathcal{Q}}$ is the number of auxiliary basis functions. DF reduces ERI storage but does not by itself reduce the scaling of the correlated exchange diagrams such as SOX, SOSEX or G3W2, which remain exchange-dominated.

THC~\cite{Martinez:LS-THC:2012, Martinez:LS-THC:3:2012} 
 provides a further reduction by factorizing the DF three-index tensors into products of low-rank matrices defined on a grid of interpolation points.
In the least-squares THC (LS-THC)~\cite{Martinez:LS-THC:2012} or interpolative separable density fitting (ISDF) framework,~\cite{Lu:ISDF:2015} the ERIs are approximated as
\begin{equation}
     (pq|rs) \approx \sum_{PQ}X^P_p X^P_q Z^{PQ} X^Q_r X^Q_s
\end{equation}
where $X_{p}^P = \phi_p(\mathbf{r}_P)$ is the collocation matrix of AO basis functions evaluated at a set of interpolation points $\mathbf{r}_P$,  and $Z_{PQ}$ is the effective interaction kernel defined on the interpolation grid.

The central idea is to approximate orbital product functions $\phi_p(\mathbf{r})\phi_q(\mathbf{r})$ by their values on a small set of real-space points, thereby sparsifying the product space while maintaining accuracy.  
The coefficients $Z_{PQ}$ are obtained by least-squares fitting of the DF three-index tensors $V_{pq}(\mathcal{Q})$  to the separable form
\begin{equation}
    V_{pq}(\mathcal{P}) \approx \sum_{Q} X_{pQ}\, X_{qQ}\, A_{PQ}
\end{equation}
with $A_{PQ}$ determined by minimizing the least-squares residual.
The THC kernel $Z^{P Q}= \sum_C A^C_{P}\, A^{C*}_{Q},$ is then assembled so that the resulting four-index tensor best reproduces the DF/RI ERIs.

Compared to DF, which requires storage of $\mathcal{O}(n_{\mathcal{Q}} n_{\mathrm{AO}}^2)$ three-index tensors, the THC/ISDF representation reduces the scaling to $\mathcal{O}(n_{\mathrm{AO}} n_{\mu} + n_{\mu}^2)$,  where $n_{\mu}$ is the number of interpolation points. We parameterize the THC rank using the interpolation-point ratio
\begin{equation}
    \alpha_{\mathrm{Ipts}} = \frac{n_\mu}{n_{\mathrm{AO}}}
    \label{eq:nipts_ratio}
\end{equation}  
This low-rank factorization drastically reduces both memory and computational costs, while retaining accuracy, making it especially powerful in large-scale calculations involving Green's function and correlated wave-function methods. 
In practice, we find that $\alpha_{\mathrm{Ipts}}\approx10$ yields converged total energies and self-energies for the systems considered (see Sec.~\ref{sec:thc_conv}), so the THC rank remains proportional to $n_{\mathrm{AO}}$.

THC lowers the computational cost of evaluating the exchange-type diagrams, which dominate the cost of self-consistent vertex schemes. 
In a conventional AO-based implementation, these diagrams are typically quintic in system size, evaluating SOX with DF remains exchange-dominated and scales as $\mathcal{O}(n^3_{k}n^5_{AO}n_{t})$, where $n_{t}$ is the number of Matsubara grid points, $n_k$ is the number of momentum point grid.

In this work, DF and THC play distinct roles. We used DF for the static part of the self-energy $\Sigma_{\infty}$ and THC for the dynamic part of the self-energy $\tilde{\Sigma}(i\Omega_m)$. 
In a series of previous publications, we demonstrated that the usage of THC for the static self-energy $\Sigma_{\infty}$  can lead to large errors if too few interpolation points are used.~\cite{Yeh:THC-RPA:2023,Yeh:THC-GW:2024} If THC is used only for the dynamical part of the self-energy, the errors are small with relatively few interpolation points.~\cite{Yeh:THC-RPA:2023,Yeh:THC-GW:2024,Pokhilko:THC-GWSOX:2024,Pokhilko:THC-G3W2:2025}

For the SOX terms evaluated here, the leading THC contractions fall into two regimes,
$\mathcal{O}(n_tn_\mu^{2}n_{\mathrm{AO}}^{2}n_k^{2})$ and $\mathcal{O}(n_tn_\mu^{2}n_{\mathrm{AO}}^{3}n_k)$ which replace the dense five-index exchange bottleneck by products over interpolation-point indices. In practice, this reduction in arithmetic intensity and intermediate-formation cost is the enabling step for self-consistent SOX-type vertex schemes at larger system sizes.
\subsection{Spectral Function and Analytical Continuation}
The Matsubara Green's function admits a Lehmann representation
\begin{equation}
    G_{pq}(i\omega_n)=\int d\omega\frac{A_{pq}(\omega)}{i\omega_n-\omega}
    \label{eq:Lehmann}
\end{equation}
which connects the imaginary-frequency Green's function to the spectral function. In practice, the real-frequency (retarded) Green's function is obtained from the Matsubara Green's function through analytical continuation,
\begin{equation}
    G_{pq}(i\omega_n)\xrightarrow[\mathrm{analytical\ continuation}]{\mathrm{Nevanlinna}} G_{pq}(\omega)
\end{equation}
where we employed the Nevanlinna analytical continuation (NAC) method.~\cite{Fei:Nevanlinna:2021,huangRobustAnalyticContinuation2023}

The spectral function is the imaginary part of the propagator: 
\begin{equation}
A(\omega)= -\frac{1}{\pi}\mathrm{Im}\left[\mathrm{Tr}[G(\omega)]\right]
\label{eq:spectral_from_retarded}
\end{equation}
The resulting total projected density of states, $A(\omega)$, exhibits peaks associated with one-electron removal and addition processes, corresponding to ionization potentials and electron affinities, respectively.

\section{Results and Discussion}
\subsection{Computational details}
We used \textsc{PySCF}~\cite{PYSCF:2015,PYSCF:2018,PYSCF:2020,PySCF:2026} (version 2.6.2, commit hash \texttt{b6bfced}) to obtain converged unrestricted Hartree-Fock (UHF) reference solutions, which served as starting points for all subsequent THC-sc$GW$ and THC-sc$GW\Gamma_{\Sigma}$ calculations. 
Within the same framework, we generated all one- and two-electron integrals and constructed dense Becke integration grids~\cite{Becke1988} used to initialize the THC decomposition and the subsequent pruning. 
To ensure that cross-study comparisons are free of basis-set bias, we matched our orbital basis choices to those of the original studies that reported the corresponding data set: cc-pVQZ~\cite{Dunning:ccpvxz:1989,Dunning:ccpvxz:LiNaBeMg,Dunning:ccpvxz:Al-Ar} was used for the 29-molecule (referred to as $G_0W_0\Gamma$29 in this work) set of Maggio \textit{et al.},~\cite{Maggio:GWVertexCorrected:2017}, and def2-TZVPP~\cite{Schafer:def2:1992,Schafer:def2:1994,Eichkorn:def2_aux:_1997,Weigend:def2_gaussian:2003,Weigend:def2_tq:2005} was used for the original $GW$100 benchmark set and its subsequent assessment of $GW$ approaches.~\cite{van_setten_gw100_2015,caruso_benchmark_2016} 

The def2-TZVPP basis set is not available as an all-electron basis for fifth-period elements and heavier, for which effective core potentials are required. Since the present work focuses on all-electron calculations, molecules containing fifth-period or heavier element were excluded from our data set. After this exclusion, a total of 93 molecules were retained in the $GW$100 benchmark set.

The choices of basis sets are consistent with previous works.~\cite{wen_comparing_2024} DF auxiliary bases for the static part of the self-energy $\Sigma_\infty[G]$ were generated via the augmented even-tempered-basis (ETB) procedure implemented in \textsc{PySCF} with $\beta_{\mathrm{ETB}}=2.0$, while THC interpolating vectors for the dynamic part, $\tilde{\Sigma}[G](i\omega_n)$, were constructed from dense Becke integration grids following the approach of Ref.~\cite{Lee:THC:2020} with the number of the THC interpolation points per AO, $\alpha_{\mathrm{Ipts}}$ of 10. 

The THC decomposition is implemented using a legacy version developed before the open-source release of the \textsc{CoQu\'i} (Correlated Quantum \'interface) code,~\cite{CoQui::Github} which performs the least-squares fitting of DF three-index tensors to the separable ISDF/LS-THC form.
All THC-sc$GW$ and THC-sc$GW\Gamma_{\Sigma}$ calculations were conducted with the inverse temperature of $\beta = 1000\ \mathrm{a.u.}^{-1}$~(corresponding to about 316 K). For imaginary-time/frequency resolution we used high-precision intermediate-representation (IR) grids from the \textsc{sparse-ir} package~\cite{Sparse-ir:grid:2023}, characterized by the IR cutoff parameter (grid number) $\Lambda$. Detail of grid choice are addressed in the Supporting Information.

To accelerate and stabilize the convergence of the Green's function calculations, we combined frequency-dependent commutator direct inversion in the iterative subspace (CDIIS)~\cite{Pokhilko:algs:2022} with simple linear damping across iterations. Detailed parameter choices for IR grid number $\Lambda$, CDIIS, and simple damping rate are included in the Supporting Information. 

Unless mentioned otherwise, all THC-sc$GW$ and THC-sc$GW\Gamma_{\Sigma}$ calculations were converged to a threshold of $10^{-7}$~ a.u. Some isolated cases were also reported with the relaxed convergence thresholds, which are noted in the Supporting Information. Data entries that failed to converge to below $10^{-4}$ a.u. with a reasonable computational cost were considered unconverged and are not reported. 

Spectral functions were obtained from the converged sc$GW$ and sc$GW\Gamma_{\Sigma}$ Green's functions using the Nevanlinna analytical continuation (NAC)~\cite{Fei:Nevanlinna:2021,huangRobustAnalyticContinuation2023}. Unless specified, NAC employed 30{,}001 real-frequency points uniformly spanning $[-2,2]$~a.u. and a broadening parameter of $\eta=0.005$.
We report the quasiparticle peak positions directly; consequently, the occupied and unoccupied quasiparticle energies correspond to the negatives of the conventional IPs and EAs, respectively. The first IP was identified as the position of the first peak of $A(\omega)$ on the negative-frequency, electron-removal side of the spectrum, while the first EA was identified as the position of the first peak on the positive-frequency, electron-addition side.

\subsection{Convergence with respect to THC decomposition}\label{sec:thc_conv}

We assessed the convergence of the THC decomposition using the $G_0W_0\Gamma$29 data set of Ref.~\cite{Maggio:GWVertexCorrected:2017} at the THC-sc$GW$/cc-pVQZ level. Convergence is monitored with respect to the interpolation point ratio in Eq.~\eqref{eq:nipts_ratio}. To monitor the convergence with respect to the interpolation point ratio, we report the total energy in terms of a one-body static (Fock-like) reference contribution and a dynamical two-body correction,
$E_\text{tot}=E^\text{stat}_{\text{1b}}+E^\text{dyn}_\text{2b}$. The static contribution collects the nuclear-repulsion energy and all frequency-independent components of the electronic energy, 
\begin{equation}
    E^\text{stat}_{\text{1b}}= E_\text{nuc}+\mathrm{Tr}[\gamma H_{0}]+\tfrac{1}{2}\mathrm{Tr}[\gamma\Sigma_{\infty}]
    \label{eq:1b_e}
\end{equation}
whereas the dynamical (two-body) contribution arises from the frequency-dependent self-energy and is evaluated from the Galitskii-Migdal expression, 
\begin{equation}
    E^\text{dyn}_\text{2b} = \frac{2}{\beta}\mathrm{Re}\sum_{n\geq 0}\mathrm{Tr}[G(i\omega_n)\tilde{\Sigma}^{T}(i\omega_n)].
\end{equation} Although $E^\text{stat}_{\text{1b}}$ has a Hartree-Fock-like form, it is not the conventional HF energy because the density matrix
$\gamma$ is obtained from the interacting Green's function using  Eq.~\eqref{eq:gamma_from_G}  rather than from an  uncorrelated HF density.

\begin{figure}[H]
    \centering
    \includegraphics[width=\linewidth]{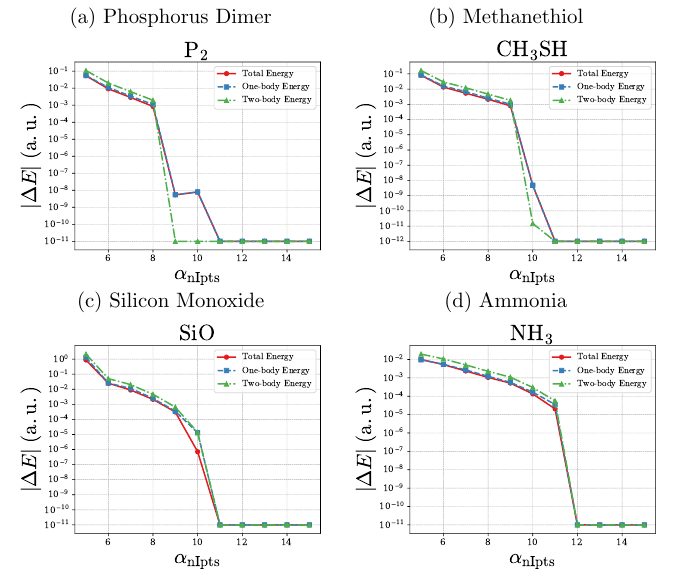}
    \caption{Convergence of $E_\text{tot}$, $E^\text{stat}_{\text{1b}}$ and $E^\text{dyn}_\text{2b}$ using the logarithmic scale, with respect to $\alpha_{\text{Ipts}}$ for selected molecules. The energy contributions are calculated at the THC-sc$GW$/cc-pVQZ level. For clarity, energy differences smaller than $10^{-12}$~a.u. are set to zero. All 29 molecules are reported in Supporting Information Figs.~S1 and S2.
    }
    \label{fig:energy_conv}
\end{figure}

In the post-HF procedures, we use THC only for the dynamical part of the self-energy. $\alpha_{\text{Ipts}}$ sets the THC rank through $n_{\mu}=\alpha_{\text{Ipts}} n_\text{AO}$ increasing both memory footprint and floating-point workload in the $GW$ and GW$\Gamma$ evaluation. Consequently, the improved accuracy obtained at larger $\alpha_\text{Ipts}n_\text{AO}$ comes at a correspondingly higher computational cost as the total energy approaches its THC convergence limit. 

In Figure~\ref{fig:energy_conv}, we illustrate how both parts of the energy, $E^\text{stat}_{1b}$ and $E^\text{dyn}_{2b}$, are affected by the number of the THC interpolation points despite only the dynamical part of the self-energy being approximated using THC. The $E^\text{stat}_{1b}$ is affected because the 1-body density matrix is evaluated from the correlated Green's function (Eq.~\eqref{eq:gamma_from_G}).  Both energy parts converge with respect to increasing $\alpha_{\mathrm{Ipts}}$ as illustrated for representative molecules drawn from the $G_0W_0\Gamma$29 data set. Over the range $\alpha_{\mathrm{Ipts}}\in[5,15]$, both the one-body and two-body energy contributions exhibit rapid convergence with negligible changes beyond a modest threshold. Quantitatively, for all representative systems, the deviation between $\alpha_{\mathrm{Ipts}}=10$ and the most tightly converged reference,($\alpha_{\mathrm{Ipts}}=15$) is well below is below $10^{-6}$~a.u for all representative cases for the energy in all representative cases.
For simplicity, energy differences smaller than $10^{-12}$~a.u. are set to zero, so the apparent vanishing of the error at $\alpha_{\mathrm{Ipts}}\geq 11$ reflects this numerical threshold rather than exact equality. Based on these observations, for most systems employed in this work,  unless otherwise noted, $\alpha_{\text{Ipts}}=10$ was used to achieve excellent convergence. This is consistent with the previous investigation of the total energy differences for intermolecular interactions\cite{Pokhilko:THC-GWSOX:2024} and magnetic interactions\cite{Pokhilko:THC-G3W2:2025}. 

\begin{figure}[H]
    \centering
    \includegraphics[width=\linewidth]{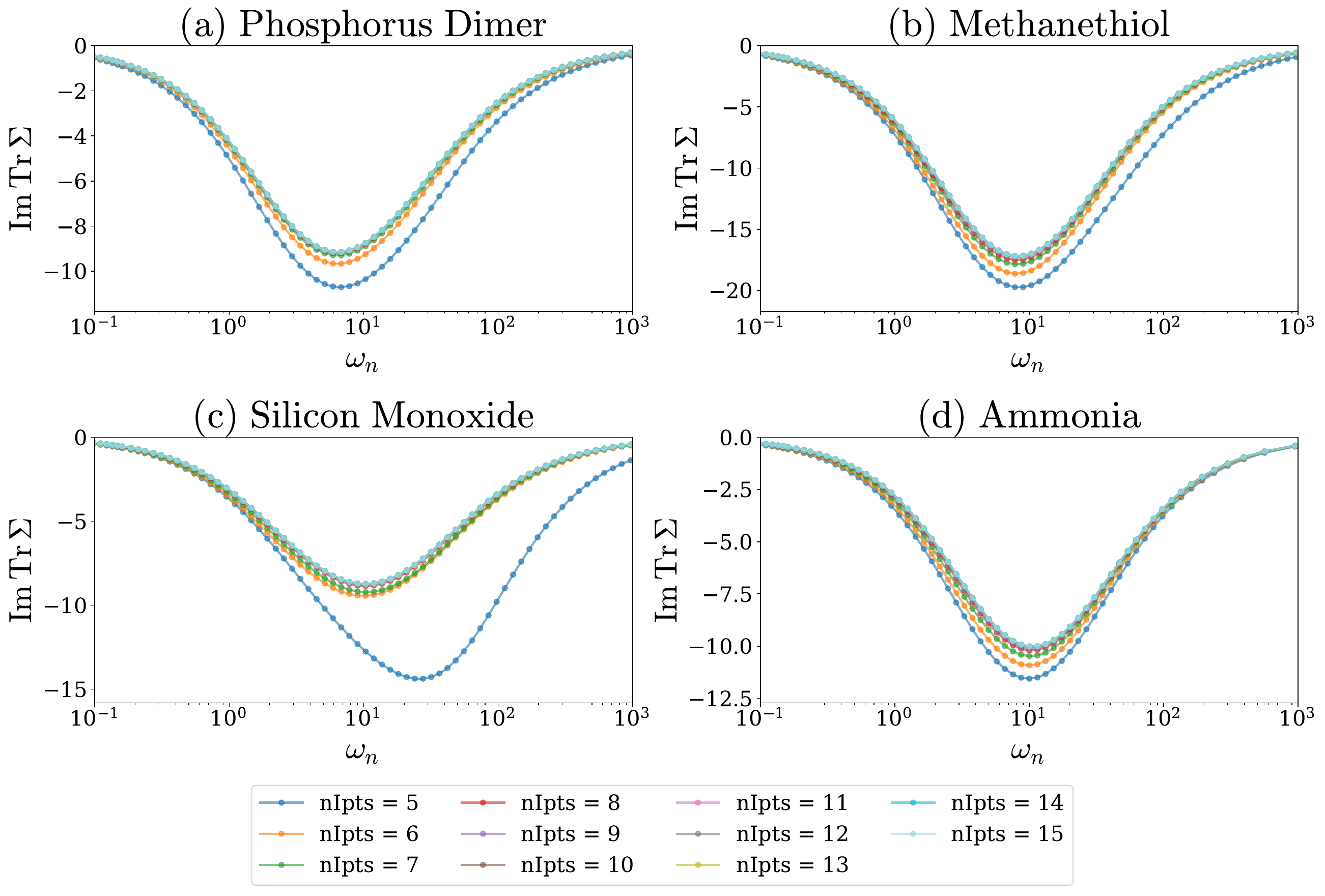}
    \caption{The convergence behavior of the trace of the imaginary part of the dynamic self-energy $\text{ImTr(}\Sigma(\omega_n))$  with respect to $\alpha_{\text{Ipts}}$ for selected molecules. Only the spin $\alpha$ component of the self-energy is shown. The calculations were performed at the THC-scGW /cc-pVQZ level.
    }
    \label{fig:dyn_self_energy_conv}
\end{figure}

In Figure~\ref{fig:dyn_self_energy_conv}, we show how the shape of the dynamic self-energy is controlled by the THC interpolation point ratio, $\alpha_{\mathrm{Ipts}}$. For very small $\alpha_{\mathrm{Ipts}}=5$ for silicon monoxide, the compression under-resolves the frequency dependence and even alters the shape of the self-energy. At moderate $\alpha_{\mathrm{Ipts}}$, the shape stabilizes and the residual error appears primarily as a nearly uniform amplitude offset, indicating that the dynamic structure is faithfully captured. By $\alpha_{\mathrm{Ipts}}=10$, the curves for all systems become visually indistinguishable on the plotted scale, achieving the same convergence behavior as for energies in Figure~\ref{fig:energy_conv}. 
\subsection{Evaluation of total energy and self-energy}
\begin{table*}[ht]
    \centering
    \resizebox{\textwidth}{!}{
    \begin{tabular}{ll|rrr|rrr|rrr|rrr}
    \hline
    \hline
    & Methods & \multicolumn{3}{c|}{Phosphorus Dimer} & \multicolumn{3}{c|}{Methanethiol} & \multicolumn{3}{c|}{Silicon Monoxide} & \multicolumn{3}{c}{Ammonia} \\
    \cline{3-14}
    &  & $E^\text{stat}_\text{1b}$ & $E^\text{dyn}_\text{2b}$ & $E_{\text{tot}}$ & $E^\text{stat}_{1b}$ & $E^\text{dyn}_{2b}$ & $E_{\text{tot}}$ & $E^\text{stat}_{1b}$ & $E^\text{dyn}_{2b}$ & $E_{\text{tot}}$ & $E^\text{stat}_{1b}$ & $E^\text{dyn}_{2b}$ & $E_{\text{tot}}$ \\
    \hline
    & THC-sc$GW$-SOX            &-681.3125&-0.4567&-681.7692&-437.5143&-0.5941&-438.1084&-363.6545&-0.4851&-364.1396&-56.0812&-0.3410&-56.4222\\
    & THC-sc$GW$-static-SOSEX   &-681.2480&-0.5826&-681.8307&-437.4416&-0.7375&-438.1792&-363.5857&-0.6205&-364.2062&-56.0388&-0.4253&-56.4641\\
    & THC-sc$GW$-dynamic-SOSEX  &-681.1995&-0.6651&-681.8646&-437.3814&-0.8376&-438.2189&-363.5414&-0.6911&-364.2326&-56.0070&-0.4768&-56.4838\\
    & THC-sc$GW$-static-$G3W2$  &-681.2066&-0.6635&-681.8701&-437.3909&-0.8376&-438.2285&-363.5429&-0.7047&-364.2477&-56.0100&-0.4826&-56.4927\\
    & THC-sc$GW$-static-2SOSEX  &-681.1822&-0.7104&-681.8927&-437.3682&-0.8820&-438.2503&-363.5159&-0.7573&-364.2732&-55.9965&-0.6083&-56.5060\\
    & THC-sc$GW$-dynamic-2SOSEX &-681.0924&-0.8641&-681.9565&-437.2533&-1.0737&-438.3270&-363.4334&-0.8896&-364.3229&-55.9358&-0.5095&-56.5441\\
    & THC-sc$GW$                &-681.1007&-0.8715&-681.9723&-437.2363&-1.1443&-438.3806&-363.4267&-0.9354&-364.3621&-55.9182&-0.6661&-56.5844\\
    \hline
    \hline
    \end{tabular}
    }
    \caption{Total energies (in a.u.) of selected molecules from the $G_0W_0\Gamma$29 data set produced by THC-sc$GW$ and THC-sc$GW\Gamma_\Sigma $/cc-pVQZ. A full list of energies from the $G_0W_0\Gamma$29 data set is presented in the Supporting Information..}
    \label{tab:vertex_total}
\end{table*}

In this section, we employ the $G_0W_0\Gamma$29 data set~\cite{Maggio:GWVertexCorrected:2017} to quantify how different vertex insertions in the self-energy affect the energy decomposition into $E^\text{stat}_{\text{1b}}$ and $E^\text{dyn}_\text{2b}$.
Throughout, THC-sc$GW$/cc-pVQZ with the stated orbital/auxiliary bases serves as the baseline, and we report the systematic shifts induced by vertex corrections in the self-energy. 
Table~\ref{tab:vertex_total} reports the total energies $E_{\mathrm{tot}}$, together with their decomposition into a static one-body energy $E^\text{stat}_\text{1b}$, as well as the dynamical two-body correlation energy $E^\text{dyn}_\text{2b}$ obtained from the Galitskii-Migdal energy expression.

Across all representative systems in Table~\ref{tab:vertex_total}, adding exchange-type vertex terms produces a well-ordered hierarchy of total energies relative to THC-sc$GW$: THC-sc$GW$-SOX $>$ THC-sc$GW$-SOSEX $>$ THC-sc$GW$-$G3W2$ $>$ THC-sc$GW$-2SOSEX $>$ THC-sc$GW$. 
This ordering is naturally rationalized using the effective-screening picture developed in our previous work,~\cite{Pokhilko:THC-G3W2:2025} where the screened interaction can be viewed qualitatively as $W \sim v/\epsilon$, with $\epsilon$ acting as an effective dielectric constant that weakens the bare interaction.

In this language, the SOX correction involves two unscreened interactions and therefore yields the largest exchange-type subtraction from the $GW$ self-energy, pushing the solution toward a weaker net dynamical self-energy and correspondingly higher (less negative) total energies. 
Replacing one of the bare interactions by $W$ (SOSEX) reduces the magnitude of this exchange correction, so SOSEX total energies fall between SOX and sc$GW$. When both interactions are screened (the $G3W2$-type exchange), the additional term is damped even further and the energies remain close to sc$GW$. 
Finally, the 2SOSEX approximation can be understood as an overcorrection of $G3W2$, since it can be seen as $G3W2$ minus SOX($\tilde{W}$, $\tilde{W}$), thereby further lowering the total energy and placing it below the $G3W2$-type exchange term, while remaining above sc$GW$ in the systems considered (however, Ref.~\cite{Pokhilko:THC-G3W2:2025} has an example for which the total energy of 2SOSEX is below scGW). 
Overall, the monotonic trends in Table~\ref{tab:vertex_total} support the interpretation that these vertex insertions tune the effective strength of dynamical correlations through screened (or partially screened) exchange.

\begin{figure}[H]
    \centering
    \includegraphics[width=\linewidth]{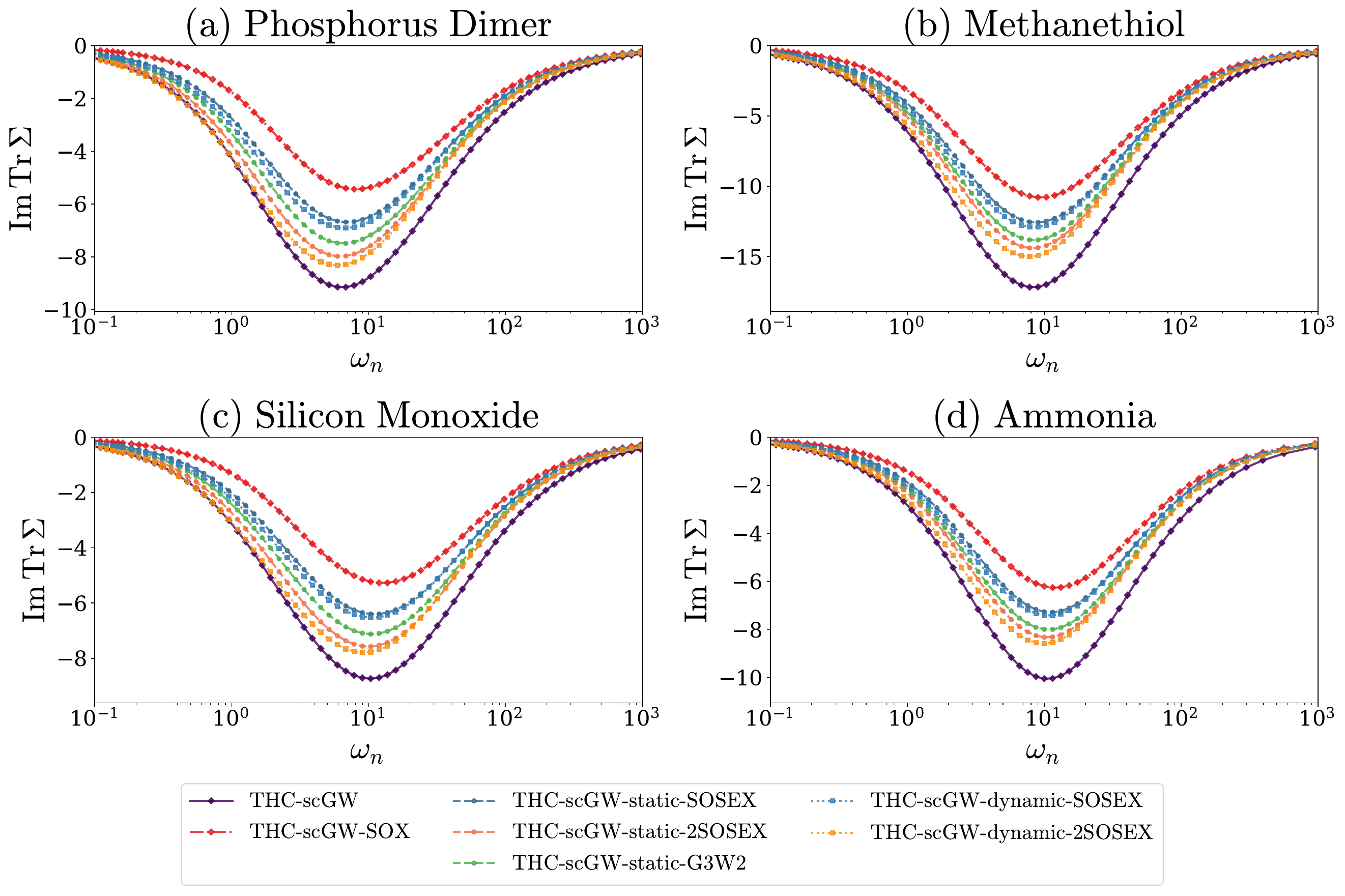}
    \caption{The trace of the imaginary part of the dynamic self-energy $\text{ImTr}[\Sigma(\omega_n)]$ as a function of frequency $\omega_n$ for various sc$GW\Gamma_\Sigma$ methods listed in the legend.} 
    \label{fig:vertex_se_imag}
\end{figure}

A useful way to rationalize the relative magnitude of static and dynamic terms is to view $\tilde{\Sigma}^{\text{SOX}(v,\tilde{W})}$ as a bosonic-frequency convolution. Schematically, one may write 
\begin{equation}
    \tilde{\Sigma}^{SOX(v,\tilde{W})}(\omega) = \sum_{\Omega} \tilde{W}(i\Omega)G(i\omega)G(i\omega)vG(i\Omega+i\omega)
\end{equation}
Since $\|\tilde{W}(i\Omega)\|$ typically decays with $|\Omega|$ and is maximal at $\Omega=0$, replacing $\tilde{W}(i\Omega_m)$ by its static value $\tilde{W}(0)$ often yields a larger screened-exchange contribution in norm. This heuristic concerns the size of the $\tilde{W}$-driven term itself; it does not imply a universal ordering for the distance of the full self-energy from a reference such as $\tilde{\Sigma}^{\mathrm{SOX}(v,v)}$, because the latter depends on cancellations between distinct frequency-dependent contributions and can differ between the real and imaginary parts.

\begin{figure}[H]
    \centering
    \includegraphics[width=\linewidth]{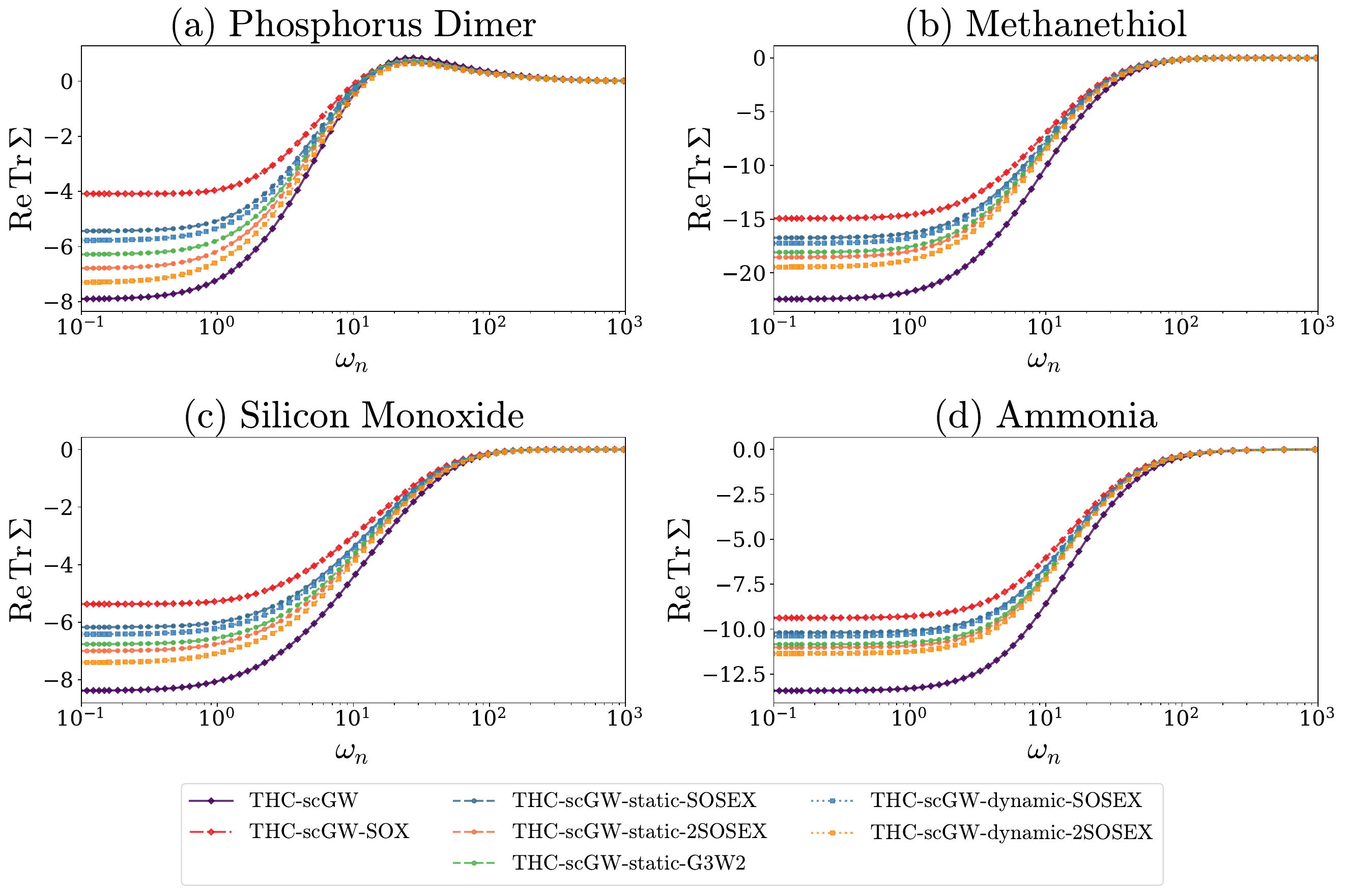}
    \caption{The trace of the real part of the dynamic self-energy $\text{ReTr}(\Sigma(\omega_n))$ as a function of frequency $\omega_n$ for various sc$GW\Gamma_\Sigma$ methods listed in the legend.}
    \label{fig:vertex_se_real}
\end{figure}

In Figures~\ref{fig:vertex_se_imag} and~\ref{fig:vertex_se_real}, we compare $\mathrm{ImTr}[\Sigma(i\omega_n)]$ and $\mathrm{ReTr}[\Sigma(i\omega_n)]$ across vertex variants for representative molecules.
Across all methods, the dominant effect is an approximately frequency-uniform vertical displacement of $\mathrm{Tr}[\Sigma(i\omega_n)]$ over a broad Matsubara range, while the curvature is only weakly modified.
Thus, to the leading order, the vertex terms renormalize the overall magnitude of dynamical correlations rather than introducing strong frequency-selective distortions.
Within these self-energy diagnostics, it can occur that the static SOSEX curve lies closer to SOX than the dynamical SOSEX curve over the Matsubara range shown.
This does not contradict the heuristic above: the dynamical $\tilde{W}(i\Omega)$ reweights the convolution over $\Omega$, which can change the degree of cancellation between $\tilde{\Sigma}^{\mathrm{SOX}(v,v)}$ and the $\tilde{W}$-driven contribution, moving the net curve either toward or away from SOX depending on frequency window and channel.

The Galitskii-Migdal total energy depends on frequency sums/integrals of matrix contractions involving both $G$ and $\Sigma$. Consequently, the phase/sign structure of the integrand, matrix-element weighting, and cancellations across frequency all matter. For this reason, absolute-value or norm heuristics about a part of $\Sigma$ do not reliably predict the ordering of correlation energies or total energies.

Empirically, Table~\ref{tab:vertex_total} shows that dynamical SOSEX yields systematically lower (more negative) $E_{\mathrm{tot}}$ than static SOSEX, with the same trend reflected in $E^\text{dyn}_{2b}$, while $E^\text{stat}_{1b}$ shifts in the opposite direction. This underscores that static/dynamical screening primarily redistributes correlation across bosonic frequencies; the GM energy integrates over that redistributed structure and can amplify or suppress cancellations differently than self-energy trace diagnostics or one-particle observables such as ionization potentials.

\subsection{First IP prediction: sc$GW$ and sc$GW\Gamma_{\Sigma}$ in $G_0W_0\Gamma$29 set}
\begin{table}[ht]
\resizebox{\linewidth}{!}{
    \centering
    \begin{tabular}{llrr}
    \hline
    \hline
    & Methods        & \multicolumn{1}{l}{MAE to $\Delta$CCSD(T)$^a$} & \multicolumn{1}{l}{MAE to experiments$^b$} \\
    \hline
    & HF-$\varepsilon^{HOMO}$            & 0.66($\pm$0.47)               & 0.78($\pm$0.50) \\
    & $G_0W_0$@HF$^c$& 0.49($\pm$0.28)& 0.65($\pm$0.36)\\ 
    &$G_0W_0\Gamma^{(NL)}$@HF$^d$ & 0.18($\pm$0.20) & 0.37($\pm$0.32) \\
    & DF-sc$GW$$^c$&  0.24($\pm$0.20)& 0.30($\pm$0.27)\\
    \hline
    &THC-sc$GW$     & 0.24($\pm$0.20)                          & 0.30($\pm$0.27) \\
    & THC-sc$GW$-static-2SOSEX  & 0.26($\pm$0.20)               & 0.32($\pm$0.24)                                   \\
    & THC-sc$GW$-dynamic-2SOSEX & 0.18($\pm$0.16)               & 0.29($\pm$0.28)                                   \\
    & THC-sc$GW$-static-$G3W2$  & 0.22($\pm$0.18)               & 0.30($\pm$0.25)                                   \\
    & THC-sc$GW$-static-SOSEX   & 0.25($\pm$0.18)               & 0.36($\pm$0.24)                                   \\
    & THC-sc$GW$-dynamic-SOSEX  & 0.25($\pm$0.20)               & 0.39($\pm$0.26)                                  \\
    & THC-sc$GW$-SOX            & 0.38($\pm$0.34)               & 0.54($\pm$0.32)                                   \\
    \hline
    \hline
    \end{tabular}
    }
    \caption{Mean absolute errors (MAEs) listed in eV  from DF-sc$GW$, THC-sc$GW$, and THC-sc$GW\Gamma_{\Sigma}$ relative to $\Delta$CCSD(T)$^a$ and experiment$^b$ evaluated for the first ionization potentials (IPs) in the $G_0W_0\Gamma$29 data set with the cc-pVQZ basis.  Values in parentheses are standard deviations of the absolute errors. $^a$Benchmarks from Bruneval \textit{et al.}~\cite{brunevalBenchmarkingStartingPoints2013} $^b$Experimental data compiled by Maggio \textit{et al.}~\cite{Maggio:GWVertexCorrected:2017} See the Supporting Information for the full list of experimental references. $^c$Calculated from data reported in Ref.~\cite{wen_comparing_2024} $^d$Calculated with data reported in Ref.~\cite{Maggio:GWVertexCorrected:2017}.
    }
    \label{tab:MAE_29}

\end{table}

In Table~\ref{tab:MAE_29}, we report mean absolute errors (MAEs) in eV for the first ionization potentials (IPs) evaluated for systems listed in the $G_0W_0\Gamma$29 set~\cite{Maggio:GWVertexCorrected:2017}. MAEs were obtained from benchmarking against both $\Delta$CCSD(T)~\cite{brunevalBenchmarkingStartingPoints2013} and experiment~\cite{Maggio:GWVertexCorrected:2017}
As a reference, we first report the Hartree-Fock (HF) Koopmans estimates, which provide an unscreened baseline and yield comparatively large MAEs (0.66 eV vs $\Delta$CCSD(T) and 0.78 eV vs experiment). We then include one-shot $G_0W_0$@HF to illustrate the impact of screening at the perturbative level and highlight the inherent starting-point dependence of $G_0W_0$ . A quantitative discussion of this starting-point dependence is given in Ref.~\cite{wen_comparing_2024}. 

Having established these reference points, we next turn to the fully self-consistent $GW$ baseline. DF-sc$GW$ achieves MAEs of 0.24 eV relative to $\Delta$CCSD(T) and 0.30 eV relative to experiment, and THC-sc$GW$ reproduces these values essentially exactly, including the dispersion. This near identity confirms that the THC factorization does not introduce a systematic bias in sc$GW$ IPs at the present accuracy target, thereby providing a controlled foundation for assessing vertex effects beyond sc$GW$. 

\begin{figure}[ht]
    \centering
    \includegraphics[width=0.9\linewidth]{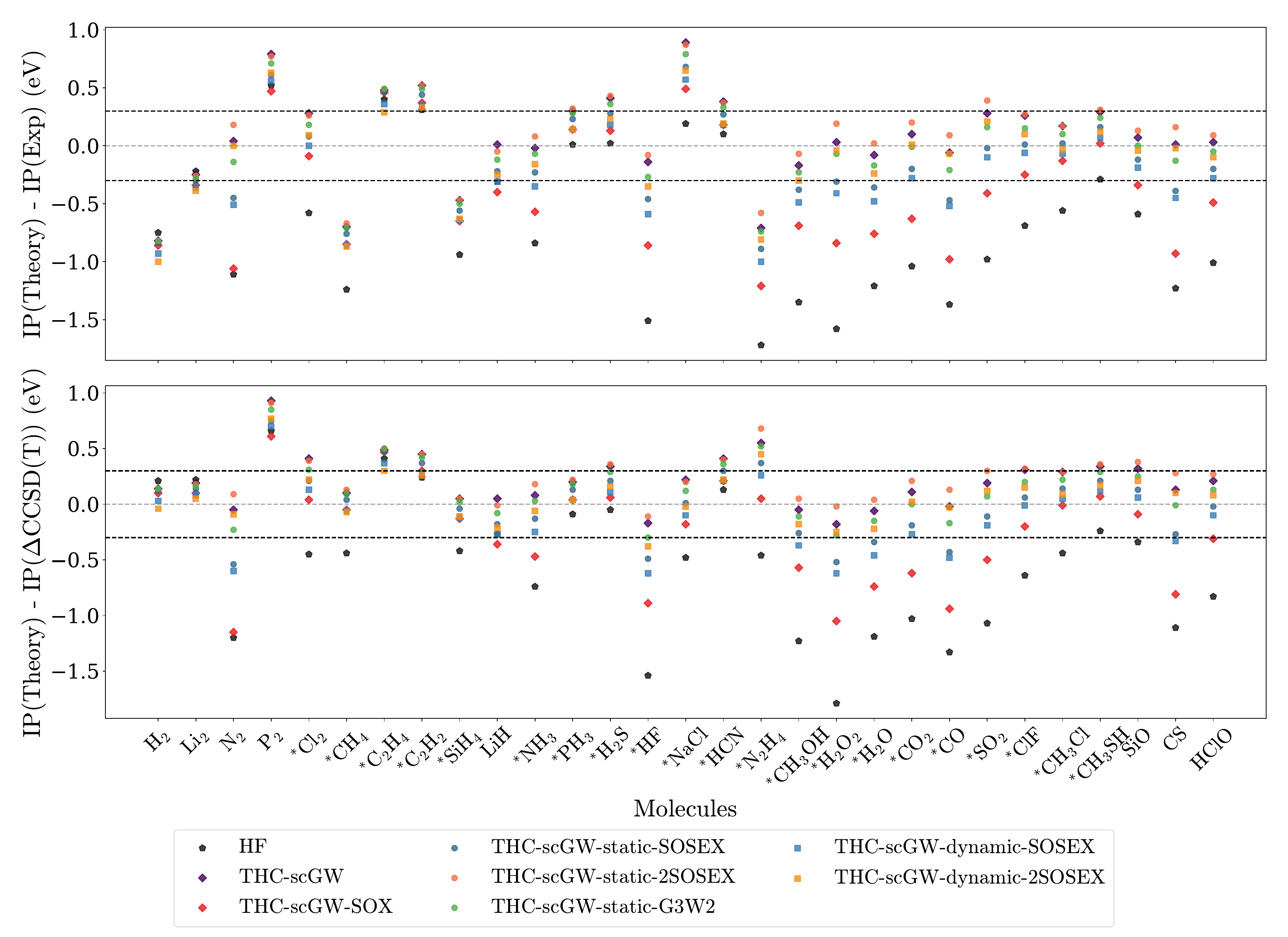}
    \caption{Upper panel: The first IP  signed errors in eV for the $G_0W_0\Gamma$29 data set evaluated relative to $\Delta$CCSD(T).
    Lower panel: The first IP  signed errors in eV for the $G_0W_0\Gamma$29 data set evaluated relative to experimental values.
   Comparisons in cc-pVQZ basis set are made for THC-sc$GW$ and all the vertex schemes discussed in this work. Experimental values marked * for respective molecular cases correspond to vertical ionization energies. 
   }
    \label{fig:All_IP_29}
\end{figure}

With this baseline in place, we finally assess self-consistent vertex corrections built on top of THC-sc$GW$. We interpret the performance of the self-consistent vertex-corrected variants in terms of the effective screening entering the vertex diagrams. In the notation introduced above, Eq.~\eqref{eq:pol_W}, makes clear that the largest vertex subtraction arises when both interactions are bare.
Accordingly, the SOX correction, $\tilde{\Sigma}^{\text{SOX}(v,v)}$, shows the largest degradation of accuracy, indicating that the strong vertex subtraction overcorrects the sc$GW$ baseline for many molecules in this set. 
Replacing one bare interaction by a screened one in Eq.~\eqref{eq:sosex}, SOSEX reduces the magnitude of the subtraction; accordingly, both static and dynamic SOSEX schemes partially mitigate the SOX shift but do not improve upon sc$GW$ on average. When both interactions are screened, as in $G3W2$, the vertex term is expected to be small and the results remain close to sc$GW$, consistent with the minor changes observed in the MAEs. Finally, 2SOSEX can exhibit more nuanced behavior because sufficiently strong screening may change the net sign of the vertex contribution; in the present benchmark, dynamic-2SOSEX is the only variant providing a modest net improvement over sc$GW$.

Figure~\ref{fig:All_IP_29} complements the aggregate statistics by showing molecule-resolved residuals, defined as $\mathrm{IP}(\text{method})-\mathrm{IP}(\text{ref})$, relative to $\Delta$CCSD(T) and experiment. 
Under this convention, positive (negative) residuals correspond to overestimation (underestimation) of the IP.
SOX produces the most pronounced systematic displacement away from sc$GW$, consistent with a large vertex subtraction when both interactions are unscreened and with the concomitant drift toward the HF (Koopmans) trend. SOSEX variants yield intermediate shifts, while $G3W2$ remains close to sc$GW$ as expected for a predominantly screened vertex term. The 2SOSEX schemes show the smallest net bias among the vertex-corrected methods, with dynamic-2SOSEX tracking sc$GW$ most closely for most molecules; moreover, because sufficiently strong screening can reverse the net sign of the 2SOSEX vertex contribution, the correction may add to (rather than subtract from) $\tilde{\Sigma}^{GW}$, explaining why static-2SOSEX can in some cases yield IPs larger than the corresponding sc$GW$ values. 

To assess when the static approximation is adequate, we compare, for each molecule and each scheme, the ionization potential obtained with the fully dynamical screened interaction, $\mathrm{IP}^{\mathrm{dynamic}}$, to its static counterpart, $\mathrm{IP}^{\mathrm{static}}$ (i.e., the ionization potential evaluated within the static approximation). We then quantify the static--dynamic discrepancy as
\begin{equation}
\delta_{\mathrm{SD}} \equiv \left|\mathrm{IP}^{\mathrm{dynamic}}-\mathrm{IP}^{\mathrm{static}}\right|
\end{equation}
Thus, $\delta_{\mathrm{SD}}$ directly measures the difference between the dynamical and static variations of the IP. Across the $G_0W_0\Gamma$29 set, we find $\langle\delta_{\mathrm{SD}}\rangle = 0.09 \pm 0.02$~eV for SOSEX and $0.19 \pm 0.03$~eV for 2SOSEX, consistent with the fact that 2SOSEX contains two $\tilde{W}$-dependent contributions whereas SOSEX contains only one. This provides a practical guideline: the static limit is typically adequate for SOSEX at the present accuracy scale, while dynamical effects are more consequential for 2SOSEX.

\begin{figure}[H]
    \centering
    \includegraphics[width=0.8\linewidth]{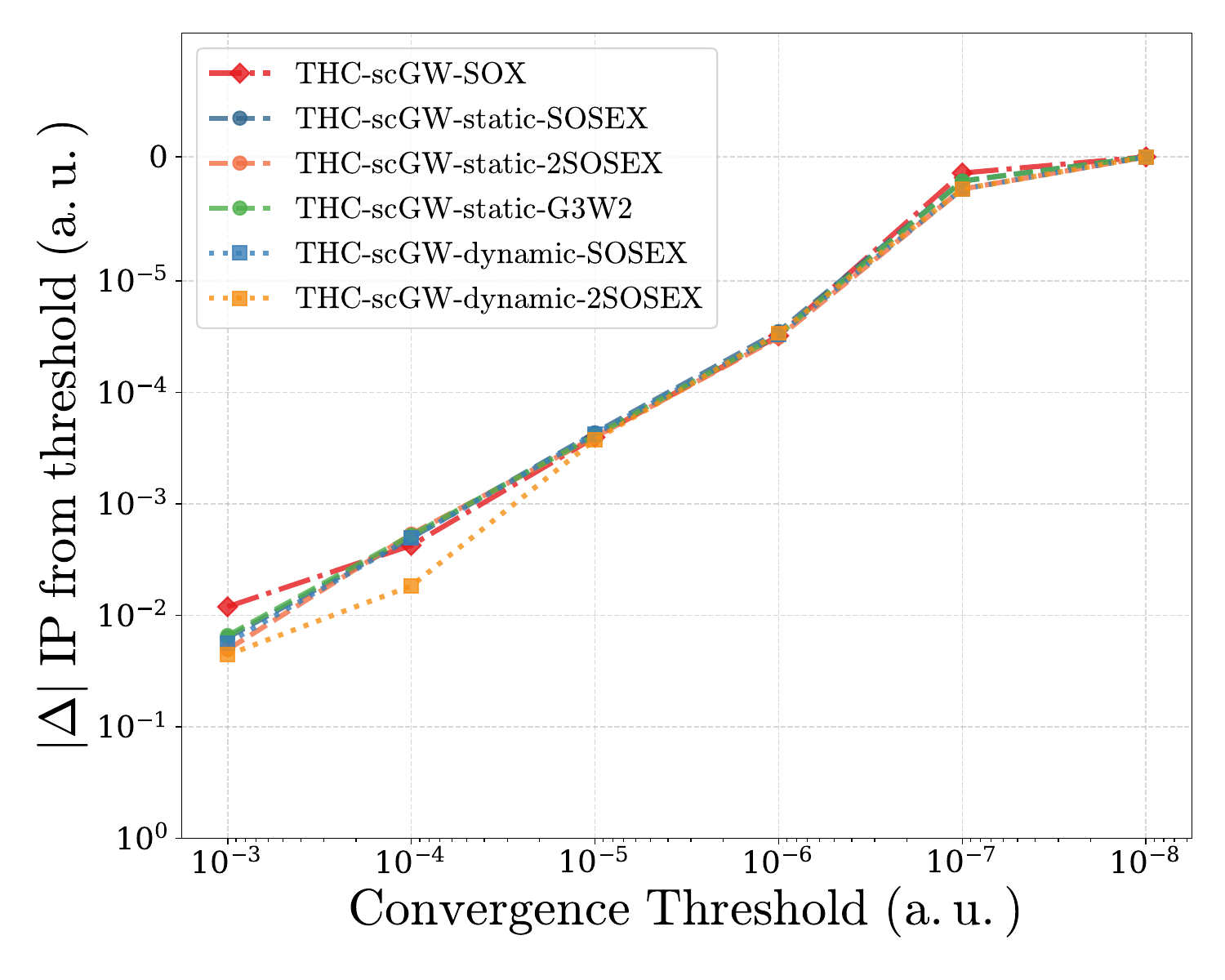}
    \caption{The convergence of the first IP value for water produced for all six THC-sc$GW\Gamma_{\Sigma}$ variants using different convergence thresholds. On the y-axis, we report the absolute difference of IP when compared with the one obtained with  $10^{-7}$ a.u convergence threshold. Both the x- and y-axes are in logarithmic scale.}
    \label{fig:thershold}
\end{figure}

\subsection{First IP prediction: sc$GW$ and sc$GW\Gamma_{\Sigma}$ in $GW$100 set}

We next evaluate THC-sc$GW$ and the THC-sc$GW\Gamma_{\Sigma}$ variants on the larger and chemically more diverse \textit{GW}100 benchmark set (def2-TZVPP). Beyond providing a more stringent test of transferability, $GW$100 data set allows us to assess whether the qualitative ordering of vertex corrections inferred above persists across chemically distinct families and to isolate the regimes in which static screening becomes unreliable. In what follows, we (i) justify the convergence protocol used for production runs, (ii) document numerical stability issues associated with SOX self-consistency and their manifestation as IR/sparse-sampling leakage, and (iii) analyze the method ordering and static-dynamic differences by chemical group, emphasizing the underlying screening physics.

The $GW$100 data set spans substantially larger one-particle spaces and polarizabilities than the $G_0W_0\Gamma29$ set, increasing the cost per self-consistency iteration and making numerical stability central. Since IPs are reported to two decimal places, a strict uniform $10^{-7}$~a.u. self-consistency threshold is unnecessarily strict for routine production. To balance wall-clock cost against the target IP precision, we thus adopt size-dependent thresholds: for systems with $250 \le N_{\mathrm{AO}} \le 300$ atomic orbitals we use $10^{-6}$~a.u., and for $N_{\mathrm{AO}} \ge 300$ we use $10^{-5}$~a.u. and dropped the dynamical screening, while retaining $10^{-7}$~a.u. for smaller systems. These criteria and exceptions are summarized in the Supporting Information.

We verified that relaxing the self-consistency threshold does not  affect the predicted ionization potentials (IPs). The reported IPs in eV, which are listed up to two decimal places, remain unchanged for the chosen convergence thresholds. In Figure.~\ref{fig:thershold} we compare the first IP of water computed with all six THC-sc$GW\Gamma_{\Sigma}$ variants using convergence thresholds from $10^{-2}$ to $10^{-7}$ a.u. For all the variants, the absolute deviations induced by looser thresholds are small. Once the threshold is tighter than $10^{-5}$ a.u. $\approx 2.72\times10^{-4}$ eV, each IP differs by less than 0.01 eV from the tightly converged results. Therefore, using $10^{-5}$--$10^{-6}$ a.u. thresholds preserves the reported two-decimal-place IP precision in the $GW$100 data set.

\begin{table}[ht!]
    \centering
    \begin{tabular}{llr}
    \hline
    \hline
    & Methods        & \multicolumn{1}{l}{MAE to $\Delta$CCSD(T)$^a$} \\
    \hline
    & $G_0W_0$@HF$^b$ & 0.35($\pm$0.22) \\
    &$G_0W_0\Gamma_0^{(1)}$@HF$^c$ & 0.54($\pm$0.29) \\
    & $G_0W_0$@PBE$^b$ & 0.62($\pm$0.29) \\
    & $G_0W_0\Gamma_0^{(1)}$@PBE$^c$ & 0.20($\pm$0.26) \\
    & DF-sc$GW$$^b$      & 0.29($\pm$0.22) \\
    \hline
    & THC-sc$GW$     & 0.29($\pm$0.22) \\
    & THC-sc$GW$-static-2SOSEX  & 0.30($\pm$0.21) \\
    & THC-sc$GW$-dynamic-2SOSEX & 0.22($\pm$0.17) \\
    & THC-sc$GW$-static-$G3W2$  & 0.27($\pm$0.20) \\
    & THC-sc$GW$-static-SOSEX   & 0.29($\pm$0.20) \\
    & THC-sc$GW$-dynamic-SOSEX  & 0.28($\pm$0.20) \\
    & THC-sc$GW$-SOX            & 0.35($\pm$0.30) \\
    \hline
    \hline
    \end{tabular}
     \caption{Mean absolute errors (MAEs) and standard deviations (SDs) of the absolute errors, in eV, for first IPs in the \textit{GW}100 set relative to $\Delta$CCSD(T) benchmarks (def2-TZVPP). $^a$The $\Delta$CCSD(T) benchmark data reported in Ref.~\cite{Krause2015} $^b$Calculated with data reported in Ref.~\cite{wen_comparing_2024}, $^cG_0W_0\Gamma_0^{(1)}$ data reported in Ref.~\cite{Rinke:G0W0Gamma0:2021}}
    \label{tab:MAE_GW100}
\end{table}

We represent imaginary-time and Matsubara-frequency quantities using the intermediate representation (IR) with sparse sampling~\cite{liSparseSamplingApproach2020}. To monitor numerical consistency during the self-consistent loop, we evaluate the \emph{spectral leakage} of the IR expansion after each forward/back Fourier transform, defined here as the ratio of the last to the first IR coefficient, following the convention used in CoQui.~\cite{CoQui::Github}
A growing leakage indicates that the current representation is failing to capture the short-time/high-frequency constraints implied by causality and moment structure. In practice, noticeable leakage is observed predominantly in SOX iterations, while the same IR grid remains well behaved for sc$GW$ and the screened vertex variants. This strongly suggests that the issue is not insufficient time/frequency resolution, but rather the intrinsic sensitivity of SOX-like self-consistency (qualitatively similar to the known problems of GF2-type iterations), in which an unscreened exchange-like correction amplifies small inconsistencies in the loop.

\begin{figure}[H]
    \centering
    \includegraphics[width=1.0\linewidth]{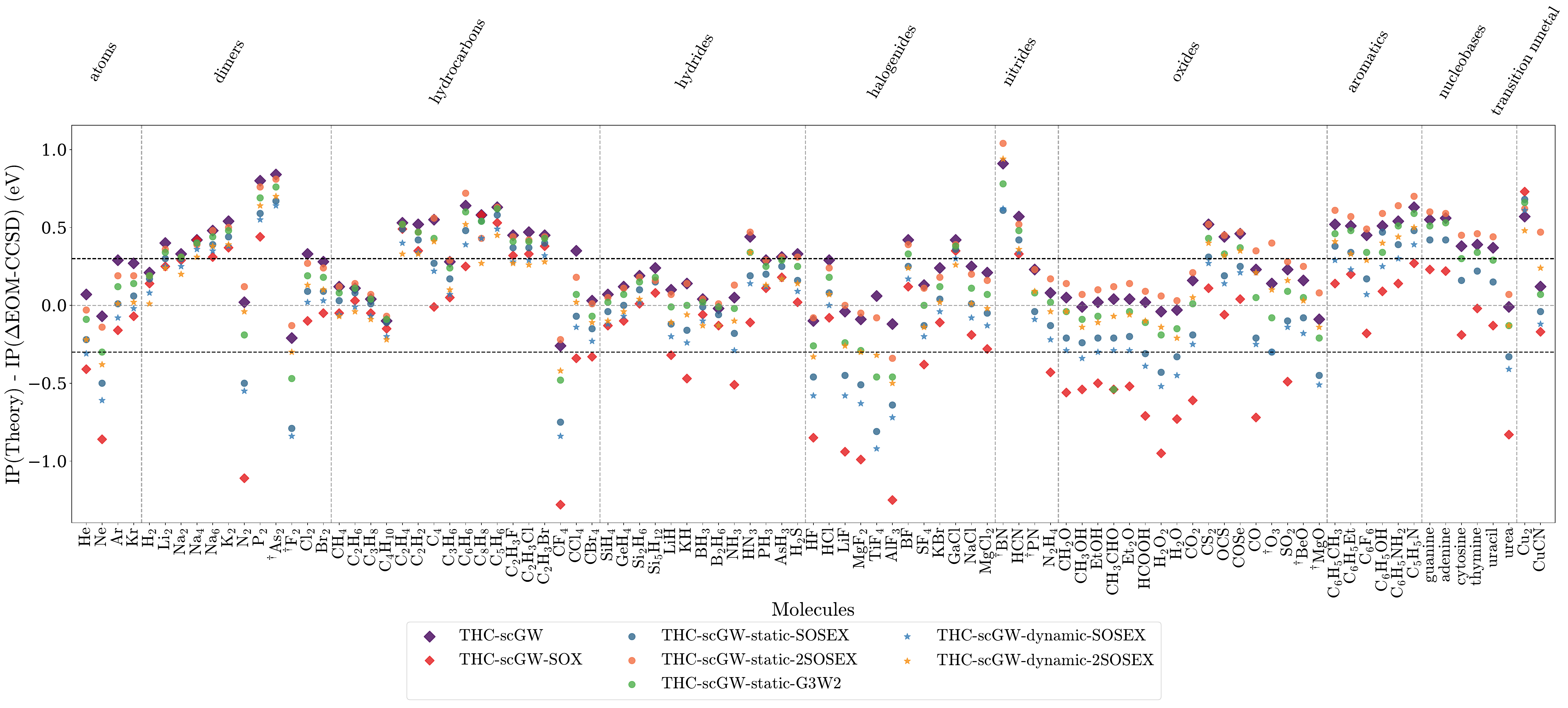}
    \caption{The first IP signed errors calculated by THC-sc$GW$ and six THC-sc$GW\Gamma_{\Sigma}$ variants for the $GW$100 set. All molecules are categorized into ten groups. The $\Delta$CCSD(T)~\cite{Krause2015} results were used as the benchmark reference. Dagger symbol $\dagger$ showcase the failure of convergence in SOX. The horizontal lines indicate errors of $\pm 0.3$ eV relative to the reference values}
    \label{fig:IP_GW_100}
\end{figure}

In Table~\ref{tab:MAE_GW100}, we summarize the mean absolute errors for the first IPs in the $GW$100 data set relative to $\Delta$CCSD(T) in the def2-TZVPP basis. As in $G_0W_0\Gamma29$, DF-sc$GW$ and THC-sc$GW$ yield indistinguishable accuracy, confirming that THC does not bias the self-consistent $GW$ baseline even for larger systems. More importantly, the qualitative ordering of vertex corrections follows the same screening logic discussed above. The SOX correction produces the largest exchange-like subtraction and thus the largest coherent displacement of IPs away from sc$GW$.

The $GW$100 data set also highlights the special role of 2SOSEX. Because 2SOSEX contains two exchange-like terms in which $\tilde{W}$ enters in both orderings, its static/dynamic discrepancy is naturally larger than for SOSEX and often roughly additive. Moreover, for sufficiently strong dielectric screening the net 2SOSEX correction can change sign, effectively corresponding to stronger screening than sc$GW$ and thereby reducing the IP magnitude. This mechanism explains why 2SOSEX can outperform SOSEX on average in large, polarizable systems, and why the dynamic variant gives the smallest IP MAE among the self-consistent variants in the present benchmark. Since the vertex is included only in the self-energy while the polarization remains at the RPA level, this improvement should be interpreted as a property- and benchmark-dependent result rather than as evidence for a generally balanced vertex-corrected treatment.

In Figure~\ref{fig:IP_GW_100}, we resolve the IP residuals on a molecule-by-molecule basis and organize the $GW$100 data set into ten chemical families. The dominant ordering of vertex schemes follows the screening-controlled hierarchy established above ($GW$-SOX $\rightarrow$ $GW$-SOSEX$\rightarrow$ $GW\text{-}G3W2$ $\rightarrow$ $GW$-2SOSEX $\rightarrow$ $GW$, with dynamic treatments typically producing smaller shifts than static ones). The remaining, family-specific deviations arise because different chemical environments reshape the effective screening entering the vertex: in particular, (i) the polarizability and spatial localization of the frontier orbitals modulate the magnitude of the $\tilde{W}$-driven exchange correction, and (ii) in systems with strongly structured low-frequency response (most prominently extended $\pi$ systems and heavy-element substitution), static screening can overweight $\tilde{W}(0)$ and thereby distort the balance between exchange-like subtraction and dynamical correlation. 

\begin{figure}[H]
    \centering
    \includegraphics[width=\linewidth]{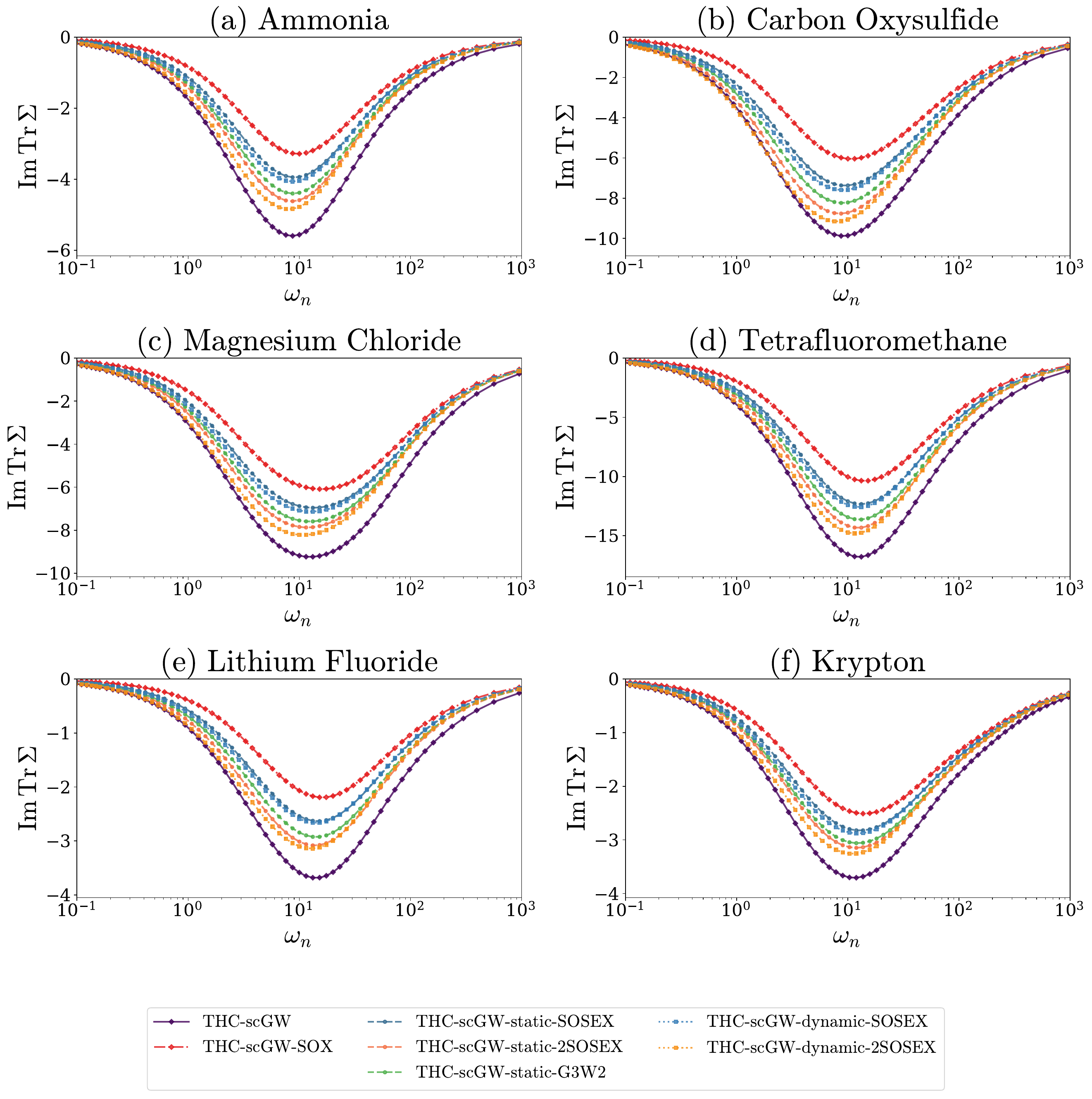}
    \caption{Imaginary part of the trace self-energy, $\mathrm{ImTr}[\Sigma(i\omega_n)]$, for six selected molecules: ammonia, carbon oxysulfide, magnesium chloride, tetrafluoromethane, lithium fluoride, and krypton. Across these representative systems, the imaginary part preserves a robust screening-controlled ordering over Matsubara frequency.}
    \label{fig:self-energy-imag-selected}
\end{figure}
To connect these IP-level trends to the underlying many-body quantities, we analyze the real and imaginary parts of the self-energy across chemical families. The screening hierarchy is most consistently reflected in the imaginary part: throughout the $GW$100 families considered, $\mathrm{ImTr}[\Sigma(i\omega_n)]$ preserves a stable ordering over Matsubara frequency, with the magnitude of the imaginary-part displacement decreasing along the same sequence above. This robustness is consistent with the strong causality constraints on $\mathrm{Im}\,\Sigma$ on the imaginary axis, which make it less susceptible to frequency-dependent cancellations than the real part. Representative imaginary-part curves for the six selected molecules are shown in Figure.~\ref{fig:self-energy-imag-selected}.

\begin{figure}[H]
    \centering
    \includegraphics[width=\linewidth]{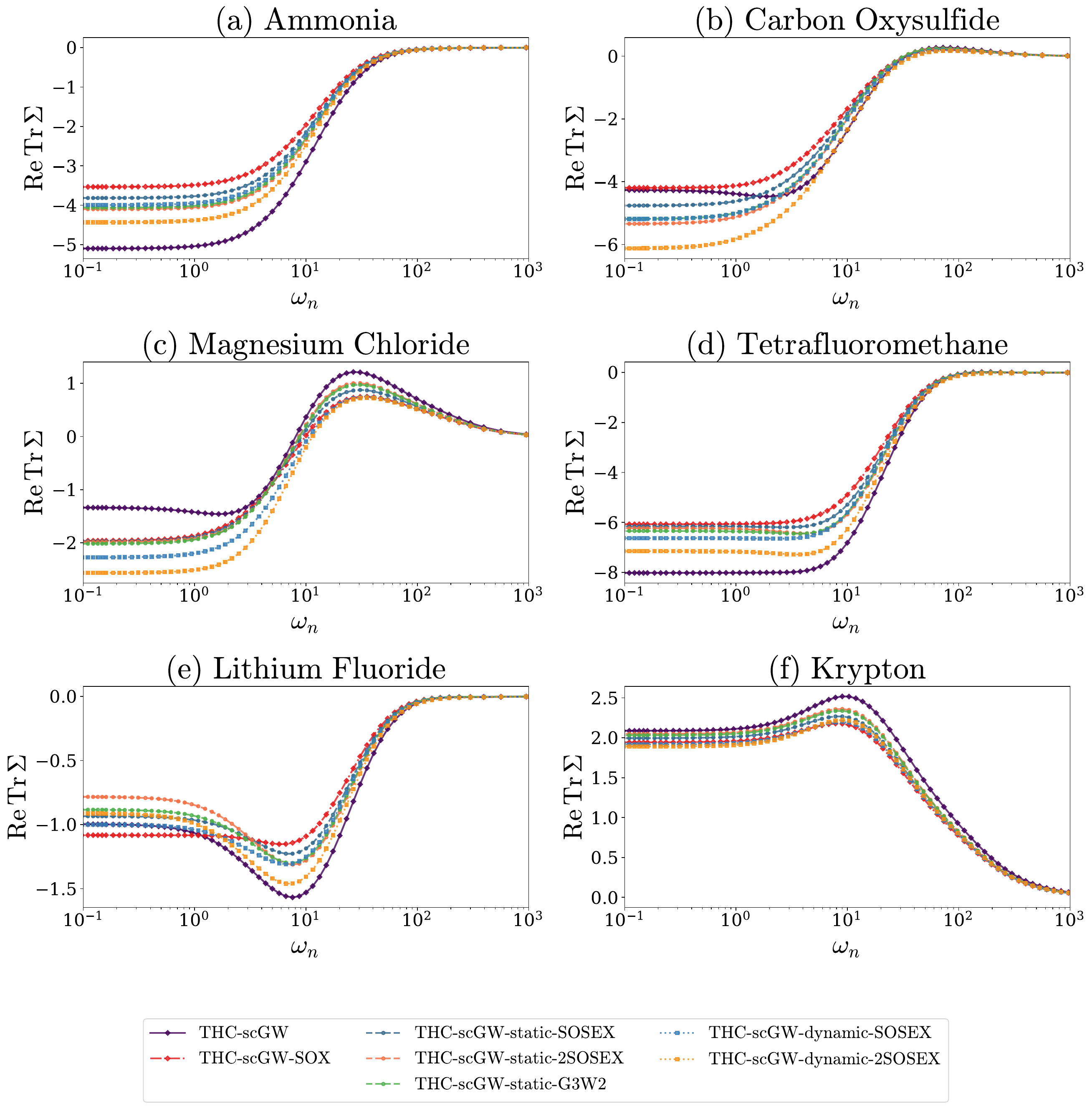}
    \caption{Real part of the trace self-energy, $\mathrm{ReTr}[\Sigma(i\omega_n)]$, for the same six selected molecules: ammonia, carbon oxysulfide, magnesium chloride, tetrafluoromethane, lithium fluoride, and krypton. In contrast to the imaginary part, the real part shows family-dependent low-frequency displacements, crossings, and curvature changes while recovering the usual ordering at larger Matsubara frequency.}
    \label{fig:self-energy-real-selected}
\end{figure}

For most molecules, the hierarchy of $\mathrm{ReTr}[\Sigma(i\omega_n)]$ closely tracks that of $\mathrm{ImTr}[\Sigma(i\omega_n)]$ across the full Matsubara range. In this dominant regime, the various vertex-corrected schemes primarily act as near-rigid vertical offsets relative to sc$GW$: the curves remain well separated and exhibit few, if any, crossings. This behavior is characteristic of hydrocarbon, oxide, and nitride families, where differences between static and dynamic variants are comparatively modest and the overall ordering is stable. The corresponding real-part curves for the same representative set are displayed in Figure.~\ref{fig:self-energy-real-selected}, which highlights both the dominant near-rigid-offset behavior and the low-frequency rearrangements discussed below.

A second regime, comprises systems in which $\mathrm{ReTr}[\Sigma(i\omega_n)]$ displays a method-dependent displacement-or even a crossing-at low frequencies, before recovering the usual ordering at larger frequencies. Within this class, three recurring patterns emerge. First, some molecules show a low-frequency displacement of sc$GW$ with the direction and magnitude of the shift depending systematically on the vertex construction. Pronounced SOX-driven shifts are typical for carbon sulfides (e.g., carbon oxysulfide and carbon disulfide) and for several chlorinated species, while phosphine trends toward $G3W2$-like behavior; depending on screening strength, analogous low-frequency displacements can also occur for SOSEX and 2SOSEX.

Second, in several chloride/ionic systems-including MgCl$_2$, NaCl, HCl, and KH-sc$GW$ can cross the SOX real-part at low frequency. This points to a particularly delicate cancellation in strongly ionic environments. The low-frequency real-part level shift is governed by competition between screened correlation, already present in $GW$, and an exchange-like subtraction that is amplified in SOX, with the cancellation most acute at small frequency.

Third, a distinct “static-dynamic splitting” relative to $G3W2$ is observed in selected cases, with static variants lying systematically above the $G3W2$ real-part curve and the corresponding dynamic variants lying below it over a low-to-intermediate Matsubara window. Tetrafluoromethane provides a clear example. This behavior is consistent with the convolution-based picture. When $|\tilde{W}(i\Omega_m)|$ is strongly peaked at $\Omega_m=0$, the static approximation effectively overweights the most strongly screened component, whereas the dynamic treatment samples a frequency-averaged interaction and therefore produces a reduced real-part shift.

Beyond these low-frequency reattachments to the “normal” hierarchy, one finds a third regime in which the real-part curve itself changes shape rather than undergoing a near-uniform offset. Several fluorides (e.g., MgF$_2$, LiF, AlF$_3$, HF) fall into this category. Essentially, in all screened vertex variants, $\mathrm{ReTr}[\Sigma(i\omega_n)]$ exhibits a pronounced curvature change, while SOX remains comparatively rigid. This contrast indicates that real-part renormalization in these strongly polar/ionic systems is controlled by structured, frequency-dependent screening to which screened vertex corrections are sensitive; the unscreened SOX contribution lacks this dynamical structure and therefore does not reproduce the same curvature.

Finally, in a small subset of systems, even the relative ordering of sc$GW$ and SOX in $\mathrm{ReTr}[\Sigma(i\omega_n)]$ can invert. Krypton is an example where the placement of sc$GW$ versus SOX in the real part deviates from the canonical pattern despite a stable hierarchy in $\mathrm{ImTr}[\Sigma(i\omega_n)]$. An even more extreme behavior appears in Na$_x$ clusters, where dynamical screened variants can decay rapidly in $\mathrm{ReTr}[\Sigma(i\omega_n)]$ and drop below SOX over part of the Matsubara range. Importantly, this fast-decay, “dropping-below-SOX” behavior is a hallmark of cluster/diatomic cases (e.g., Na$_x$, P$_2$) and highly screened molecules (including CF$_4$), where the real-part correction is dominated by low-frequency screening structure and enhanced cancellation effects.

These regimes clarify why ionization-potential orderings remain broadly screening-controlled-consistent with the imaginary part and with the high-frequency real-part tails-while family-specific exceptions arise in the low-frequency real part. Because quasiparticle level shifts  depend on the real part of the relevant orbital-resolved self-energy $\mathrm{Re}[\Sigma(i\omega_n)]$ , low-frequency displacements and crossings observed in the real part of traced self-energy $\mathrm{ReTr}\Sigma(i\omega_n)$ may help rationalize the family-dependent deviations in IP ordering observed in $GW$100, even in cases where $\mathrm{ImTr}[\Sigma(i\omega_n)]$ preserves a uniform hierarchy across chemical families.

\subsection{Cross-method benchmarks of charged excitations}
We benchmark first ionization potentials and electron affinities against EOM-CCSD.~\cite{LangeBerkelbach:EOMGW:2018} Unlike Table~\ref{tab:MAE_GW100}, which assesses IPs relative to $\Delta$CCSD(T)~\cite{Krause2015}, Table~\ref{tab:ea-gw100} uses EOM-CCSD as the common reference for both quantities. This permits a controlled comparison of one-shot $G_0W_0$~\cite{wen_comparing_2024} and $G_0W_0\Gamma_0^{(1)}$~\cite{Rinke:G0W0Gamma0:2021} with the self-consistent $GW$ and self-energy-corrected variants.

\begin{figure}[H]
    \centering
    \includegraphics[width=1.0\linewidth]{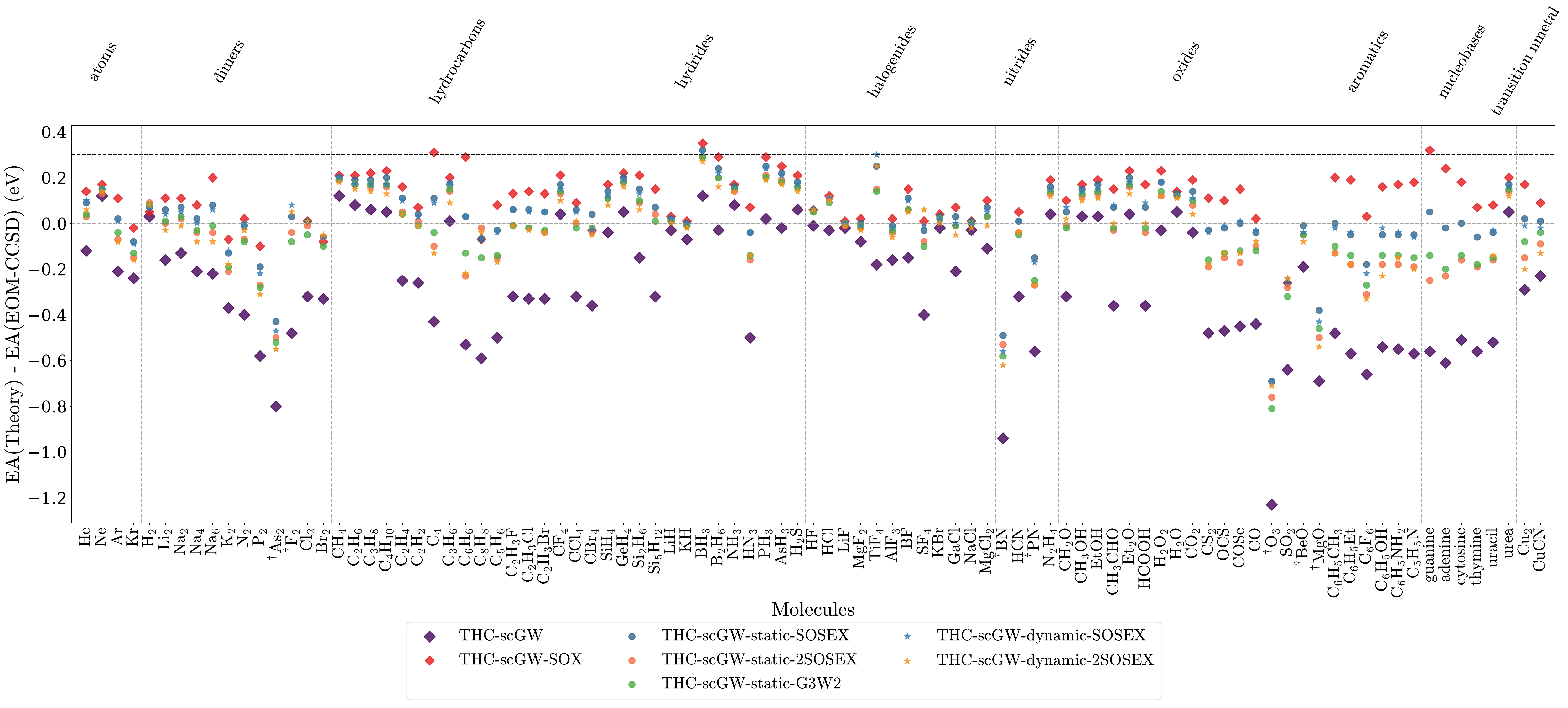}
    \caption{The first EA signed errors calculated by THC-sc$GW$ and six THC-sc$GW\Gamma_{\Sigma}$ variants for the $GW$100 set. All molecules are categorized into ten groups. The EOM-CCSD~\cite{LangeBerkelbach:EOMGW:2018} EA values results were used as the reference. Dagger symbol $\dagger$ showcase the failure of convergence in SOX. The horizontal lines indicate errors of $\pm 0.3$ eV relative to the reference values}
    \label{fig:EA_GW_100}
\end{figure}

This comparison requires an important qualification. For many molecules in this subset, no stable anionic state exists: the negative ion is unbound, and its physical characterization requires a continuum or resonance treatment.~\cite{Jagau:resonances:2017,Zuev:CAP-EOMCC:2014,Mukherjee:pyrrole-resonances:2022,Almeida:pyrimidine-DEA:2013} The reported discrete electron-addition states therefore have no direct connection to experimental electron affinities and should instead be viewed as model systems restricted by the basis set for examining the response of the different self-energy-only vertex corrections to addition of one electron. We do not form or analyze fundamental gaps from these data: combining a physical removal energy with an EA that does not correspond to a stable anionic state would not yield the true fundamental gap and could appear accurate for the wrong reason. Previous molecular $GW$ benchmarks have likewise shown that relative method performance can differ between IPs and nominal EAs.~\cite{knight_accurate_2016} The molecule-resolved IP errors are provided in the Supporting Information, while the corresponding EA errors are shown in Figure.~\ref{fig:EA_GW_100}.

\begin{table}[ht!]
    \centering
    \resizebox{\columnwidth}{!}{
    \begin{tabular}{llrrrr}
    \hline
    \hline
     & Methods & \multicolumn{1}{l}{IP MAE (SD)$^a$} & \multicolumn{1}{l}{IP MAPE (\%)$^a$} & \multicolumn{1}{l}{EA MAE (SD)$^a$} & \multicolumn{1}{l}{EA MAPE (\%)$^a$} \\
    \hline
     & $G_0W_0$@HF$^b$                       & 0.34($\pm$0.24) & 3.1 & 0.15($\pm$0.14) & 22.1 \\
     & $G_0W_0\Gamma_0^{(1)}$@HF$^c$         & 0.52($\pm$0.31) & 4.7 & 0.26($\pm$0.14) & 34.4 \\
     & $G_0W_0$@PBE$^b$                      & 0.64($\pm$0.30) & 5.9 & 0.26($\pm$0.17) & 48.3 \\
     & $G_0W_0\Gamma_0^{(1)}$@PBE$^c$        & 0.24($\pm$0.29) & 2.5 & 0.20($\pm$0.14) & 52.0 \\
     & DF-sc$GW$$^b$                         & 0.33($\pm$0.22) & 3.4 & 0.27($\pm$0.23) & 48.1 \\
    \hline
     & THC-sc$GW$                             & 0.33($\pm$0.22) & 3.5 & 0.28($\pm$0.24) & 51.0 \\
     & THC-sc$GW$-static-2SOSEX               & 0.33($\pm$0.21) & 3.4 & 0.13($\pm$0.12) & 20.9 \\
     & THC-sc$GW$-dynamic-2SOSEX              & 0.25($\pm$0.20) & 2.5 & 0.13($\pm$0.13) & 22.8 \\
     & THC-sc$GW$-static-$G3W2$               & 0.30($\pm$0.21) & 3.1 & 0.13($\pm$0.13) & 20.2 \\
     & THC-sc$GW$-static-SOSEX                & 0.30($\pm$0.21) & 3.1 & 0.11($\pm$0.11) & 16.0 \\
     & THC-sc$GW$-dynamic-SOSEX               & 0.30($\pm$0.23) & 2.9 & 0.11($\pm$0.12) & 17.2 \\
     & THC-sc$GW$-SOX                         & 0.36($\pm$0.33) & 3.4 & 0.14($\pm$0.08) & 15.6 \\
    \hline
    \hline
    \end{tabular}
    }
\caption{Mean absolute errors (MAEs) and standard deviations (SDs) of the absolute errors (both in eV), together with mean absolute percentage errors (MAPEs), for IPs and EAs relative to EOM-CCSD. $^a$The EOM-CCSD IP and EA reference values are from Ref.~\cite{LangeBerkelbach:EOMGW:2018}.
 $^b$Calculated with data reported in Ref.~\cite{wen_comparing_2024}
 $^c$Calculated using the one-shot $G_0W_0\Gamma_0^{(1)}$ results reported in Ref.~\cite{Rinke:G0W0Gamma0:2021}. For the corresponding IPs, the difference between the $\Delta$CCSD(T) and EOM-CCSD references has an MAE of 0.07~eV and a standard deviation of 0.10~eV.}
\label{tab:ea-gw100}
\end{table}

Within the IP sector, changing the reference from $\Delta$CCSD(T) to EOM-CCSD increases several absolute MAEs modestly but preserves the qualitative ordering among the sc$GW\Gamma_\Sigma$ methods, dynamic-2SOSEX remains the most accurate sc$GW\Gamma_\Sigma$ variant, whereas the one-shot $G_0W_0\Gamma_0^{(1)}$@PBE result is comparable. Within the EA sector, all tested sc$GW\Gamma_\Sigma$ variants reduce the aggregate MAE relative to sc$GW$, and the SOSEX variants give the smallest aggregate MAEs in Table~\ref{tab:ea-gw100}. Figure~\ref{fig:EA_GW_100}, however, shows that the signed deviations and the apparent improvement vary substantially among groups of compounds. The EA statistics must therefore be interpreted within individual chemical groups rather than as a uniform improvement in experimental electron affinities.

Because the chemical groups contain states with different physical character and energy scales, error magnitudes should not be compared directly across groups; the molecule-resolved signed errors provide the more informative diagnostic. At the aggregate level, the EA MAPEs (15.6--52.0\%) are substantially larger than the IP MAPEs (2.5--5.9\%), showing that electron addition is the more difficult benchmark.

The EAs span three regimes: genuinely bound anions, metastable anions with localized electron-addition states due to the confined basis set, and unbound systems in which the excess electron is far from the molecule. The last regime explains why several methods can appear deceptively accurate. If sufficiently diffuse functions are added for an unbound system, the excess electron moves progressively farther away and its orbital overlaps and two-electron integrals with the molecule approach zero. The magnitude of the nuclear--electron attraction is reduced as well, while the kinetic-energy expectation value $\langle\phi_{\mathrm{diff}}|\hat{T}|\phi_{\mathrm{diff}}\rangle$ and the magnitude of $\langle\phi_{\mathrm{diff}}|\hat{V}_{\mathrm{ne}}|\phi_{\mathrm{diff}}\rangle$ both approach zero. This zero-coupling limit is closely related to the established continuum-orbital approach, in which a single Gaussian function with an exponent of order $10^{-18}$ has negligible overlap with the molecular basis and produces a zero-energy orbital.~\cite{StantonGauss:continuum-orbital:1999,Alag:continuum-orbital:2022} For an added electron occupying a similarly decoupled orbital, the corresponding molecular Fock matrix and connected self-energy blocks vanish. The connected $GW$ and self-energy-corrected approximations considered here therefore yield EAs that approach zero in this limit. Consequently, the agreement between theories near zero is an asymptotic result that is not sensitive to the nature of the vertex approximation. Thus, this agreement  does not demonstrate that the vertex approximation accurately describes an anion.

The group-resolved signed errors in Figure.~\ref{fig:EA_GW_100} expose the different regimes and clarify the aggregate statistics. Relative to EOM-CCSD, sc$GW$ produces predominantly negative errors in the groups of dimers, hydrocarbons, hydrides, nitrides, oxides, aromatics, and nucleobases, whereas SOX generally shifts the errors in the positive direction; the screened vertex variants usually lie between these limits. The resulting MAE reduction therefore reflects different balances of over- and underestimation across groups rather than a uniform shift toward the reference. Of the 93 EOM-CCSD reference EAs, 68 are positive and 25 are negative. All members of the atomic and silane groups (4 of 4 in each), 10 of 11 hydrocarbons, and all 12 aromatic/nucleobase systems have positive reported EAs. Small saturated hydrocarbons show little separation among the methods because the excess electron is weakly coupled to the molecule, whereas larger aromatic systems and nucleobases can retain a localized $\pi^*$-like resonance and display much larger vertex-dependent shifts.~\cite{Mukherjee:pyrrole-resonances:2022,Almeida:pyrimidine-DEA:2013} In contrast, the cluster/diatomic group contains 9 negative values out of 12, and the hydride/halide group contains 7 out of 21; polar examples such as KH, LiF, and MgF$_2$ require a dedicated analysis of possible dipole-bound states.

\section{Conclusion}
In this work, we assessed selected self-energy-only corrections to sc$GW$ for molecular charged excitation energies, including IPs and EAs. The calculations are enabled by the THC decomposition of the Coulomb repulsion integrals with a moderate rank ($\alpha_\text{Ipts}=10$) achieving an excellent cost-accuracy balance without introducing systematic bias in total energies or first charged excitation energies. The work does not introduce new diagram topologies beyond the established SOX-, SOSEX-, 2SOSEX-, and $G3W2$-type self-energy approximations; rather, the usual Feynman lines are replaced by bold Green's-function lines obtained from the fully self-consistent Dyson equation.

Self-consistent vertex correction into the self-energy produce a well-ordered and physically interpretable hierarchy of total energy and imaginary part of self-energy . The signed values are ordered from least to most negative as the magnitude of self-energy is ordered from the least negative SOX $>$ SOSEX $>$ G3W2 $>$ 2SOSEX $>$ sc$GW$ ; equivalently, the absolute magnitude increases in the opposite direction. This ordering is naturally explained by the effective-screening picture: SOX$(v,v)$ contains two bare interactions and therefore gives the largest exchange subtraction, resulting in the weakest net dynamical self-energy, and finally yielding the highest (least negative) total energy. SOSEX$(v,W)$ replaces one interaction by a screened one, reducing the exchange subtraction and moving both self-energies and total energies closer towards scGW. $G3W2(W,W)$ screens both interactions, yielding the smallest exchange correction resulting in self-energies that are very close to scGW. 2SOSEX can overcorrect the screened-exchange contribution and may approach or even cross scGW depending on screening strength. The systematic monotonic trends across chemically different systems support the dielectric-screening interpretation of the self-energy shifts, while the resulting accuracy improvements remain dependent on the property, reference method, and self-energy-only approximation being assessed.

We also observed consistent differences between static and dynamic SOSEX and 2SOSEX that arise exclusively due to the treatment of the screened interaction $\tilde{W}$. Since $||\tilde{W}(i\Omega)||$ is typically maximal at $\Omega=0$,
the dynamical screened-exchange term samples, on average, a weaker interaction than the static approximation. Consequently, the static variants produce larger self-energy displacements $\tilde{\Sigma}^{\mathrm{SOX}(v,\tilde{W})}$ making the total self-energy less negative and reducing its absolute magnitude: $\tilde{\Sigma}^\text{SOSEX}=\tilde{\Sigma}^{\text{SOX}(v,W)} = \tilde{\Sigma}^{\text{SOX}(v,v)} + \tilde{\Sigma}^{\text{SOX}(v,\tilde{W})}$. Dynamic variants yield systematically lower (more negative) $\text{ImTr}[\Sigma(\omega_n)]$ self-energies. The total energies depend on frequency integrals involving phases and cancellations between $G$ and  $\Sigma$ so energetic trends cannot be inferred from the norms of self-energies alone. However, in many cases they trace the behavior of the self-energy.

Based on the numerical observations done in this work, we conclude that all vertex corrections performed here predominantly act as an approximately frequency-uniform vertical renormalization of Tr$\Sigma(i\omega)$ rather than strongly reshaping its frequency dependence.

Moreover, adding the vertex corrections usually results in  convergence difficulties for self-consistent loops that are consistent with difficulties in converging methods such as Green's function second order (GF2)\cite{Pokhilko:algs:2022}. Such convergence difficulties can be mitigated with heating-cooling methods\cite{Pokhilko:algs:2022} or homotopy continuation.~\cite{Pokhilko:homotopy:2025} Additional details are provided in the Supporting Information.

We observed the following trends in the accuracy of first IPs when comparing them against $\Delta$CCSD(T) and experiment. For the $G_0W_0\Gamma$29 set, Hartree-Fock Koopmans values show significantly larger errors, confirming that dynamical screening is essential. Both $G_0W_0$@HF and $G_0W_0\Gamma^{\mathrm{(NL)}}$@HF yield IPs with significantly higher errors than THC-sc$GW$. Fully self-consistent THC-sc$GW$-SOX  degrades IP accuracy most strongly. Both static and dynamic THC-sc$GW$-SOSEX partially mitigate THC-sc$GW$-SOX but do not improve over THC-sc$GW$. 
Both THC-sc$GW$-dynamic-2SOSEX and THC-sc$GW$-static-G3W2 remain very close to sc$GW$.

For the $GW$100 IPs relative to $\Delta$CCSD(T), the conclusion remains essentially the same; however, both THC-sc$GW$ -dynamic-2SOSEX and THC-sc$GW$-static-G3W2 yield very minor improvements over THC-sc$GW$. This improvement, however, comes with a significantly increased computational cost of the vertex-corrected methods compared to THC-sc$GW$.

With EOM-CCSD as a common reference, dynamic-2SOSEX remains the most accurate self-consistent variant for IPs. For the EA, every tested sc$GW\Gamma_\Sigma$ variant reduces the aggregate MAE relative to sc$GW$, with the static and dynamic SOSEX variants giving the smallest MAE of 0.11~eV. Nevertheless, the EA percentage errors remain substantially larger than the IP errors. This contrast reflects the heterogeneous character of the added-electron states---bound, metastable, or weakly coupled---and prevents the lower aggregate MAEs from being interpreted as improved predictions of experimental electron affinities. Instead, this benchmark measures how the self-energy-only vertex corrections respond to electron addition and highlights the need for a dedicated continuum or resonance treatment when the anion is not stable.

Overall, the tested self-energy-only corrections do not provide a uniform improvement over sc$GW$. Their accuracy depends on the property, reference, and screening treatment. THC nevertheless provides an efficient framework for exploring better-balanced diagrammatic approximations, including schemes that treat vertex effects consistently in both the self-energy and polarization.

\begin{acknowledgement}
M. Wang, M. Wen, P. P., and D. Z. were supported from the Center for Scalable Predictive Methods for Excitations and Correlated Phenomena (SPEC), which is funded by the U.S. Department of Energy (DOE), Office of Science, Basic Energy Sciences (BES), Chemical Sciences, Geosciences, and Biosciences Division (CSGB), as part of the Computational Chemical Sciences (CCS) program under FWP 70942 at Pacific Northwest National Laboratory  (PNNL), a multi-program national laboratory operated for DOE by Battelle. 
The Flatiron Institute is a division of the Simons Foundation.
\end{acknowledgement}

\begin{suppinfo}
The Supplementary Material collects the numerical and methodological details that support the main-text discussion. It includes convergence tests of the THC interpolation-point ratio and decomposed total-energy contributions, full molecule-by-molecule first-IP tables for the $G_0W_0\Gamma29$ and $GW$100 benchmark sets together with the experimental references used for comparison, the molecule-resolved numerical data underlying the $GW$100 IP and EA error plots and the combined machine-readable workbook, and the IR-grid, damping, CDIIS, and convergence-threshold settings used in the self-consistent calculations.
\end{suppinfo}

\bibliography{refs/ab_initio, refs/abbr, refs/alg, refs/other_ref, refs/analysis, refs/basis, refs/cc, refs/crit_temp, refs/dft, refs/eff_ham, refs/finite_size, refs/gf2, refs/grid, refs/group, refs/gw, refs/gw100, refs/nuclear, refs/programs, refs/rdm, refs/ri_cholesky, refs/so, refs/textbooks, refs/thc, refs/tt_tci,refs/theorems, refs/intermol_int, refs/mol_magnets}

\end{document}


\maketitle
\newpage

\section{Convergence of the THC decomposition with respect to the number of interpolation vectors}\label{sec:convergence}

We use HF/cc-pVQZ as the starting point for the subsequent THC-\scgw{} calculations in order to assess the convergence of the tensor-hypercontraction (THC) decomposition. We define the interpolation-point ratio as $\alpha_\text{Ipts} = \frac{n_{\mu}}{n_{\mathrm{AO}}}$, where $n_{\mu}$ is the number of THC auxiliary functions and $n_{\mathrm{AO}}$ is the total number of atomic orbitals.

Table~\ref{tab:nIpts} summarizes the convergence of the total energies (in Hartrees) as a function of $\alpha_\text{Ipts}$ for the 29-molecule set. We consider three representative values, $\alpha_\text{Ipts}=7$, 10, and 13. The electronic energy is decomposed into a one-body contribution associated with the static self-energy, $\Sigma^{\infty}$, and a two-body contribution associated with the dynamic self-energy, $\tilde{\Sigma}$.

\begin{figure}[p]
\centering
\resizebox{0.95\textwidth}{!}{%
\begin{tabular}{@{}ccc@{}}
\includegraphics[width=0.33\linewidth]{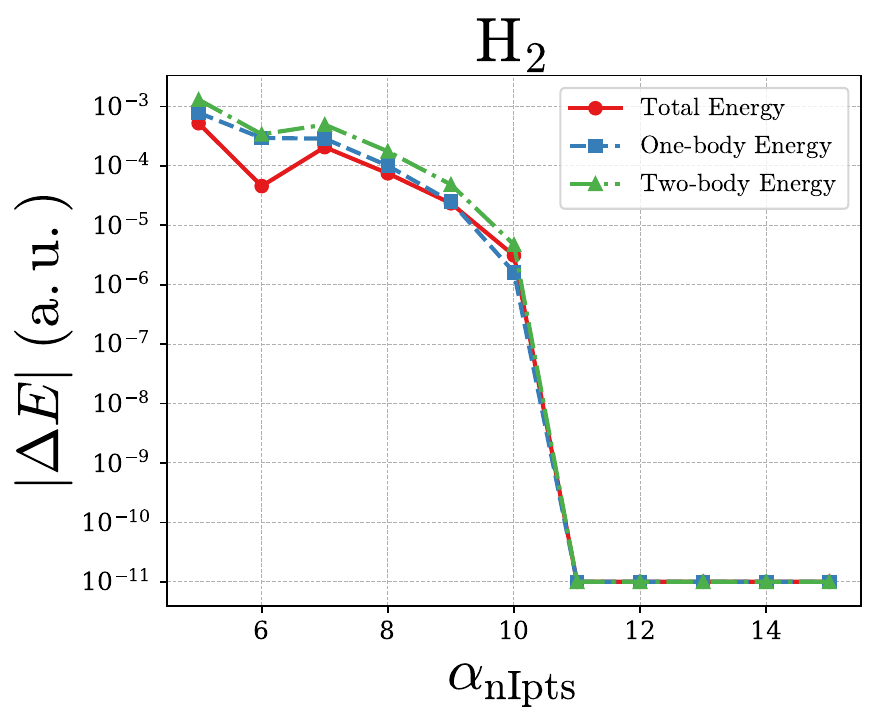} & 
\includegraphics[width=0.33\linewidth]{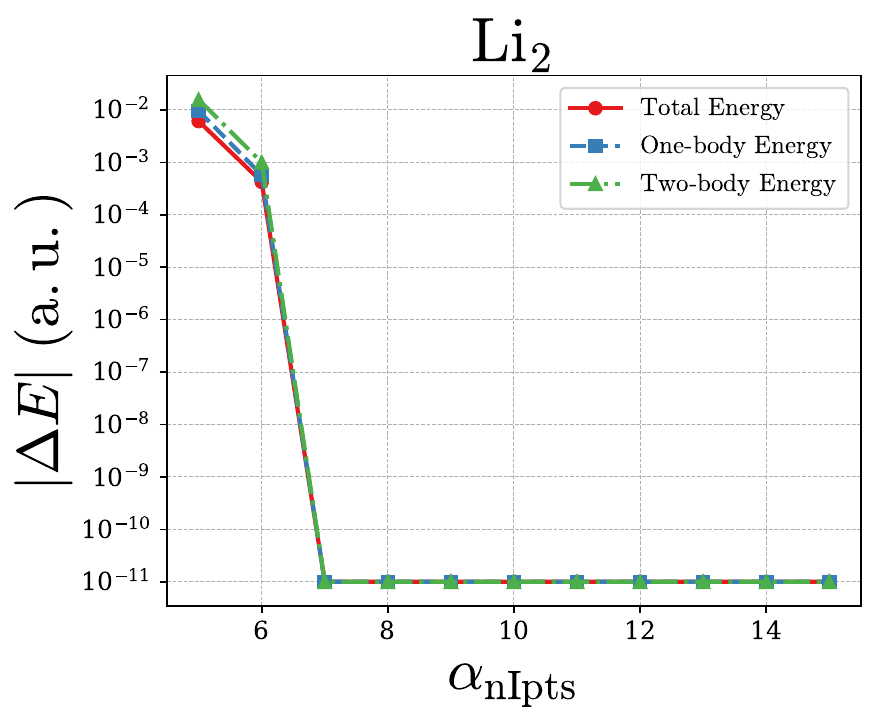}& 
\includegraphics[width=0.33\linewidth]{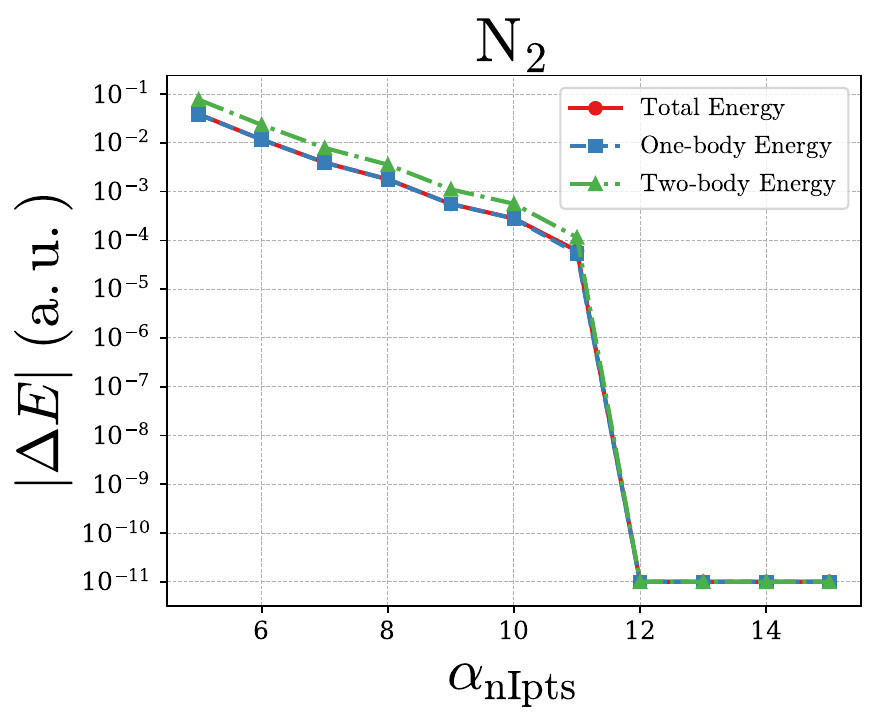}\\
\includegraphics[width=0.33\linewidth]{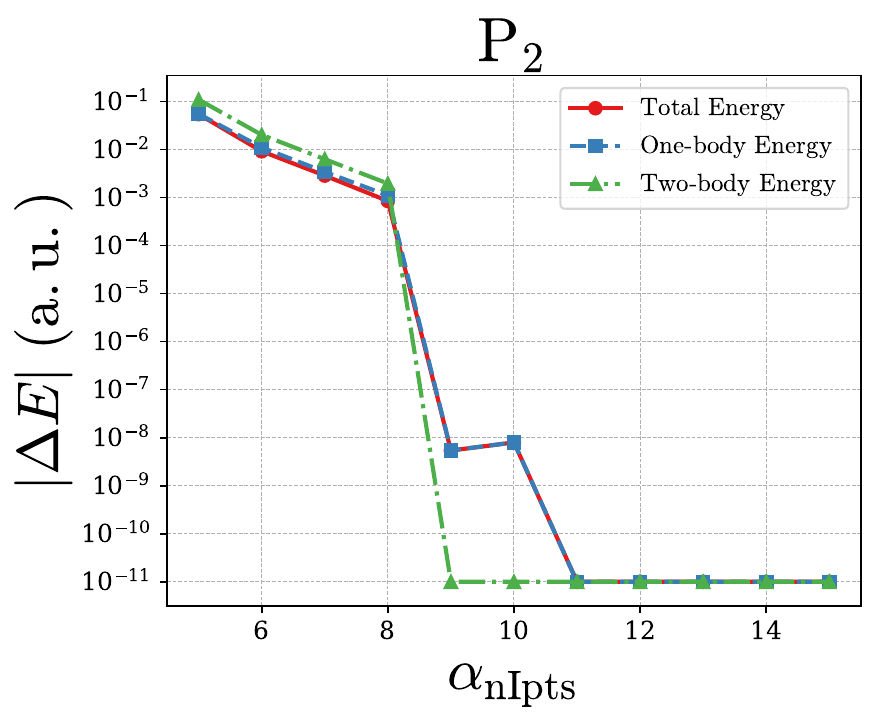} & 
\includegraphics[width=0.33\linewidth]{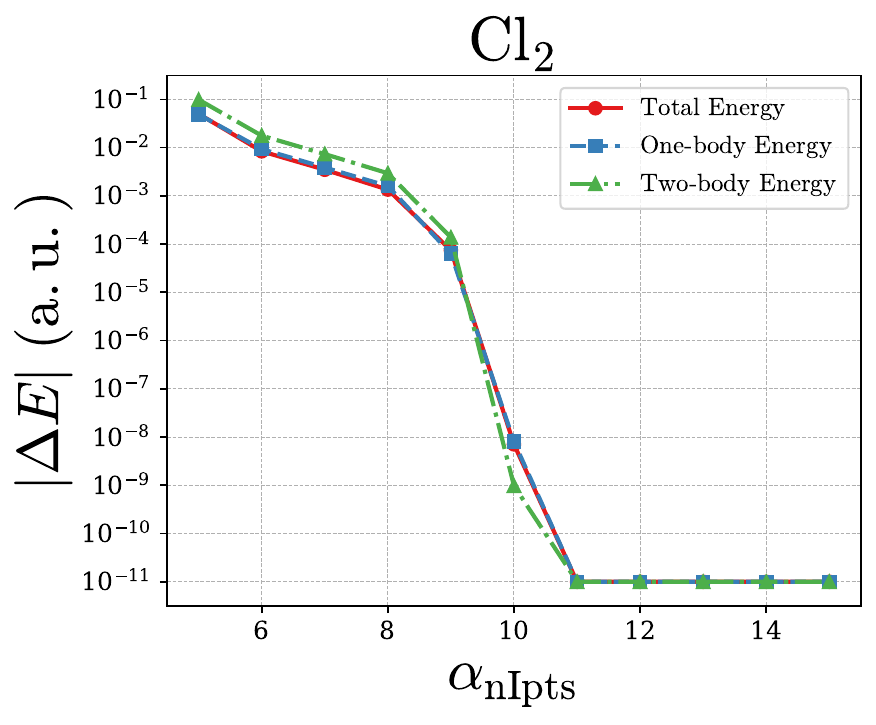}& 
\includegraphics[width=0.33\linewidth]{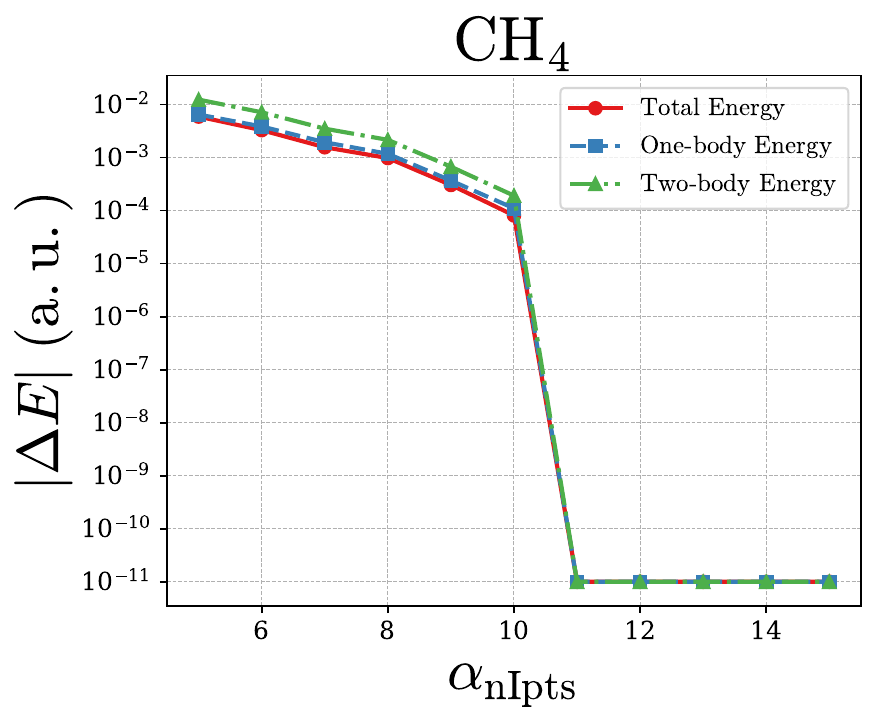}\\
\includegraphics[width=0.33\linewidth]{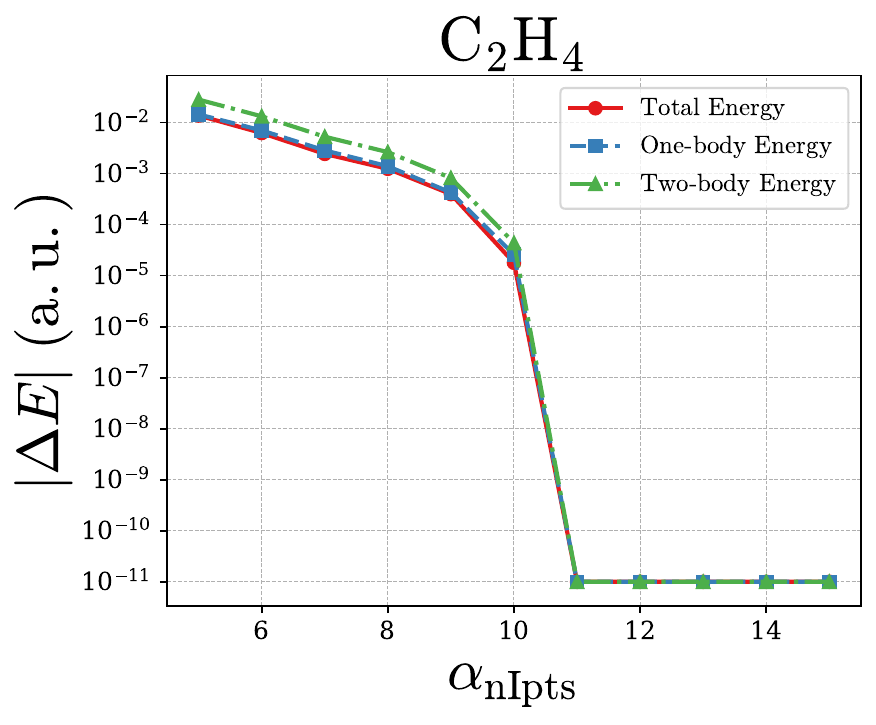} & 
\includegraphics[width=0.33\linewidth]{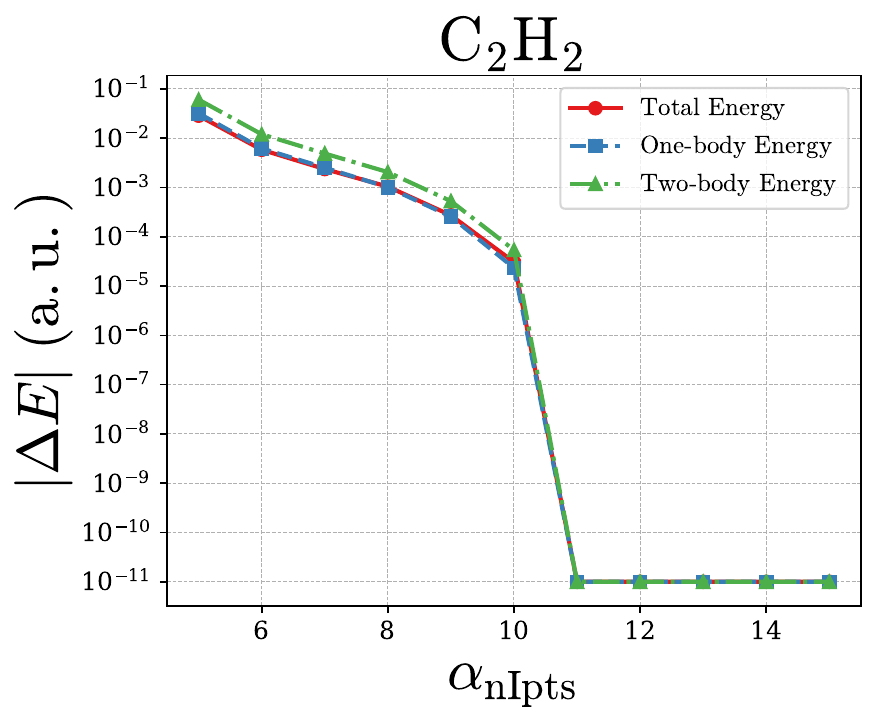}& 
\includegraphics[width=0.33\linewidth]{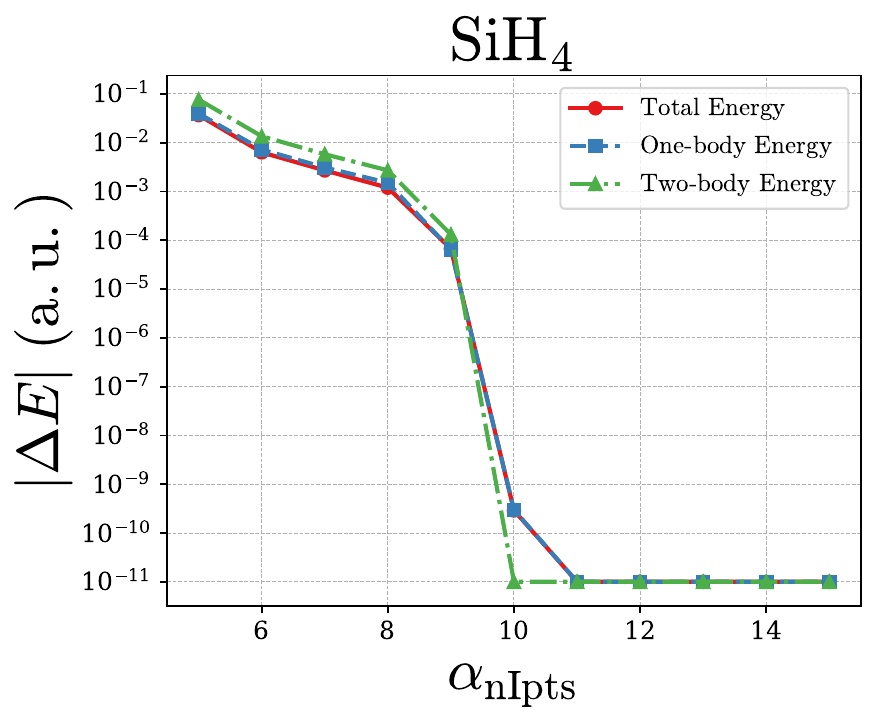}\\
\includegraphics[width=0.33\linewidth]{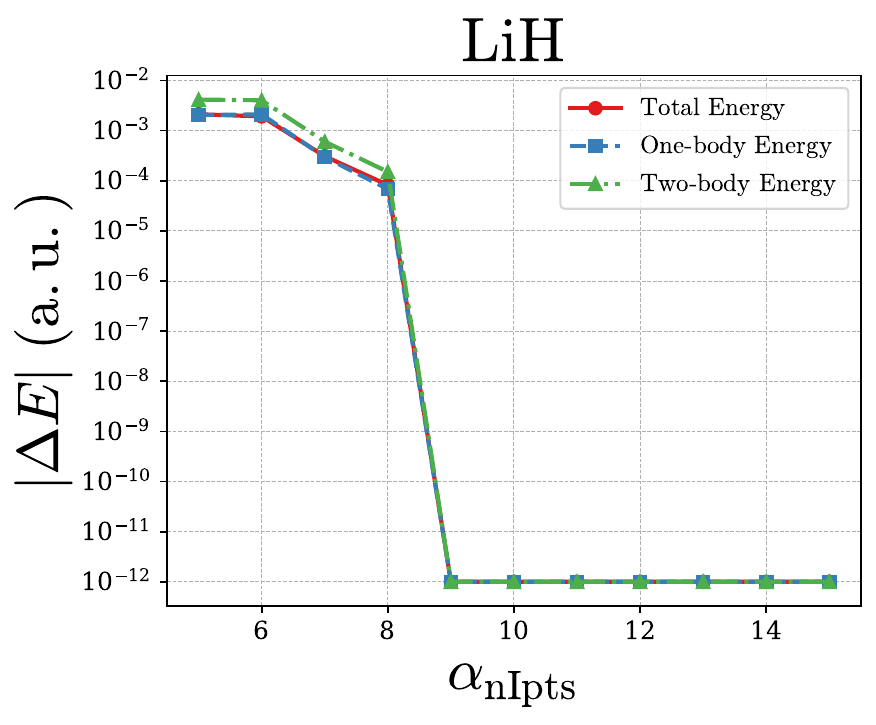} & 
\includegraphics[width=0.33\linewidth]{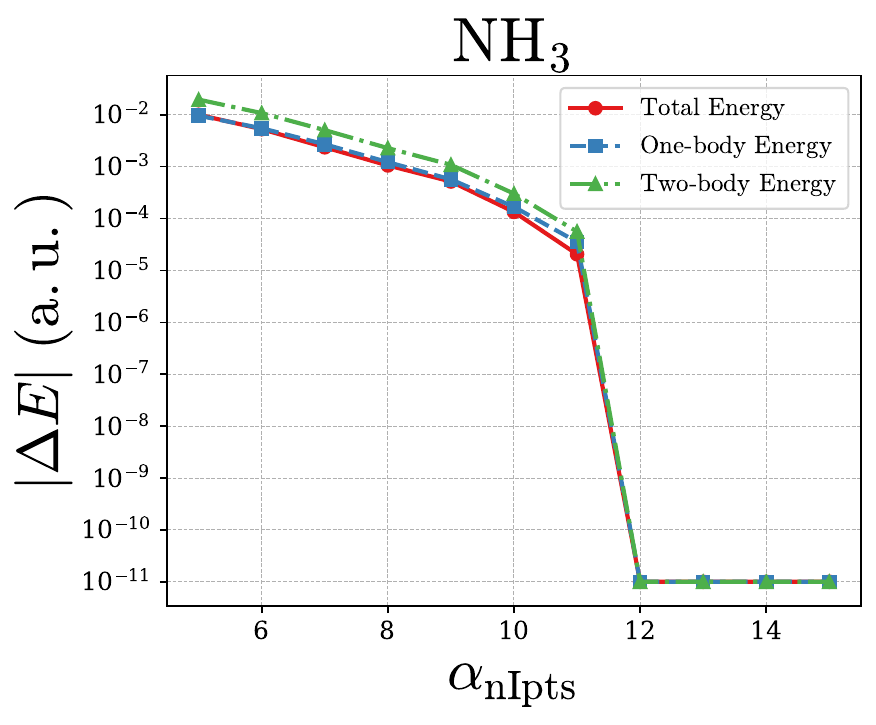}& 
\includegraphics[width=0.33\linewidth]{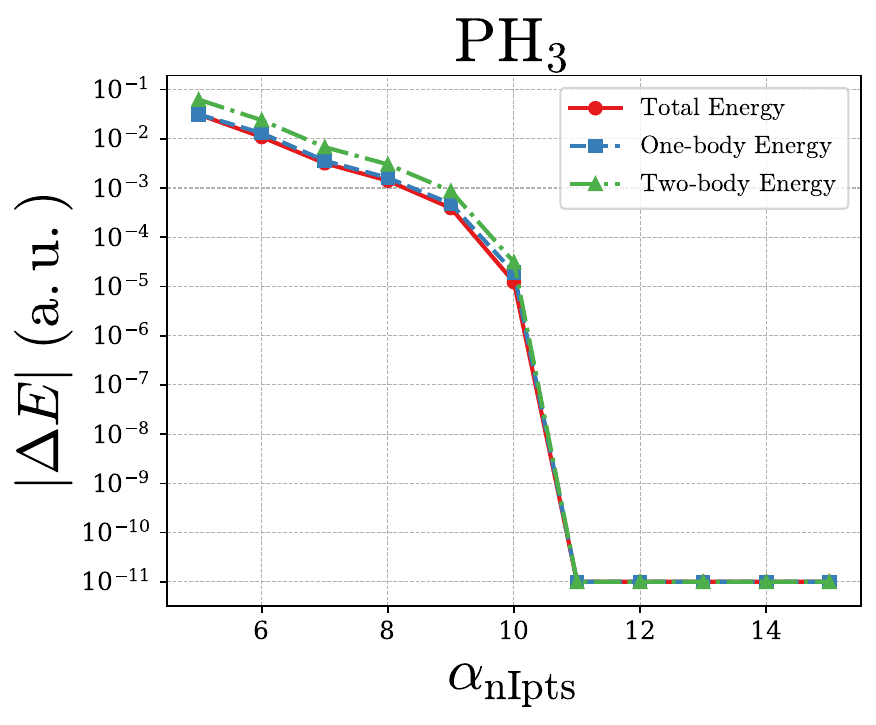}\\
\includegraphics[width=0.33\linewidth]{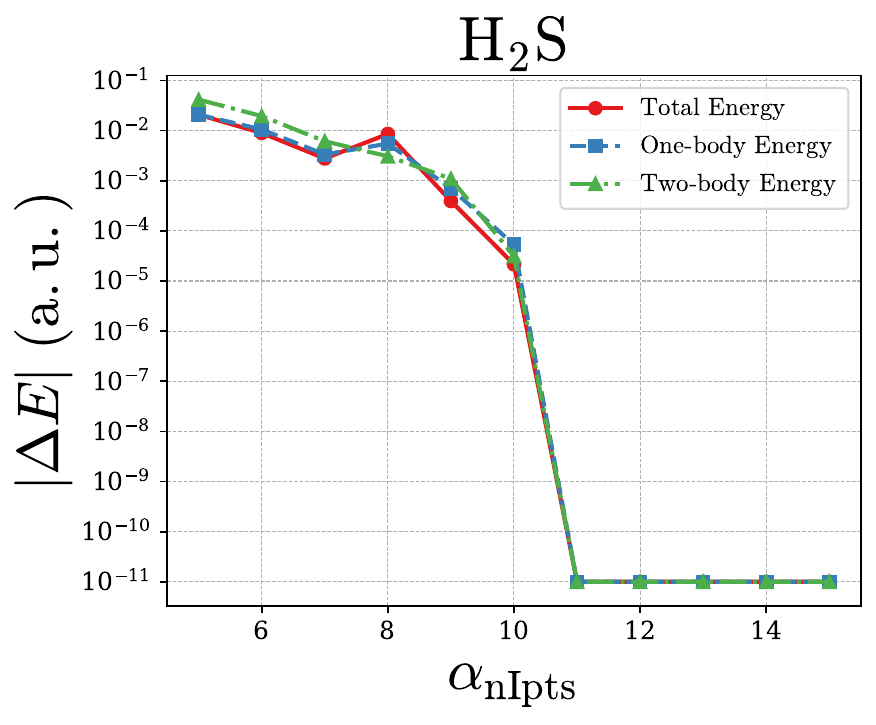} & 
\includegraphics[width=0.33\linewidth]{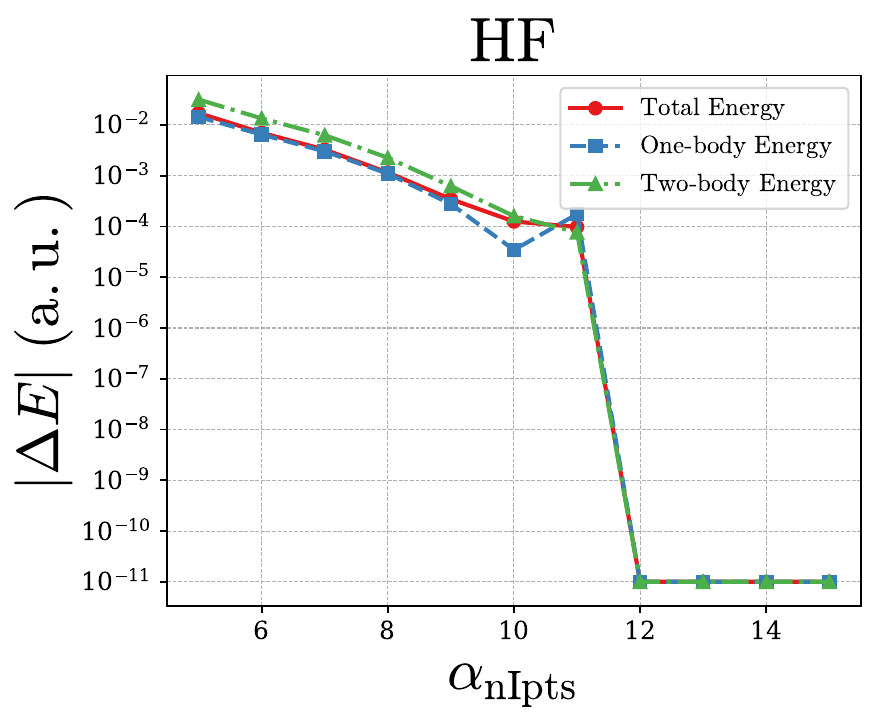}& 
\includegraphics[width=0.33\linewidth]{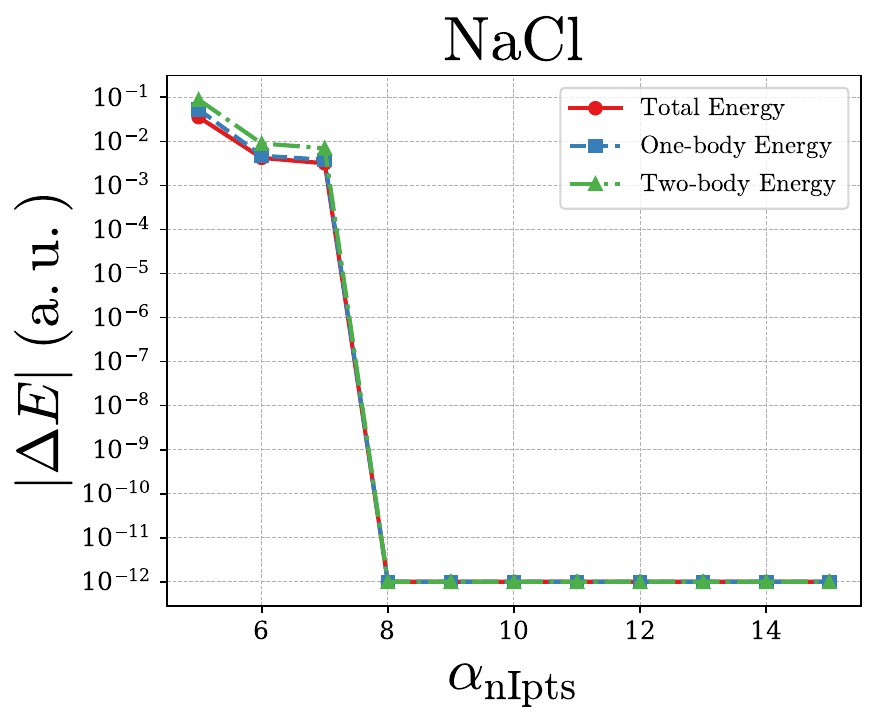}
\end{tabular}
}
\caption{Total-energy convergence for molecules 1--15 as $\alpha_\text{Ipts}$ increases from 5 to 15.}
\label{fig:nIpts_tote1}
\end{figure}

\begin{figure}[p]
\centering
\resizebox{0.95\textwidth}{!}{%
\begin{tabular}{@{}ccc@{}}
\includegraphics[width=0.33\linewidth]{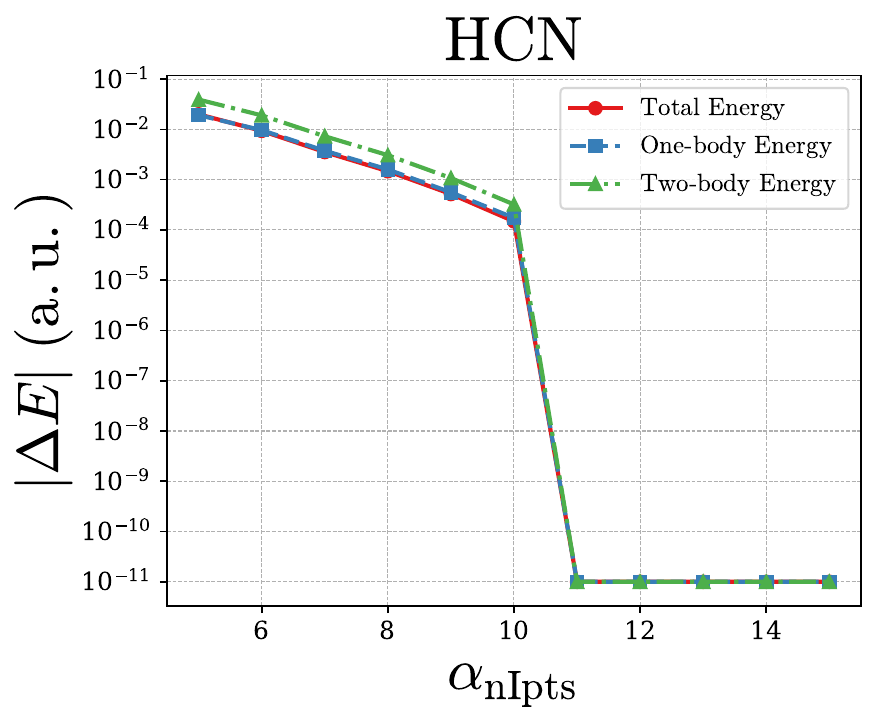} & 
\includegraphics[width=0.33\linewidth]{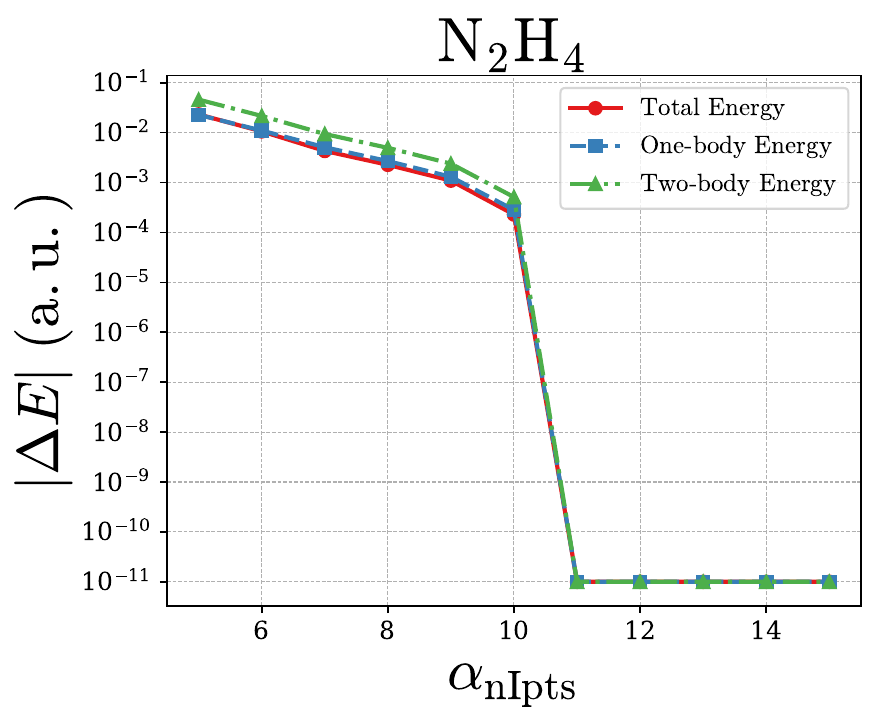}& 
\includegraphics[width=0.33\linewidth]{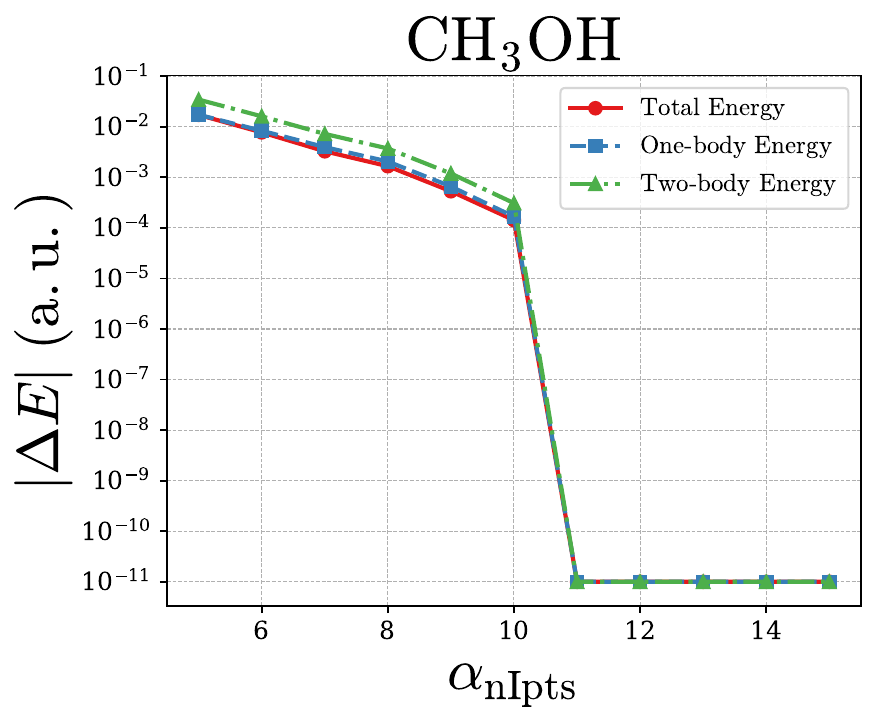}\\
\includegraphics[width=0.33\linewidth]{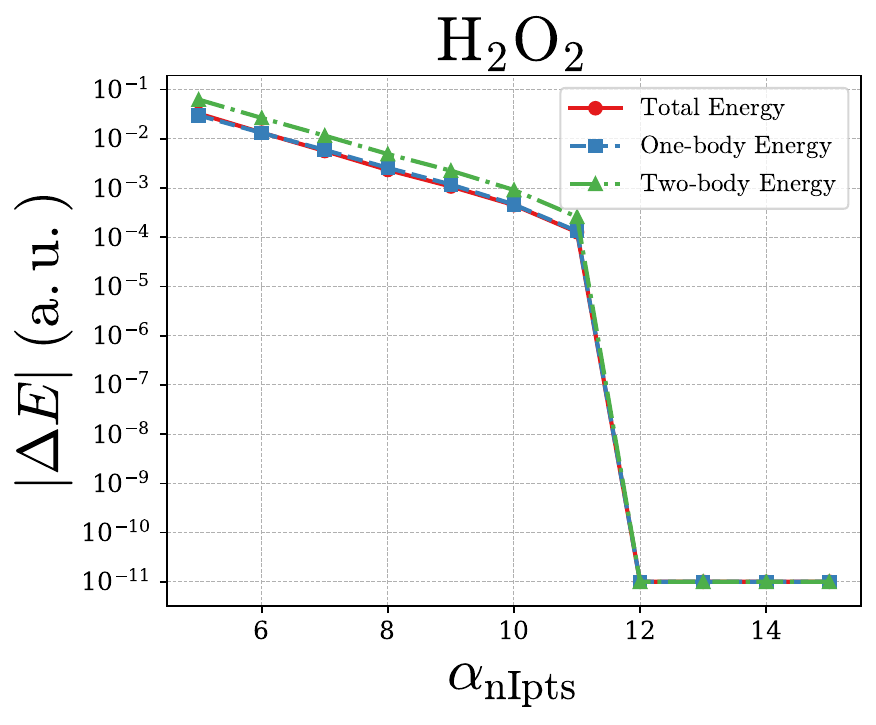} & 
\includegraphics[width=0.33\linewidth]{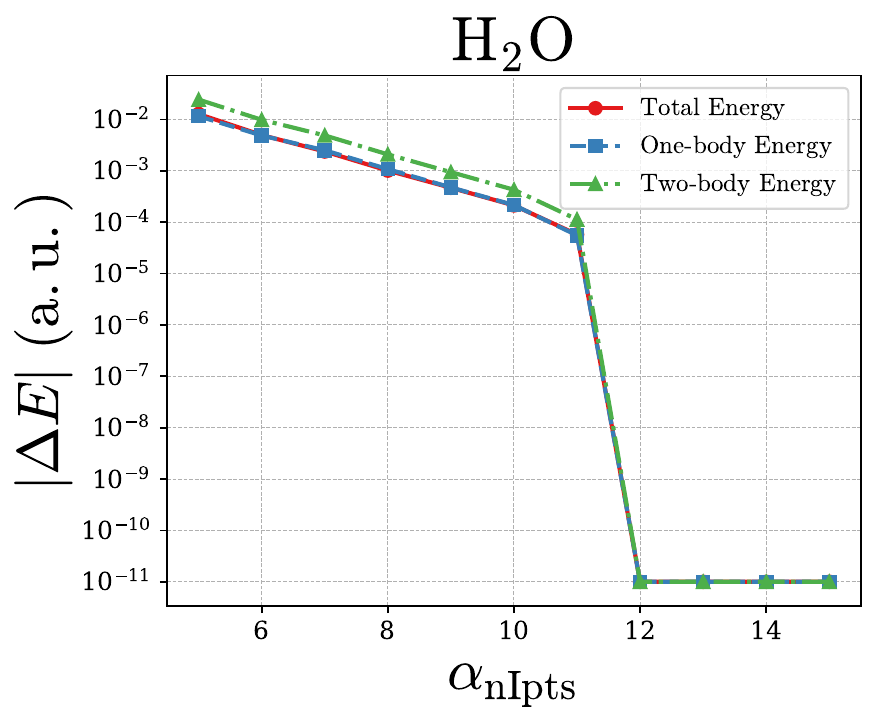}& 
\includegraphics[width=0.33\linewidth]{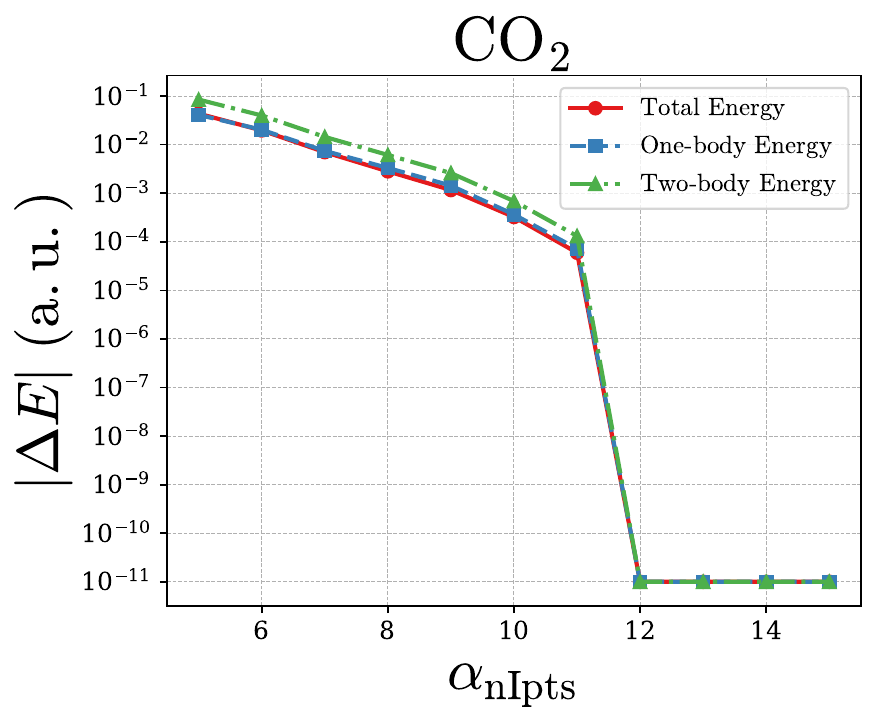}\\
\includegraphics[width=0.33\linewidth]{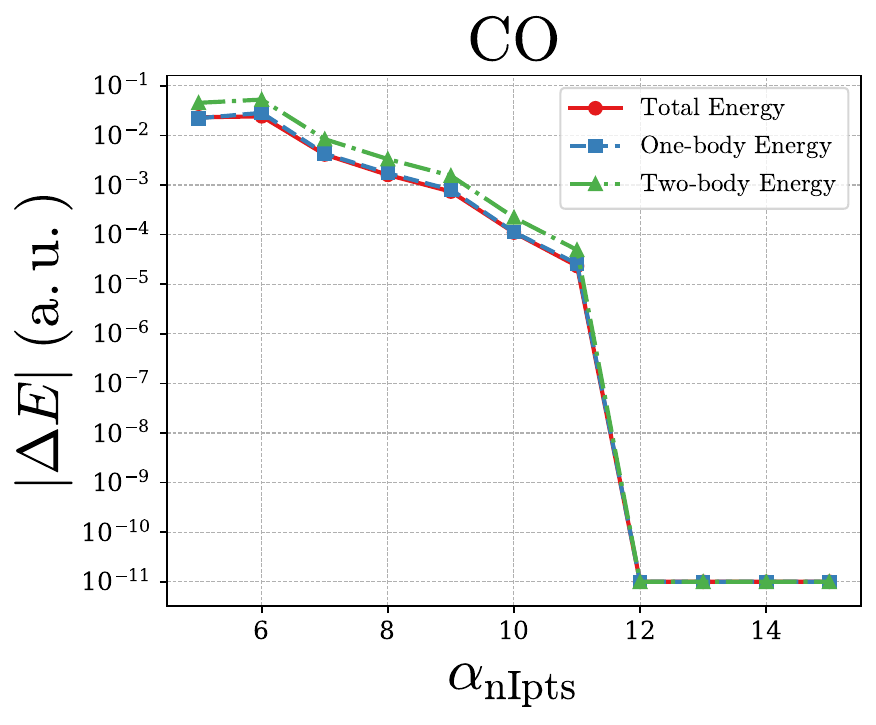} & 
\includegraphics[width=0.33\linewidth]{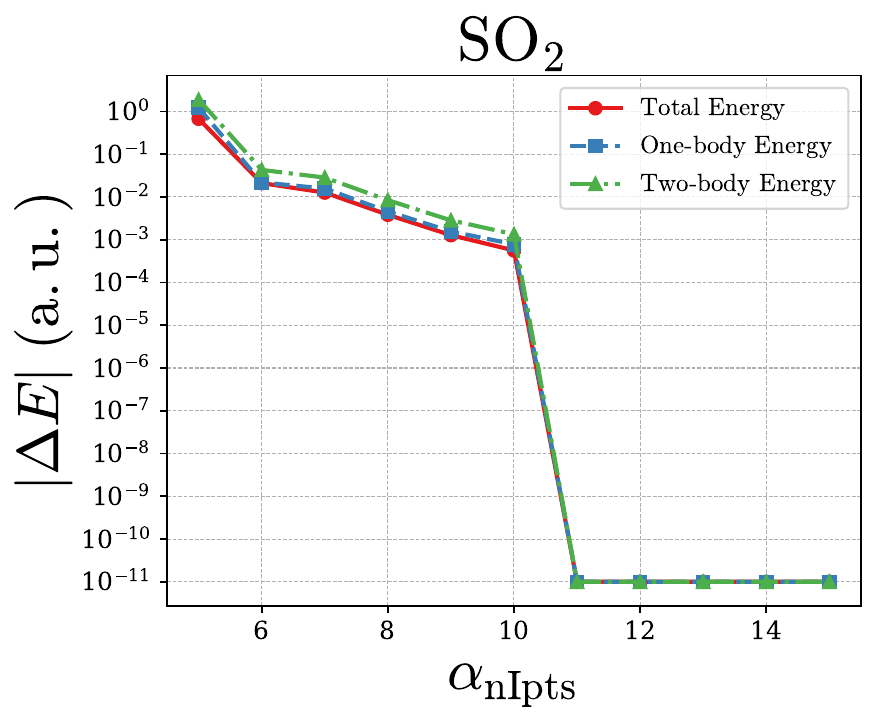}& 
\includegraphics[width=0.33\linewidth]{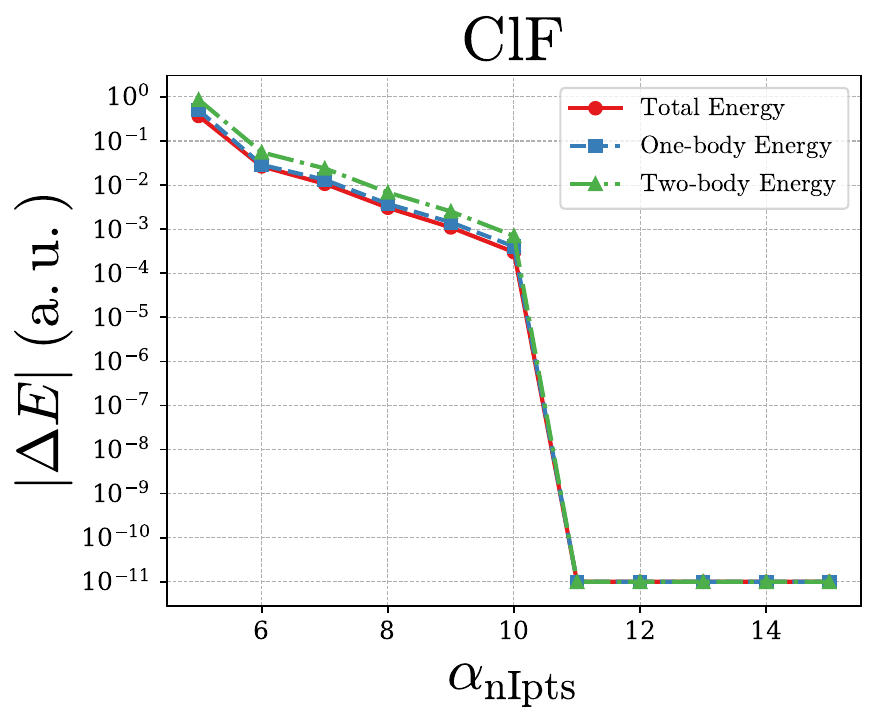}\\
\includegraphics[width=0.33\linewidth]{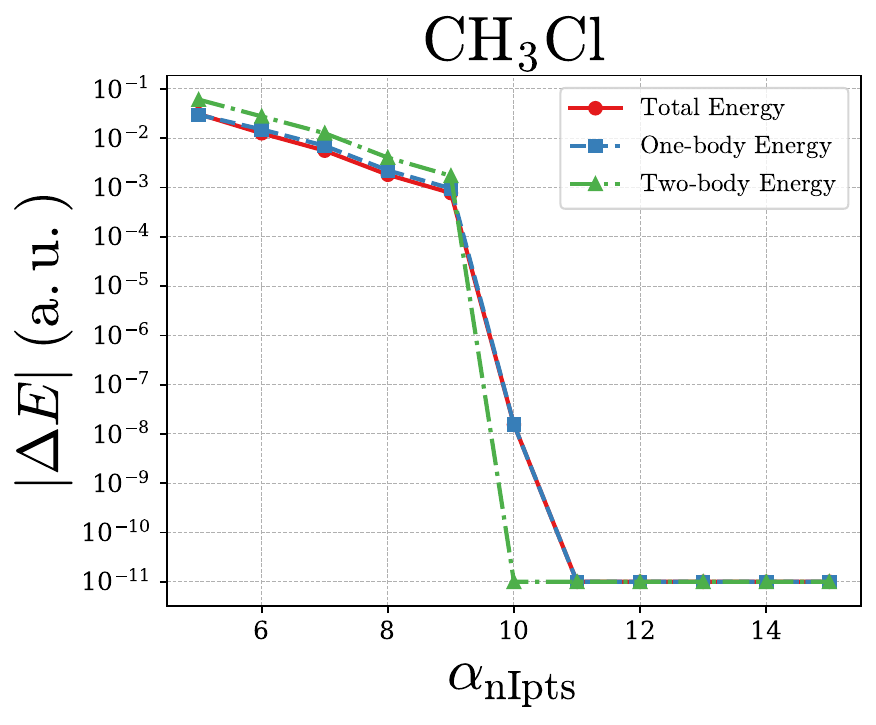} & 
\includegraphics[width=0.33\linewidth]{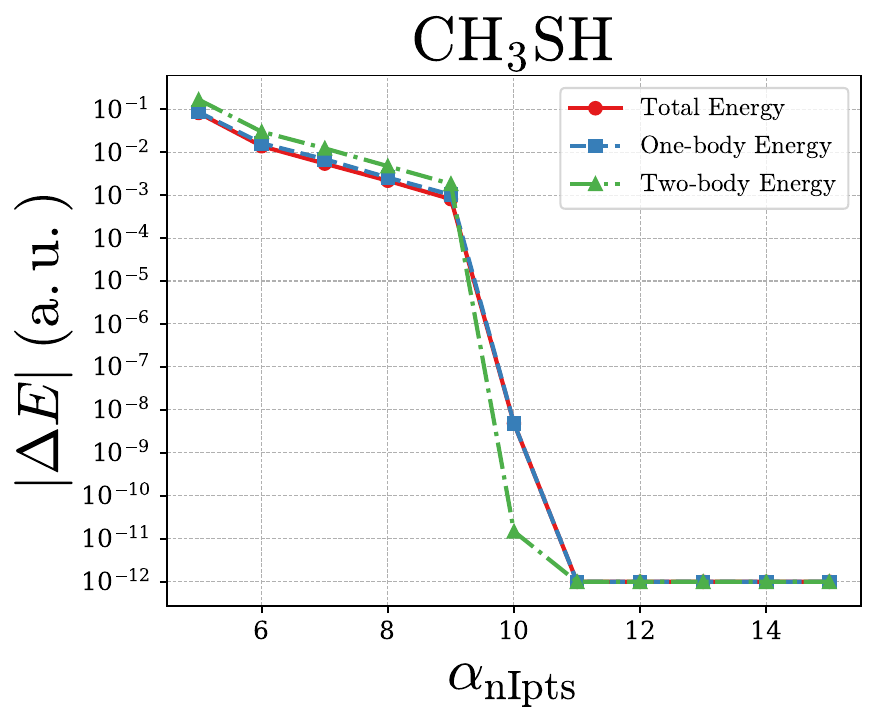}& 
\includegraphics[width=0.33\linewidth]{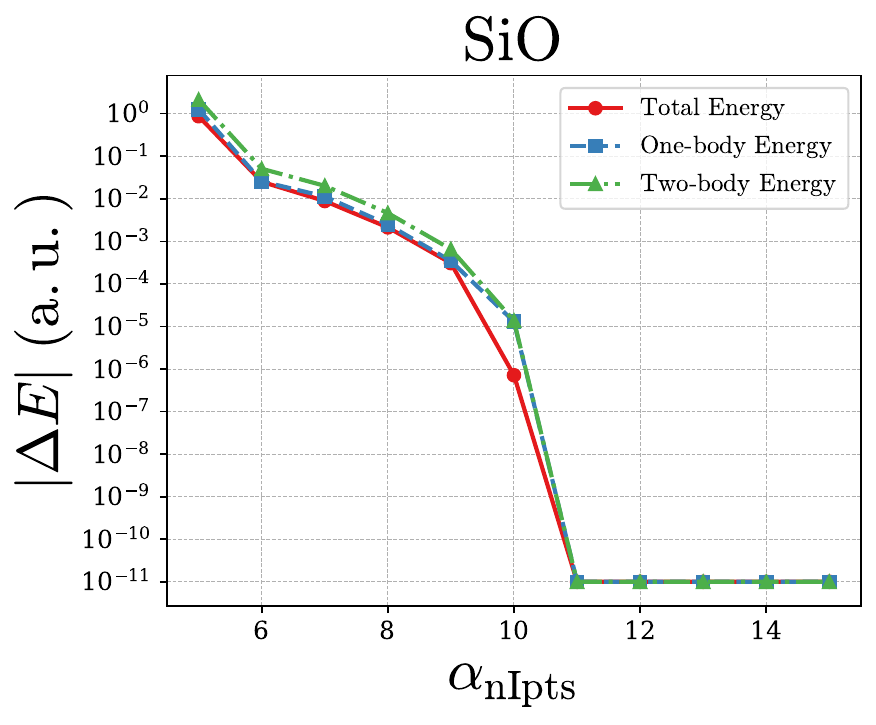}\\
\includegraphics[width=0.33\linewidth]{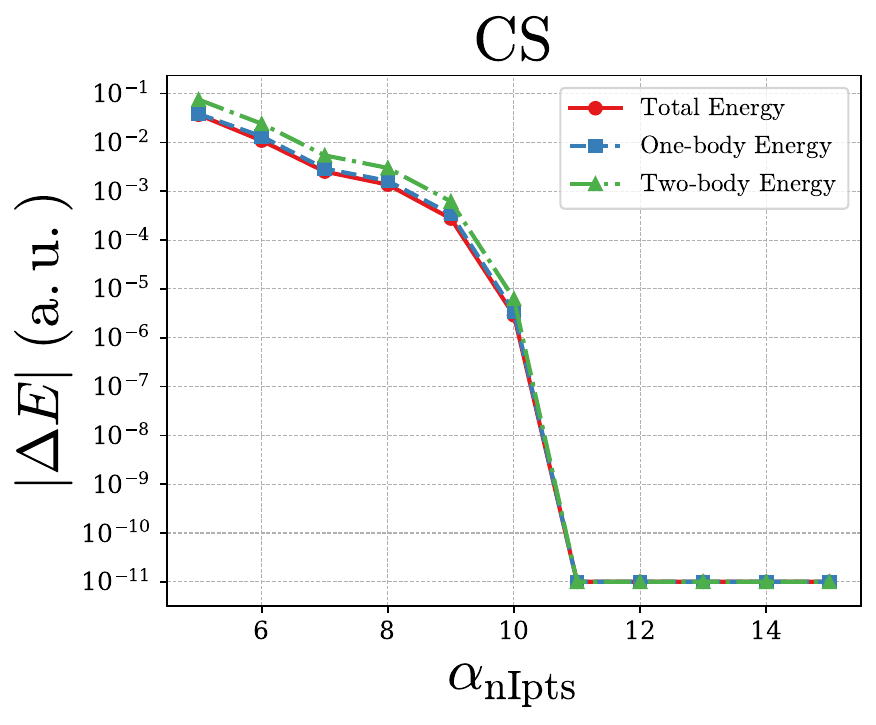} & & 
\includegraphics[width=0.33\linewidth]{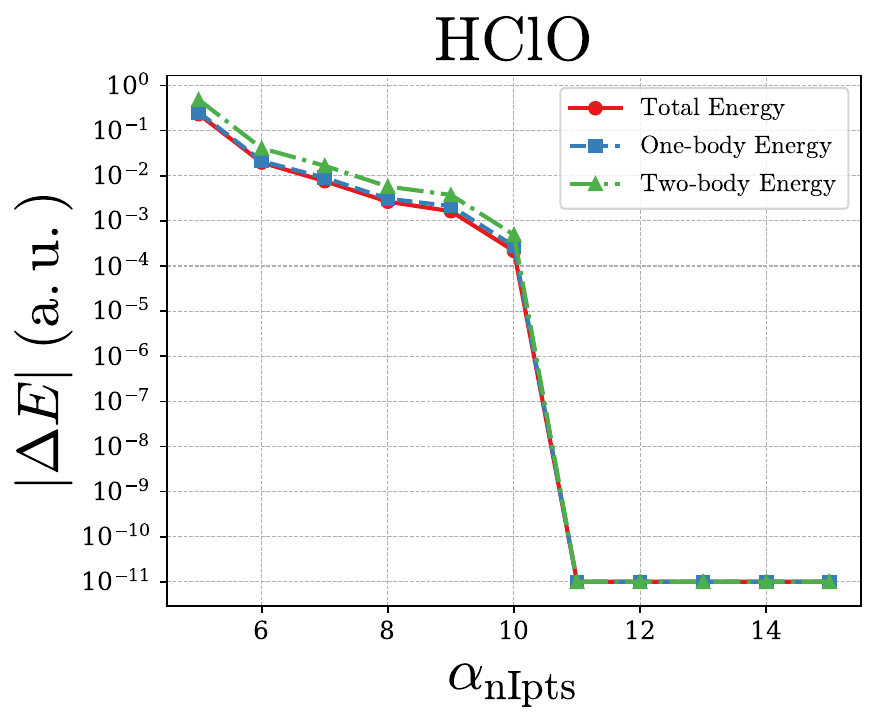}
\end{tabular}
}
\caption{Total-energy convergence for molecules 16--29 as $\alpha_\text{Ipts}$ increases from 5 to 15.}
\label{fig:nIpts_tote2}
\end{figure}

The one-body contribution is defined as
\begin{equation}
    E^\text{stat}_{\text{1b}}= E_\text{nuc}+\mathrm{Tr}[\gamma H_{0}]+\tfrac{1}{2}\mathrm{Tr}[\gamma\Sigma_{\infty}]
    \label{eq:1b_e}
\end{equation}
where \(E_{\mathrm{Nuc}}\) denotes the nuclear-repulsion energy.
\begin{equation}
    E^\text{dyn}_\text{2b} = \frac{2}{\beta}\mathrm{Re}\sum_{n\geq 0}\mathrm{Tr}[G(i\omega_n)\tilde{\Sigma}^{T}(i\omega_n)].
\end{equation} 
The total electronic energy is then written as
\begin{equation}
    E_\text{tot}=E^\text{stat}_{\text{1b}}+E^\text{dyn}_\text{2b}
\end{equation}
\begingroup
\small
\setlength{\tabcolsep}{3pt}
\renewcommand{\arraystretch}{0.7}
\begin{longtable}{ll rr rrr}
\caption{Total energies (Ha) for the 29-molecule set at $\alpha_\text{Ipts}=7$, 10, and 13.}
\label{tab:nIpts}\\
\hline
\hline
Index & Molecule Name & Molecule & $\alpha_\text{Ipts}$ & $E^{stat}_{\mathrm{1b}}$ (Ha) & $E^{dyn}_{\mathrm{2b}}$ (Ha) & $E_{\mathrm{tot}}$ (Ha)\\
\hline
\endfirsthead
\multicolumn{7}{c}{\tablename\ \thetable\ -- continued from previous page} \\
\hline
\hline
Index & Molecule Name & Molecule & $\alpha_\text{Ipts}$ & $E^{stat}_{\mathrm{1b}}$ (Ha) & $E^{dyn}_{\mathrm{2b}}$ & $E_{\mathrm{Tot}}$ (Ha)\\
\hline
\endhead
\hline
\multicolumn{7}{r}{Continued on next page} \\
\endfoot
\hline
\hline
\endlastfoot
1 & hydrogen & $\mathrm{H_2}$ & 7 & -1.084636717 & -0.105154036 & -1.189790753\\
&&& 10 & -1.084921811 & -0.104665740 & -1.189587551\\
&&& 13 & -1.084923425 & -0.104661024 & -1.189584449\\
\hline
2 & lithium dimer & $\mathrm{Li_2}$ & 7 & -14.791986046 & -0.170004260 & -14.961990306\\
&&& 10 & -14.791986046 & -0.170004260 & -14.961990306\\
&&& 13 & -14.791986046 & -0.170004260 & -14.961990306\\
\hline
3 & nitrogen & $\mathrm{N_2}$ & 7 & -108.543797072 & -0.989053462 & -109.532850534\\
&&& 10 & -108.547446517 & -0.981765066 & -109.529211582\\
&&& 13 & -108.547725461 & -0.981203980 & -109.528929442\\
\hline
4 & phosphorus dimer & $\mathrm{P_2}$ & 7 & -681.097340866 & -0.877761912 & -681.975102778\\
&&& 10 & -681.100746180 & -0.871506454 & -681.972252634\\
&&& 13 & -681.100746188 & -0.871506454 & -681.972252642\\
\hline
5 & chlorine & $\mathrm{Cl_2}$ & 7 & -918.447509487 & -1.222024268 & -919.669533755\\
&&& 10 & -918.451366807 & -1.214693291 & -919.666060099\\
&&& 13 & -918.451366799 & -1.214693292 & -919.666060092\\
\hline
6 & methane & $\mathrm{CH_4}$ & 7 & -39.932208741 & -0.613793428 & -40.546002170\\
&&& 10 & -39.934044694 & -0.610471327 & -40.544516021\\
&&& 13 & -39.934155095 & -0.610279755 & -40.544434850\\
\hline
7 & ethylene & $\mathrm{C_2H_4}$ & 7 & -77.604268493 & -1.008768194 & -78.613036687\\
&&& 10 & -77.607090279 & -1.003527704 & -78.610617984\\
&&& 13 & -77.607116093 & -1.003483991 & -78.610600084\\
\hline
8 & ethyne & $\mathrm{C_2H_2}$ & 7 & -76.448466472 & -0.888899186 & -77.337365658\\
&&& 10 & -76.450965350 & -0.884063744 & -77.335029094\\
&&& 13 & -76.450988690 & -0.884009669 & -77.334998359\\
\hline
9 & silane & $\mathrm{SiH_4}$ & 7 & -291.008361619 & -0.551642384 & -291.560004004\\
&&& 10 & -291.011468151 & -0.545832701 & -291.557300851\\
&&& 13 & -291.011468150 & -0.545832701 & -291.557300851\\
\hline
10 & lithium hydride & $\mathrm{LiH}$ & 7 & -7.921040919 & -0.142087518 & -8.063128437\\
&&& 10 & -7.921336389 & -0.141483826 & -8.062820215\\
&&& 13 & -7.921336389 & -0.141483826 & -8.062820215\\
\hline
11 & ammonia & $\mathrm{NH_3}$ & 7 & -55.915708394 & -0.670887737 & -56.586596131\\
&&& 10 & -55.918238388 & -0.666134212 & -56.584372600\\
&&& 13 & -55.918405825 & -0.665832720 & -56.584238545\\
\hline
12 & phosphine & $\mathrm{PH_3}$ & 7 & -342.215108892 & -0.597859043 & -342.812967935\\
&&& 10 & -342.218670148 & -0.591124380 & -342.809794529\\
&&& 13 & -342.218689158 & -0.591093180 & -342.809782338\\
\hline
13 & hydrogen sulfide & $\mathrm{H_2S}$ & 7 & -398.425916929 & -0.634581085 & -399.060498014\\
&&& 10 & -398.429205408 & -0.628493764 & -399.057699172\\
&&& 13 & -398.429258365 & -0.628462269 & -399.057720634\\
\hline
14 & hydrogen fluoride & $\mathrm{HF}$ & 7 & -99.726833679 & -0.747093315 & -100.473926994\\
&&& 10 & -99.729810782 & -0.740987819 & -100.470798601\\
&&& 13 & -99.729845323 & -0.740828002 & -100.470673325\\
\hline
15 & sodium chloride & $\mathrm{NaCl}$ & 7 & -621.146177436 & -0.680443222 & -621.826620658\\
&&& 10 & -621.149984615 & -0.673469025 & -621.823453640\\
&&& 13 & -621.149984615 & -0.673469025 & -621.823453640\\
\hline
16 & hydrogen cyanide & $\mathrm{HCN}$ & 7 & -92.486024370 & -0.943314295 & -93.429338665\\
&&& 10 & -92.489609269 & -0.936336549 & -93.425945818\\
&&& 13 & -92.489782460 & -0.936015438 & -93.425797898\\
\hline
17 & hydrazine & $\mathrm{N_2H_4}$ & 7 & -110.654955938 & -1.245977337 & -111.900933275\\
&&& 10 & -110.659761682 & -1.237107113 & -111.896868795\\
&&& 13 & -110.660043778 & -1.236596561 & -111.896640340\\
\hline
18 & methanol & $\mathrm{CH_3OH}$ & 7 & -114.535108901 & -1.227279038 & -115.762387939\\
&&& 10 & -114.538889931 & -1.220377093 & -115.759267025\\
&&& 13 & -114.539056015 & -1.220069152 & -115.759125168\\
\hline
19 & hydrogen peroxide & $\mathrm{H_2O_2}$ & 7 & -150.235068068 & -1.339991853 & -151.575059922\\
&&& 10 & -150.240552376 & -1.329319640 & -151.569872017\\
&&& 13 & -150.241014727 & -1.328407680 & -151.569422407\\
\hline
20 & water & $\mathrm{H_2O}$ & 7 & -75.739287146 & -0.714560434 & -76.453847579\\
&&& 10 & -75.741579981 & -0.710102944 & -76.451682925\\
&&& 13 & -75.741795777 & -0.709675039 & -76.451470817\\
\hline
21 & carbon dioxide & $\mathrm{CO_2}$ & 7 & -186.988364032 & -1.610188516 & -188.598552548\\
&&& 10 & -186.995427927 & -1.596452196 & -188.591880124\\
&&& 13 & -186.995791606 & -1.595764740 & -188.591556346\\
\hline
22 & carbon monoxide & $\mathrm{CO}$ & 7 & -112.281345285 & -0.992300746 & -113.273646031\\
&&& 10 & -112.285504026 & -0.984134277 & -113.269638303\\
&&& 13 & -112.285615989 & -0.983912324 & -113.269528313\\
\hline
23 & sulfur dioxide & $\mathrm{SO_2}$ & 7 & -546.496205115 & -1.793754960 & -548.289960075\\
&&& 10 & -546.511157831 & -1.766702310 & -548.277860141\\
&&& 13 & -546.511938404 & -1.765360445 & -548.277298849\\
\hline
24 & chlorine fluoride & $\mathrm{ClF}$ & 7 & -558.307091988 & -1.324975187 & -559.632067174\\
&&& 10 & -558.319880823 & -1.301800118 & -559.621680941\\
&&& 13 & -558.320284106 & -1.301095827 & -559.621379933\\
\hline
25 & chloromethane & $\mathrm{CH_3Cl}$ & 7 & -498.611094695 & -1.178301853 & -499.789396547\\
&&& 10 & -498.618120060 & -1.165682221 & -499.783802281\\
&&& 13 & -498.618120075 & -1.165682221 & -499.783802297\\
\hline
26 & methanethiol & $\mathrm{CH_3SH}$ & 7 & -437.229502936 & -1.156526879 & -438.386029815\\
&&& 10 & -437.236324161 & -1.144317347 & -438.380641508\\
&&& 13 & -437.236324166 & -1.144317347 & -438.380641513\\
\hline
27 & silicon monoxide & $\mathrm{SiO}$ & 7 & -363.415389543 & -0.955510647 & -364.370900190\\
&&& 10 & -363.426681421 & -0.935368931 & -364.362050352\\
&&& 13 & -363.426694340 & -0.935355298 & -364.362049637\\
\hline
28 & carbon monosulfide & $\mathrm{CS}$ & 7 & -434.943586897 & -0.910997092 & -435.854583989\\
&&& 10 & -434.946482679 & -0.905577244 & -435.852059923\\
&&& 13 & -434.946486064 & -0.905570989 & -435.852057053\\
\hline
29 & hypochlorous acid & $\mathrm{HClO}$ & 7 & -534.340778675 & -1.288272701 & -535.629051377\\
&&& 10 & -534.349534877 & -1.272137454 & -535.621672332\\
&&& 13 & -534.349811698 & -1.271644832 & -535.621456530\\
\end{longtable}
\endgroup
Figures ~\ref{fig:nIpts_tote1} and ~\ref{fig:nIpts_tote2} showcase the convergence of total energies, in Hartree, as the number of interpolation vectors increases from $\alpha_\text{Ipts}$=5 to $\alpha_\text{Ipts}$=15 for the 29-molecule data set.

\section{Vertex-corrected ionization potentials}
\subsection{29-molecule set}
Using HF/cc-pVQZ and $\alpha_\text{Ipts}=10$ as the reference setup for the 29-molecule set, we computed a series of fully self-consistent vertex-corrected THC-sc$GW\Gamma_\Sigma$ schemes: scGW-SOX, scGW-SOSEX, scGW-2SOSEX, and scG3W2.

The screened interaction was treated either statically, $W(i\Omega_m=0)$, or dynamically through the full $W(i\Omega_m)$. The corresponding first ionization potentials are summarized below.

\begin{table}[H]
\resizebox{\textwidth}{!}{
    \begin{tabular}{cllcrrrrrcrrrrrrrcr}
    \hline 
    \hline
    \multicolumn{3}{c}{Integral Type} && \multicolumn{5}{c}{DF} && \multicolumn{7}{c}{THC}&& \\
    \cline{1-3}\cline{5-9}\cline{11-17}
    Index &Molecule Name & Molecule &&HF &CCSD(T)~\cite{Maggio:GWVertexCorrected:2017} &\gowo@HF~\cite{Ming_gw100} &\gowov@HF~\cite{Maggio:GWVertexCorrected:2017} &\scgw~\cite{Ming_gw100} &&\scgw &\scgw-SOX &\scgw-stat-SOSEX &\scgw-stat-2SOSEX & \scgtwt &\scgw-dyn-SOSEX & \scgw-dyn-2SOSEX &&Experiment$^*$\\
    \hline
     1&     hydrogen&           $\mathrm{H_2}$   &&    -16.18&    -16.39&    -16.59&    -16.25&    -16.24& &    -16.25&    -16.29&    -16.27&    -16.25&    -16.25&    -16.36&    -16.43&&    -15.42~\cite{mccormack_measurement_1989}              \\
     2&     lithium dimer&      $\mathrm{Li_2}$  &&     -4.95&     -5.17&     -5.38&     -5.16&     -4.94& &     -4.98&     -5.07&     -5.03&     -4.99&     -5.01&     -5.09&     -5.12&&     -4.73~\cite{dugourd_measurements_1992}               \\
     3&     nitrogen&           $\mathrm{N_2}$   &&    -16.69&    -15.49&    -16.56&    -16.06&    -15.57& &    -15.54&    -16.64&    -16.03&    -15.40&    -15.72&    -16.09&    -15.58&&    -15.58~\cite{trickl_stateselective_1989}              \\
     4&     phosphorus dimer&   $\mathrm{P_2}$   &&    -10.10&    -10.76&    -10.71&    -10.92&     -9.85& &     -9.83&    -10.15&    -10.01&     -9.85&     -9.91&    -10.06&     -9.99&&    -10.62~\cite{bulgin_hei_1976}                         \\
     5&     chlorine&           $\mathrm{Cl_2}$  &&    -12.07&    -11.62&    -11.98&    -11.52&    -11.24& &    -11.21&    -11.58&    -11.41&    -11.23&    -11.31&    -11.49&    -11.40&&    \textit{-11.49}~\cite{dyke_ionization_1984}           \\
     6&     methane&            $\mathrm{CH_4}$  &&    -14.84&    -14.40&    -14.92&    -14.37&    -14.32& &    -14.30&    -14.45&    -14.36&    -14.27&    -14.31&    -14.46&    -14.47&&    \textit{-13.60}~\cite{bieri_304-nm_1980}              \\
     7&     ethylene&           $\mathrm{C_2H_4}$&&    -10.28&    -10.69&    -10.88&    -10.49&    -10.18& &    -10.20&    -10.22&    -10.22&    -10.19&    -10.19&    -10.32&    -10.39&&    \textit{-10.68}~\cite{bieri_304-nm_1980}              \\
     8&     ethyne&             $\mathrm{C_2H_2}$&&    -11.18&    -11.42&    -11.77&    -11.26&    -10.96& &    -10.97&    -11.12&    -11.05&    -10.97&    -11.00&    -11.14&    -11.16&&    \textit{-11.49}~\cite{bieri_304-nm_1980}              \\
     9&     silane&             $\mathrm{SiH_4}$ &&    -13.24&    -12.82&    -13.35&    -12.94&    -12.80& &    -12.77&    -12.95&    -12.86&    -12.77&    -12.80&    -12.94&    -12.93&&    \textit{-12.30}~\cite{roberge_far_1978}               \\
    10&     lithium hydride&    $\mathrm{LiH}$   &&     -8.21&     -7.94&     -8.28&     -7.87&     -7.94& &     -7.89&     -8.30&     -8.12&     -7.95&     -8.02&     -8.21&     -8.15&&    -7.90~\cite{lias2025ion}                              \\
    11&    ammonia&             $\mathrm{NH_3}$  &&    -11.66&    -10.92&    -11.40&    -11.04&    -10.86& &    -10.84&    -11.39&    -11.05&    -10.74&    -10.89&    -11.17&    -10.98&&    \textit{-10.82}~\cite{baumgaertel_photoelectron_1989} \\
    12&    phosphine&           $\mathrm{PH_3}$  &&    -10.58&    -10.49&    -10.92&    -10.47&    -10.28& &    -10.29&    -10.45&    -10.36&    -10.27&    -10.31&    -10.45&    -10.45&&    \textit{-10.59}~\cite{cowley_lewis_1982}              \\
    13&    hydrogen sulfide&    $\mathrm{H_2S}$  &&    -10.48&    -10.43&    -10.69&    -10.39&    -10.09& &    -10.09&    -10.37&    -10.22&    -10.07&    -10.14&    -10.32&    -10.27&&    \textit{-10.50}~\cite{bieri_304-nm_1982}              \\
    14&    hydrogen fluoride&   $\mathrm{HF}$    &&    -17.63&    -16.09&    -16.49&    -15.72&    -16.26& &    -16.26&    -16.98&    -16.58&    -16.20&    -16.39&    -16.71&    -16.47&&    \textit{-16.12}~\cite{banna_molecular_1975}           \\
    15&    sodium chloride&     $\mathrm{NaCl}$  &&     -9.61&     -9.13&     -9.43&     -9.06&     -8.93& &     -8.91&     -9.31&     -9.12&     -8.93&     -9.01&     -9.23&     -9.15&&    \textit{ -9.80}~\cite{potts_photoelectron_1977}       \\
    16&    hydrogen cyanide&    $\mathrm{HCN}$   &&    -13.51&    -13.64&    -14.07&    -13.40&    -13.22& &    -13.23&    -13.43&    -13.34&    -13.24&    -13.28&    -13.43&    -13.42&&    \textit{-13.61}~\cite{kreile_experimental_1982}       \\
    17&    hydrazine&           $\mathrm{N_2H_4}$&&    -10.70&    -10.24&    -10.38&    -10.28&     -9.71& &     -9.69&    -10.19&     -9.87&     -9.56&     -9.72&     -9.98&     -9.79&&    \textit{ -8.98}~\cite{vovna_1975}                     \\
    18&    methanol&            $\mathrm{CH_3OH}$&&    -12.31&    -11.08&    -11.79&    -11.06&    -11.16& &    -11.13&    -11.65&    -11.34&    -11.03&    -11.19&    -11.45&    -11.26&&    \textit{-10.96}~\cite{vorobev_mass_1989}              \\
    19&    hydrogen peroxide&   $\mathrm{H_2O_2}$&&    -13.28&    -11.49&    -12.32&    -11.39&    -11.66& &    -11.67&    -12.54&    -12.01&    -11.51&    -11.77&    -12.11&    -11.74&&    \textit{-11.70}~\cite{ashmore_study_1977}             \\
    20&    water&               $\mathrm{H_2O}$  &&    -13.83&    -12.64&    -13.11&    -12.55&    -12.73& &    -12.70&    -13.38&    -12.98&    -12.60&    -12.79&    -13.10&    -12.86&&    \textit{-12.62}~\cite{kimura_handbook_1981}           \\
    21&    carbon dioxide&      $\mathrm{CO_2}$  &&    -14.81&    -13.78&    -14.46&    -13.83&    -13.66& &    -13.67&    -14.40&    -13.97&    -13.57&    -13.78&    -14.05&    -13.76&&    \textit{-13.77}~\cite{eland_photoionization_1977}     \\
    22&    carbon monoxide&     $\mathrm{CO}$    &&    -15.38&    -14.05&    -15.25&    -14.71&    -14.08& &    -14.07&    -14.99&    -14.48&    -13.92&    -14.22&    -14.53&    -14.08&&    \textit{-14.01}~\cite{potts_observation_1974}         \\
    23&    sulfur dioxide&      $\mathrm{SO_2}$  &&    -13.48&    -12.41&    -13.16&    -12.64&    -12.21& &    -12.22&    -12.91&    -12.52&	 -12.11&	-12.34&	   -12.60&	  -12.29&&    \textit{-12.50}~\cite{kimura_handbook_1981}           \\
    24&    chlorine fluoride&   $\mathrm{ClF}$   &&    -13.46&    -12.82&    -13.29&    -13.14&    -12.49& &    -12.51&    -13.02&    -12.76&    -12.50&    -12.62&    -12.83&    -12.67&&    \textit{-12.77}~\cite{dyke_ionization_1984}           \\
    25&    chloromethane&       $\mathrm{CH_3Cl}$&&    -11.85&    -11.41&    -11.76&    -11.37&    -11.16& &    -11.12&    -11.42&    -11.27&    -11.12&    -11.19&    -11.36&    -11.32&&    \textit{-11.29}~\cite{kimura_handbook_1981}          \\
    26&    methanethiol&        $\mathrm{CH_3SH}$&&     -9.73&     -9.49&     -9.84&     -9.62&     -9.16& &     -9.15&     -9.42&	   -9.28&	  -9.13&	 -9.20&     -9.37&	   -9.32&&    \textit{ -9.44}~\cite{cradock_photoelectron_1972}     \\
    27&    silicon monoxide&    $\mathrm{SiO}$   &&    -11.89&    -11.55&    -12.04&    -11.63&    -11.25& &    -11.23&    -11.64&    -11.42&    -11.17&    -11.30&    -11.49&    -11.34&&    -11.30~\cite{nakasgawa_mass_1981}                     \\
    28&    carbon monosulfide&  $\mathrm{CS}$    &&    -12.56&    -11.45&    -12.55&    -12.29&    -11.34& &    -11.32&    -12.26&    -11.72&    -11.17&    -11.46&    -11.78&    -11.35&&    -11.33~\cite{king_photoelectron_1972}                 \\
    29&    hypochlorous acid&  	$\mathrm{HClO}$  &&    -12.13&    -11.30&    -11.84&    -11.41&    -11.10& &    -11.09&    -11.61&    -11.32&    -11.03&    -11.17&    -11.40&    -11.22&&    -11.12~\cite{colbourne_photoelectron_1978}            \\
    \hline
    \multicolumn{3}{c}{MAE from experiment}      &&      0.78&      0.23&      0.65&      0.37&      0.30& &      0.30&      0.54&      0.37&      0.32&      0.30&      0.39&      0.29&&\\ 
    \hline
    \hline
    \end{tabular}
}
\caption{First ionization potentials (eV) calculated using HF, CCSD(T), \gowo@HF, \gowov@HF, \scgw, THC-\scgw, SOX, static-SOSEX, static-2SOSEX, static-\gtwt, dynamic-SOSEX, and dynamic-2SOSEX in the PW (for \gowov@HF only) and cc-pVQZ basis sets. Vertical IP values are shown in italics.}
\label{tab:ip_all}
\end{table}

\subsection{$GW$100 data set}
The def2-TZVPP basis set is not available as an all-electron basis for fifth-period elements and heavier, for which effective core potentials are required. Since the present work focuses on all-electron calculations, molecules containing fifth-period or heavier element were excluded from our dataset. After this exclusion, a total of 93 molecules were retained in the $GW$100 benchmark set.

$GW$100 ionization energies and electron affinities are reported in eV and preserve the sign convention used in the manuscript. The $GW$100 worksheets contain the $GW$100 index, CAS identifier, number of atomic orbitals, molecule name and calculated values. The marker \texttt{*} denotes an unavailable numerical result. The IP and EA worksheets should be interpreted separately because the removal and addition states have qualitatively different orbital character. For many molecules, the discrete electron-addition state does not correspond to a stable experimental anion; consequently, no fundamental gaps are formed or analyzed from these data.

\begin{landscape}
\begin{table}[htbp]
\centering
\caption{First IPs (in eV) calculated by sc$GW$ and sc$GW\Gamma_\Sigma$ methods for the $GW$100 test set. $^*$THC-sc$GW\Gamma$-SOX results that exhibit severe leakage in the self-consistent loop were not reported. $^{**}$Cases that fail to converge under the most tolerant threshold of $< 10^{-4}$ a.u. within a reasonable computational time.}
\label{tab:GW100}
\resizebox{\textwidth}{!}{
\begin{tabular}{llrlrrrrrrrrrrrr}
\hline
\hline
& & & & & & & & \multicolumn{6}{c}{THC-sc$GW\Gamma_\Sigma$} & & \\
\cline{8-14}
GW100 Index&CAS $\#$ &$N_{\mathrm{AO}}$ & Molecule&G0W0@PBE&G0W0@HF&DF-scGW&THC-scGW&THC-scGW-SOX&THC-scGW-static-SOSEX&THC-scGW-static-2SOSEX&THC-scGW-static-G3W2&THC-scGW-dynamic-SOSEX&THC-scGW-dynamic-2SOSEX&EOM-CCSD~\cite{LangeBerkelbach:EOMGW:2018}&CCSD(T)~\cite{brunevalGWMiracleManyBody2021}\\
1&7440-59-7&14&helium&-23.75&-24.58&-24.41&-24.44&-24.92&-24.73&-24.54&-24.60&-24.82&-24.73&-24.51&-24.51\\
2&7440-01-9&31&neon&-20.49&-21.39&-21.39&-21.39&-22.18&-21.82&-21.46&-21.62&-21.93&-21.70&-21.21&-21.32\\
3&7440-37-1&42&argon&-15.02&-15.76&-15.26&-15.25&-15.70&-15.53&-15.35&-15.42&-15.62&-15.53&-15.63&-15.54\\
4&7439-90-9&48&krypton&-13.37&-14.00&-13.64&-13.67&-14.01&-13.88&-13.75&-13.80&-13.96&-13.92&-13.98&-13.94\\
6&1333-74-0&28&hydrogen&-15.90&-16.48&-16.15&-16.19&-16.26&-16.23&-16.21&-16.21&-16.32&-16.39&-16.40&-16.40\\
7&14452-59-6&38&lithium dimer&-5.05&-5.32&-4.97&-4.87&-5.02&-4.97&-4.91&-4.93&-5.02&-5.03&-5.27&-5.27\\
8&25681-79-2&64&sodium dimer&-4.92&-4.97&-4.63&-4.62&-4.66&-4.65&-4.64&-4.64&-4.70&-4.75&-4.94&-4.95\\
9&39297-86-4&128&sodium tetramer&-4.11&-4.29&-3.83&-3.82&-3.82&-3.84&-3.85&-3.84&-3.88&-3.93&-4.25&-4.24\\
10&39297-88-6&192&sodium hexamer&-4.25&-4.47&-3.96&-3.91&-4.08&-4.00&-3.91&-3.95&-4.04&-4.01&-4.37&-4.39\\
11&25681-80-5&66&potassium dimer&-3.94&-4.06&-3.73&-3.53&-3.70&-3.63&-3.57&-3.59&-3.69&-3.68&-4.08&-4.07\\
13&7727-37-9&62&nitrogen&-14.74&-16.35&-15.45&-15.46&-16.59&-15.98&-15.36&-15.67&-16.03&-15.52&-15.60&-15.48\\
14&12185-09-0&84&phosphorus dimer&-10.05&-10.55&-9.73&-9.73&-10.09&-9.94&-9.77&-9.84&-9.98&-9.89&-10.59&-10.53\\
15&23878-46-8&96&arsenic dimer&-9.34&-9.74&-9.05&-9.01&*&-9.18&-9.04&-9.09&-9.21&-9.15&-9.91&-9.85\\
16&7782-41-4&62&fluorine&-14.83&-16.31&-15.80&-15.79&*&-16.37&-15.71&-16.05&-16.42&-15.88&-15.53&-15.58\\
17&7782-50-5&84&chlorine&-10.90&-11.77&-11.08&-11.08&-11.51&-11.32&-11.14&-11.22&-11.39&-11.28&-11.46&-11.41\\
18&7726-95-6&96&bromine&-9.98&-10.75&-10.21&-10.24&-10.57&-10.43&-10.28&-10.34&-10.49&-10.42&-10.54&-10.52\\
20&74-82-8&87&methane&-13.91&-14.77&-14.25&-14.25&-14.42&-14.34&-14.25&-14.29&-14.43&-14.44&-14.38&-14.37\\
21&74-84-0&146&ethane&-12.32&-13.18&-12.62&-12.60&-12.68&-12.63&-12.57&-12.60&-12.72&-12.75&-12.71&-12.71\\
22&74-98-6&205&propane&-11.66&-12.63&-12.03&-11.99&-12.08&-12.02&-11.96&-11.99&-12.10&-12.12&-12.05&-12.03\\
23&106-97-8&264&butane&-11.23&-12.19&-11.65&-11.67&-11.72&-11.68&-11.64&-11.66&-11.77&-11.79&-11.56&-11.57\\
24&74-85-1&118&ethylene&-10.25&-10.76&-10.14&-10.14&-10.18&-10.18&-10.15&-10.15&-10.27&-10.34&-10.69&-10.67\\
25&74-86-2&90&ethyne/acetylene&-10.94&-11.59&-10.89&-10.90&-11.07&-11.00&-10.95&-10.95&-11.08&-11.09&-11.55&-11.42\\
26&12184-80-4&124&tetracarbon&-10.64&-11.62&-10.69&-10.69&-11.25&-10.97&-10.68&-10.81&-11.02&-10.83&-11.27&-11.24\\
27&75-19-4&177&cyclopropane&-10.46&-11.30&-10.59&-10.59&-10.82&-10.70&-10.58&-10.63&-10.80&-10.77&-10.85&-10.87\\
28&71-43-2&270&benzene&-8.86&-9.53&-8.72&-8.72&-9.11&-8.88&-8.64&-8.76&-8.97&-8.84&-9.32&-9.36\\
29&629-20-9&360&cyclooctatetraene&-7.93&-8.67&-7.82&-7.82&-7.82&-7.86&-7.97&-7.86&-7.97&-8.13&-8.40&-8.40\\
30&542-92-7&239&cyclopentadiene&-8.23&-8.86&-8.07&-8.08&-8.18&-8.13&-8.08&-8.09&-8.22&-8.26&-8.69&-8.71\\
31&75-02-5&135&vinyl fluoride&-10.05&-10.82&-10.10&-10.10&-10.23&-10.18&-10.11&-10.14&-10.27&-10.28&-10.60&-10.55\\
32&75-01-4&146&vinyl chloride&-9.60&-10.36&-9.62&-9.62&-9.76&-9.72&-9.67&-9.68&-9.80&-9.83&-10.13&-10.09\\
33&593-60-2&152&vinyl bromide&-8.83&-9.46&-8.84&-8.82&-8.89&-8.87&-8.83&-8.84&-8.95&-8.99&-9.29&-9.27\\
35&75-73-0&155&tetrafluoromethane&-15.18&-16.85&-16.55&-16.49&-17.51&-16.98&-16.45&-16.71&-17.07&-16.65&-16.24&-16.23\\
36&56-23-5&199&tetrachloromethane&-10.77&-12.01&-11.14&-11.15&-11.84&-11.57&-11.32&-11.43&-11.64&-11.48&-11.60&-11.50\\
37&558-13-4&223&tetrabromomethane&-9.67&-10.77&-10.10&-10.45&-10.74&-10.56&-10.40&-10.48&-10.64&-10.52&-10.48&-10.41\\
39&7803-62-5&98&silane&-12.28&-13.25&-12.73&-12.73&-12.93&-12.84&-12.75&-12.78&-12.92&-12.90&-12.84&-12.80\\
40&7782-65-2&104&germane&-11.99&-12.88&-12.37&-12.38&-12.60&-12.50&-12.39&-12.43&-12.57&-12.54&-12.53&-12.50\\
41&1590-87-0&168&disilane&-10.21&-11.11&-10.44&-10.46&-10.64&-10.55&-10.47&-10.50&-10.63&-10.61&-10.71&-10.65\\
42&14868-53-2&378&pentasilane&-8.81&-9.82&-9.04&-9.03&-9.19&-9.12&-9.11&-9.09&*&*&-9.36&-9.27\\
43&7580-67-8&33&lithium hydride&-7.02&-8.17&-7.89&-7.86&-8.28&-8.08&-7.89&-7.97&-8.16&-8.07&-7.96&-7.96\\
44&7693-26-7&47&potassium hydride&-4.81&-6.29&-6.03&-5.99&-6.60&-6.29&-5.99&-6.13&-6.37&-6.19&-6.13&-6.13\\
45&13283-31-3&73&borane&-12.84&-13.67&-13.18&-13.23&-13.33&-13.28&-13.23&-13.25&-13.37&-13.40&-13.31&-13.27\\
46&19287-45-7&146&diborane&-11.79&-12.82&-12.22&-12.27&-12.38&-12.31&-12.24&-12.26&-12.38&-12.38&-12.29&-12.25\\
47&7664-41-7&73&ammonia&-10.29&-11.19&-10.77&-10.76&-11.32&-10.99&-10.68&-10.83&-11.10&-10.91&-10.77&-10.81\\
48&7782-79-8&107&hydrogen azide&-10.24&-11.09&-10.23&-10.24&-10.79&-10.49&-10.21&-10.34&-10.54&-10.34&-10.72&-10.68\\
49&7803-51-2&84&phosphine&-10.20&-10.81&-10.23&-10.23&-10.41&-10.32&-10.23&-10.27&-10.40&-10.39&-10.57&-10.52\\
50&7784-42-1&90&arsine&-10.04&-10.58&-10.08&-10.09&-10.22&-10.15&-10.09&-10.11&-10.23&-10.23&-10.42&-10.40\\
51&7783-06-4&70&hydrogen sulfide&-9.92&-10.52&-9.99&-9.98&-10.29&-10.15&-10.00&-10.06&-10.22&-10.17&-10.35&-10.31\\
52&7647-01-0&45&hydrogen fluoride&-15.29&-16.22&-16.13&-16.13&-16.88&-16.49&-16.11&-16.29&-16.61&-16.36&-15.90&-16.03\\
53&7664-39-3&56&hydrogen chloride&-12.13&-12.81&-12.28&-12.30&-12.67&-12.51&-12.35&-12.41&-12.59&-12.52&-12.64&-12.59\\
54&7789-24-4&50&lithium fluoride&-9.77&-11.37&-11.36&-11.36&-12.26&-11.77&-11.32&-11.56&-11.90&-11.58&-11.28&-11.32\\
55&7783-40-6&94&magnesium fluoride&-12.29&-13.79&-13.77&-13.80&-14.70&-14.22&-13.76&-14.00&-14.34&-14.01&-13.71&-13.71\\
56&7783-63-3&188&titanium fluoride&-13.77&-16.17&-15.47&-15.35&*&-16.22&-15.49&-15.87&-16.33&-15.73&-15.69&-15.41\\
57&7784-18-1&135&aluminium fluoride&-14.14&-15.63&-15.41&-15.44&-16.57&-15.96&-15.66&-15.78&-16.04&-15.82&-15.31&-15.32\\
58&13768-60-0&62&boron monofluoride&-10.43&-11.31&-10.61&-10.67&-10.97&-10.84&-10.70&-10.76&-10.92&-10.85&-11.20&-11.09\\
59&7783-60-0&166&sulfur tetrafluoride&-11.91&-13.30&-12.49&-12.46&-12.97&-12.72&-12.48&-12.59&-12.79&-12.73&-12.70&-12.59\\
60&7758-02-3&81&potassium bromide&-7.15&-8.21&-7.88&-7.89&-8.24&-8.09&-7.95&-8.01&-8.17&-8.11&-8.17&-8.13\\
61&17108-85-9&90&gallium monochloride&-9.42&-9.90&-9.34&-9.35&-9.42&-9.41&-9.38&-9.39&-9.47&-9.51&-9.79&-9.77\\
62&7647-14-5&74&sodium chloride&-7.90&-9.24&-8.77&-8.78&-9.22&-9.02&-8.83&-8.92&-9.11&-9.02&-9.12&-9.03\\
63&7786-30-3&116&magnesium chloride&-10.84&-11.93&-11.44&-11.45&-11.94&-11.71&-11.50&-11.59&-11.79&-11.68&-11.76&-11.66\\
65&10043-11-5&62&boron nitride&-11.08&-11.76&-11.03&-11.07&*&-11.37&-10.94&-11.20&-11.36&-11.04&-11.93&-11.98\\
66&74-90-8&76&hydrogen cyanide&-13.00&-13.87&-13.13&-13.15&-13.39&-13.30&-13.20&-13.24&-13.38&-13.36&-13.90&-13.72\\
67&17739-47-8&73&phosphorus mononitride&-10.99&-12.37&-11.59&-11.58&*&-11.85&-11.58&-11.73&-11.90&-11.72&-11.80&-11.81\\
68&302-01-2&118&hydrazine&-9.21&-10.17&-9.63&-9.61&-10.12&-9.82&-9.52&-9.67&-9.91&-9.73&-9.62&-9.69\\
69&50-00-0&90&formaldehyde&-10.17&-11.37&-10.82&-10.79&-11.40&-11.05&-10.70&-10.88&-11.13&-10.88&-10.78&-10.84\\
70&67-56-1&118&methanol&-10.46&-11.57&-11.04&-11.05&-11.58&-11.28&-10.97&-11.13&-11.38&-11.18&-10.18&-11.04\\
71&64-17-5&177&ethanol&-10.05&-11.29&-10.69&-10.67&-11.19&-10.90&-10.59&-10.76&-10.99&-10.80&-10.61&-10.69\\
72&75-07-0&149&acetaldehyde&-9.40&-10.81&-10.18&-10.17&-10.75&-10.42&-10.09&-10.75&-10.50&-10.28&-10.18&-10.21\\
73&60-29-7&295&ethyl ether&-9.21&-10.49&-9.81&-9.78&-10.34&-10.02&-9.68&-9.86&-10.11&-9.88&-9.75&-9.82\\
74&64-18-6&121&formic acid&-10.59&-11.95&-11.41&-11.40&-12.13&-11.73&-11.33&-11.53&-11.81&-11.52&-11.42&-11.42\\
75&7722-84-1&90&hydrogen peroxide&-10.87&-12.06&-11.54&-11.56&-12.47&-11.95&-11.46&-11.71&-12.04&-11.66&-11.39&-11.52\\
76&7732-18-5&59&water&-11.94&-12.87&-12.59&-12.60&-13.30&-12.90&-12.54&-12.72&-13.02&-12.78&-12.48&-12.57\\
77&124-38-9&93&carbon dioxide&-13.07&-14.22&-13.55&-13.55&-14.32&-13.90&-13.50&-13.70&-13.96&-13.66&-13.73&-13.71\\
78&75-15-0&115&carbon disulfide&-9.55&-10.33&-9.44&-9.46&-9.87&-9.67&-9.46&-9.55&-9.71&-9.58&-10.01&-9.98\\
79&463-58-1&104&carbon oxysulfide&-10.74&-11.55&-10.70&-10.73&-11.23&-10.98&-10.72&-10.84&-11.03&-10.85&-11.24&-11.17\\
80&1603-84-5&110&carbon oxyselenide&-10.00&-10.69&-9.98&-10.01&-10.43&-10.22&-10.00&-10.10&-10.26&-10.12&-10.50&-10.47\\
81&630-08-0&62&carbon monoxide&-13.43&-15.06&-13.97&-13.98&-14.93&-14.42&-13.86&-14.16&-14.46&-14.00&-14.37&-14.21\\
82&10028-15-6&93&ozone&-11.73&-13.54&-12.57&-12.57&*&-13.01&-12.31&-12.79&-13.01&-12.61&-12.79&-12.71\\
83&7446-09-5&104&sulfur dioxide&-11.61&-12.94&-12.03&-12.07&-12.79&-12.40&-12.02&-12.21&-12.44&-12.14&-12.37&-12.30\\
84&1304-56-9&50&beryllium monoxide&-9.16&-9.83&-9.75&-9.78&*&-10.02&-9.69&-9.89&-10.12&-9.91&-9.88&-9.94\\
85&1309-48-4&63&magnesium monoxide&-7.05&-7.95&-7.95&-8.00&*&-8.36&-7.83&-8.12&-8.42&-8.05&-8.17&-7.91\\
86&108-88-3&329&toluene&-8.49&-9.16&-8.37&-8.45&-8.83&-8.59&-8.36&-8.51&-8.68&-8.56&-8.90&-8.97\\
87&100-41-4&388&ethylbenzene&-8.43&-9.13&-8.32&-8.41&-8.72&-8.58&-8.35&-8.44&-8.69&-8.59&-8.85&-8.92\\
88&392-56-3&372&hexafluorobenzene&-9.28&-10.63&-9.50&-9.48&-10.11&-9.76&-9.44&-9.59&-9.86&-9.64&-10.15&-9.93\\
89&108-95-2&301&phenol&-8.22&-9.03&-8.17&-8.19&-8.61&-8.23&-8.11&-8.36&-8.45&-8.30&-8.69&-8.70\\
90&62-53-3&315&aniline&-7.49&-8.35&-7.52&-7.50&-7.90&-7.65&-7.40&-7.53&-7.74&-7.60&-7.98&-8.04\\
91&110-86-1&256&pyridine&-8.87&-9.91&-9.12&-9.10&-9.46&-9.25&-9.03&-9.14&-9.34&-9.23&-9.72&-9.73\\
92&73-40-5&411&guanine&-7.52&-8.44&-7.48&-7.48&-7.80&-7.61&-7.43&-7.52&*&*&-8.04&-8.03\\
93&73-24-5&380&adenine&-7.80&-8.71&-7.78&-7.77&-8.11&-7.91&-7.74&-7.80&*&*&-8.33&-8.33\\
94&71-30-7&318&cytosine&-8.08&-9.28&-8.40&-8.39&-8.96&-8.61&-8.32&-8.47&*&*&-8.78&-8.77\\
95&65-71-4&363&thymine&-8.49&-9.68&-8.70&-8.69&-9.10&-8.86&-8.62&-8.74&*&*&-9.15&-9.08\\
96&66-22-8&304&uracil&-8.86&-10.09&-9.13&-9.11&-9.61&-9.33&-9.04&-9.19&*&*&-9.57&-9.48\\
97&57-13-6&180&urea&-9.18&-10.68&-10.05&-10.06&-10.88&-10.38&-9.98&-10.18&-10.46&-10.18&-10.08&-10.05\\
99&12190-70-4&128&copper dimer&-7.61&-7.20&-6.96&-7.00&-6.84&-6.89&-6.95&-6.91&-6.96&-7.09&-7.38&-7.57\\
100&544-92-3&126&copper cyanide&-9.80&-11.29&-10.67&-10.73&-11.02&-10.89&-10.38&-10.78&-10.97&-10.61&-10.69&-10.85\\
\hline
\hline
\end{tabular}
}
\end{table}
\end{landscape}
\restoregeometry

\begin{landscape}
\begin{table}[htbp]
\centering
\caption{First EAs (in eV) calculated by sc$GW$ and sc$GW\Gamma_\Sigma$ methods for the $GW$100 test set. $^*$THC-sc$GW\Gamma$-SOX results that exhibit severe leakage in the self-consistent loop were not reported. $^{**}$Cases that fail to converge under the most tolerant threshold of $< 10^{-4}$ a.u. within a reasonable computational time.}
\label{tab:GW100_EA}
\resizebox{\textwidth}{!}{
\begin{tabular}{llrlrrrrrrrrrrr}
\hline
\hline
& & & & & & & & \multicolumn{6}{c}{THC-sc$GW\Gamma_\Sigma$} &  \\
\cline{8-14}
GW100 Index&CAS $\#$ &$N_{\mathrm{AO}}$ & Molecule&G0W0@PBE&G0W0@HF&DF-scGW&THC-scGW&THC-scGW-SOX&THC-scGW-static-SOSEX&THC-scGW-static-2SOSEX&THC-scGW-static-G3W2&THC-scGW-dynamic-SOSEX&THC-scGW-dynamic-2SOSEX&EOM-CCSD~\cite{LangeBerkelbach:EOMGW:2018}\\
1&7440-59-7&14&helium&21.81&22.11&22.12&22.10&22.36&22.31&22.25&22.26&22.32&22.28&22.22\\
2&7440-01-9&31&neon&20.71&21.19&20.95&20.96&21.01&20.99&20.97&20.99&20.99&20.98&20.84\\
3&7440-37-1&42&argon&14.57&14.79&14.62&14.52&14.84&14.75&14.66&14.69&14.74&14.65&14.73\\
4&7439-90-9&48&krypton&10.31&10.47&10.30&10.17&10.39&10.33&10.26&10.28&10.32&10.25&10.41\\
6&1333-74-0&28&hydrogen&4.31&4.30&4.26&4.25&4.27&4.29&4.31&4.30&4.29&4.30&4.22\\
7&14452-59-6&38&lithium dimer&-0.53&-0.06&-0.23&-0.28&-0.01&-0.06&-0.12&-0.11&-0.08&-0.15&-0.12\\
8&25681-79-2&64&sodium dimer&-0.55&-0.26&-0.38&-0.39&-0.15&-0.19&-0.24&-0.23&-0.21&-0.27&-0.26\\
9&39297-86-4&128&sodium tetramer&-0.28&-0.51&-0.72&-0.74&-0.45&-0.51&-0.57&-0.56&-0.53&-0.61&-0.53\\
10&39297-88-6&192&sodium hexamer&-0.05&-0.40&-0.69&-0.71&-0.29&-0.41&-0.53&-0.50&-0.43&-0.57&-0.49\\
11&25681-80-5&66&potassium dimer&-0.58&-0.33&-0.47&-0.69&-0.39&-0.45&-0.53&-0.51&-0.44&-0.50&-0.32\\
13&7727-37-9&62&nitrogen&2.76&3.04&2.62&2.65&3.07&3.04&2.98&2.97&3.05&3.02&3.05\\
14&12185-09-0&84&phosphorus dimer&-0.40&-0.27&-0.70&-0.68&-0.20&-0.29&-0.37&-0.38&-0.32&-0.41&-0.10\\
15&23878-46-8&96&arsenic dimer&-0.41&-0.40&-0.88&-0.90&*&-0.53&-0.60&-0.62&-0.57&-0.65&-0.10\\
16&7782-41-4&62&fluorine&-0.19&0.77&-0.08&-0.09&*&0.42&0.35&0.31&0.47&0.44&0.39\\
17&7782-50-5&84&chlorine&-0.38&-0.07&-0.52&-0.51&-0.18&-0.18&-0.20&-0.24&-0.18&-0.18&-0.19\\
18&7726-95-6&96&bromine&-1.03&-0.78&-1.29&-1.27&-1.02&-1.00&-1.00&-1.04&-1.00&-0.99&-0.94\\
20&74-82-8&87&methane&3.41&3.60&3.49&3.57&3.66&3.65&3.64&3.64&3.65&3.63&3.45\\
21&74-84-0&146&ethane&3.03&3.28&3.09&3.19&3.32&3.30&3.28&3.28&3.28&3.26&3.11\\
22&74-98-6&205&propane&2.83&3.13&2.88&3.01&3.17&3.14&3.11&3.12&3.12&3.09&2.95\\
23&106-97-8&264&butane&2.73&3.07&2.83&2.93&3.11&3.08&3.04&3.05&3.05&3.01&2.88\\
24&74-85-1&118&ethylene&2.31&2.76&2.41&2.38&2.79&2.74&2.68&2.67&2.73&2.67&2.63\\
25&74-86-2&90&ethyne/acetylene&3.26&3.71&3.25&3.24&3.57&3.54&3.51&3.49&3.53&3.49&3.50\\
26&12184-80-4&124&tetracarbon&-2.78&-2.25&-2.80&-2.79&-2.05&-2.25&-2.46&-2.40&-2.27&-2.49&-2.36\\
27&75-19-4&177&cyclopropane&3.31&3.64&3.40&3.47&3.66&3.63&3.60&3.61&3.64&3.55&3.46\\
28&71-43-2&270&benzene&1.31&1.80&1.29&1.25&2.07&1.81&1.55&1.65&1.81&1.56&1.78\\
29&629-20-9&360&cyclooctatetraene&0.25&0.85&0.29&0.20&0.72&0.72&0.77&0.64&0.73&0.75&0.79\\
30&542-92-7&239&cyclopentadiene&1.25&1.81&1.29&1.27&1.85&1.74&1.62&1.63&1.73&1.60&1.77\\
31&75-02-5&135&vinyl fluoride&2.46&2.87&2.50&2.48&2.93&2.86&2.79&2.79&2.86&2.79&2.80\\
32&75-01-4&146&vinyl chloride&1.75&2.21&1.81&1.79&2.26&2.18&2.10&2.10&2.17&2.09&2.12\\
33&593-60-2&152&vinyl bromide&1.68&2.12&1.72&1.69&2.15&2.07&1.98&1.99&2.07&1.98&2.02\\
35&75-73-0&155&tetrafluoromethane&4.92&5.14&4.84&4.93&5.10&5.06&5.02&5.03&5.04&4.99&4.89\\
36&56-23-5&199&tetrachloromethane&0.52&1.04&0.56&0.54&0.95&0.92&0.86&0.84&0.91&0.87&0.86\\
37&558-13-4&223&tetrabromomethane&-0.75&-0.30&-0.80&-0.85&-0.52&-0.45&-0.53&-0.51&-0.52&-0.54&-0.49\\
39&7803-62-5&98&silane&2.95&3.27&3.10&3.06&3.27&3.24&3.21&3.21&3.22&3.18&3.10\\
40&7782-65-2&104&germane&3.00&3.40&3.47&3.21&3.38&3.36&3.34&3.34&3.35&3.32&3.16\\
41&1590-87-0&168&disilane&1.92&2.51&2.19&2.12&2.48&2.42&2.36&2.37&2.40&2.33&2.27\\
42&14868-53-2&378&pentasilane&0.35&1.01&0.53&0.47&0.94&0.86&0.83&0.80&*&*&0.79\\
43&7580-67-8&33&lithium hydride&0.06&0.10&0.05&0.06&0.12&0.11&0.10&0.10&0.11&0.09&0.09\\
44&7693-26-7&47&potassium hydride&-0.05&-0.03&-0.10&-0.11&-0.03&-0.04&-0.05&-0.05&-0.04&-0.06&-0.04\\
45&13283-31-3&73&borane&0.20&0.66&0.42&0.45&0.68&0.65&0.62&0.62&0.64&0.60&0.33\\
46&19287-45-7&146&diborane&0.94&1.50&1.18&1.17&1.49&1.44&1.40&1.40&1.42&1.36&1.20\\
47&7664-41-7&73&ammonia&2.85&2.98&2.83&2.92&3.01&3.00&2.98&2.99&2.99&2.98&2.84\\
48&7782-79-8&107&hydrogen azide&1.72&2.03&1.51&1.52&2.09&1.98&1.86&1.88&1.98&1.88&2.02\\
49&7803-51-2&84&phosphine&2.88&3.13&2.94&2.97&3.24&3.20&3.16&3.15&3.19&3.14&2.95\\
50&7784-42-1&90&arsine&2.82&3.07&2.82&2.84&3.11&3.08&3.05&3.04&3.07&3.03&2.86\\
51&7783-06-4&70&hydrogen sulfide&2.77&2.95&2.80&2.85&3.00&2.97&2.95&2.95&2.96&2.93&2.79\\
52&7647-01-0&45&hydrogen fluoride&3.22&3.16&3.03&3.06&3.13&3.13&3.12&3.12&3.13&3.13&3.07\\
53&7664-39-3&56&hydrogen chloride&2.75&2.82&2.63&2.67&2.82&2.81&2.81&2.79&2.81&2.80&2.70\\
54&7789-24-4&50&lithium fluoride&0.20&-0.01&-0.04&-0.04&-0.01&-0.02&-0.03&-0.02&-0.02&-0.03&-0.02\\
55&7783-40-6&94&magnesium fluoride&-0.02&0.05&-0.11&-0.12&-0.02&-0.04&-0.06&-0.05&-0.05&-0.07&-0.04\\
56&7783-63-3&188&titanium fluoride&-0.43&-0.32&-1.26&-1.24&*&-0.81&-0.91&-0.92&-0.76&-0.81&-1.06\\
57&7784-18-1&135&aluminium fluoride&0.47&0.76&0.49&0.51&0.69&0.66&0.63&0.64&0.65&0.61&0.67\\
58&13768-60-0&62&boron monofluoride&1.36&1.62&1.36&1.36&1.66&1.62&1.57&1.57&1.61&1.56&1.51\\
59&7783-60-0&166&sulfur tetrafluoride&0.80&0.97&0.57&0.54&0.95&0.91&0.86&0.84&0.93&1.00&0.94\\
60&7758-02-3&81&potassium bromide&-0.23&-0.38&-0.48&-0.47&-0.41&-0.42&-0.43&-0.43&-0.43&-0.45&-0.45\\
61&17108-85-9&90&gallium monochloride&0.27&0.30&0.02&0.06&0.34&0.30&0.26&0.26&0.27&0.22&0.27\\
62&7647-14-5&74&sodium chloride&-0.38&-0.56&-0.60&-0.62&-0.58&-0.58&-0.59&-0.59&-0.59&-0.61&-0.59\\
63&7786-30-3&116&magnesium chloride&-0.21&-0.07&-0.29&-0.30&-0.09&-0.12&-0.16&-0.16&-0.14&-0.20&-0.19\\
65&10043-11-5&62&boron nitride&-3.68&-3.89&-4.12&-4.10&*&-3.65&-3.69&-3.74&-3.72&-3.78&-3.16\\
66&74-90-8&76&hydrogen cyanide&2.98&3.52&2.93&2.89&3.26&3.22&3.17&3.16&3.22&3.17&3.21\\
67&17739-47-8&73&phosphorus mononitride&0.17&0.30&-0.11&-0.10&*&0.31&0.19&0.21&0.29&0.19&0.46\\
68&302-01-2&118&hydrazine&2.46&2.67&2.46&2.55&2.70&2.67&2.64&2.65&2.66&2.63&2.51\\
69&50-00-0&90&formaldehyde&1.27&1.83&1.37&1.35&1.77&1.72&1.66&1.65&1.74&1.69&1.67\\
70&67-56-1&118&methanol&2.98&3.20&3.00&3.07&3.21&3.19&3.16&3.17&3.18&3.14&3.04\\
71&64-17-5&177&ethanol&2.75&3.04&2.81&2.89&3.05&3.03&2.99&3.00&3.01&2.97&2.86\\
72&75-07-0&149&acetaldehyde&1.32&2.09&1.57&1.55&2.06&1.98&1.88&1.89&1.99&1.91&1.91\\
73&60-29-7&295&ethyl ether&2.80&3.15&2.87&3.00&3.19&3.16&3.12&3.13&3.14&3.09&2.96\\
74&64-18-6&121&formic acid&2.26&3.23&2.36&2.34&2.87&2.77&2.66&2.68&2.79&2.70&2.70\\
75&7722-84-1&90&hydrogen peroxide&2.89&3.16&2.89&2.98&3.24&3.19&3.13&3.15&3.19&3.13&3.01\\
76&7732-18-5&59&water&2.94&3.01&2.87&2.93&3.02&3.01&3.00&3.00&3.00&2.99&2.88\\
77&124-38-9&93&carbon dioxide&2.92&2.97&2.72&2.76&2.99&2.94&2.88&2.90&2.91&2.84&2.80\\
78&75-15-0&115&carbon disulfide&0.15&0.14&-0.19&-0.19&0.40&0.26&0.10&0.13&0.25&0.11&0.29\\
79&463-58-1&104&carbon oxysulfide&1.59&1.73&1.39&1.38&1.95&1.83&1.70&1.72&1.84&1.72&1.85\\
80&1603-84-5&110&carbon oxyselenide&1.15&1.33&1.02&0.99&1.59&1.44&1.27&1.32&1.45&1.31&1.44\\
81&630-08-0&62&carbon monoxide&0.95&1.11&0.78&0.78&1.24&1.18&1.12&1.10&1.19&1.14&1.22\\
82&10028-15-6&93&ozone&-1.91&-2.00&-2.77&-2.75&*&-2.21&-2.28&-2.33&-2.21&-2.23&-1.52\\
83&7446-09-5&104&sulfur dioxide&-0.55&-0.54&-0.98&-0.98&-0.60&-0.60&-0.62&-0.66&-0.58&-0.58&-0.34\\
84&1304-56-9&50&beryllium monoxide&-2.07&-2.10&-2.19&-2.20&*&-2.02&-2.06&-2.06&-2.05&-2.09&-2.01\\
85&1309-48-4&63&magnesium monoxide&-1.46&-1.56&-1.96&-1.98&*&-1.67&-1.79&-1.75&-1.72&-1.83&-1.29\\
86&108-88-3&329&toluene&1.23&1.75&1.26&1.23&1.91&1.71&1.58&1.61&1.69&1.58&1.71\\
87&100-41-4&388&ethylbenzene&1.24&1.92&1.26&1.19&1.95&1.71&1.58&1.62&1.72&1.58&1.76\\
88&392-56-3&372&hexafluorobenzene&0.76&0.85&0.46&0.42&1.11&0.90&0.77&0.81&0.86&0.75&1.08\\
89&108-95-2&301&phenol&1.19&1.62&1.14&1.08&1.78&1.57&1.44&1.48&1.60&1.39&1.62\\
90&62-53-3&315&aniline&1.38&1.85&1.36&1.28&2.00&1.78&1.65&1.69&1.79&1.68&1.83\\
91&110-86-1&256&pyridine&0.75&1.25&0.73&0.67&1.42&1.19&1.05&1.09&1.18&1.04&1.24\\
92&73-40-5&411&guanine&1.03&2.03&1.12&1.01&1.89&1.62&1.32&1.43&*&*&1.57\\
93&73-24-5&380&adenine&0.72&1.36&0.71&0.67&1.52&1.26&1.05&1.08&*&*&1.28\\
94&71-30-7&318&cytosine&0.52&0.92&0.47&0.41&1.10&0.92&0.76&0.78&*&*&0.92\\
95&65-71-4&363&thymine&0.29&0.76&0.23&0.21&0.84&0.71&0.58&0.59&*&*&0.77\\
96&66-22-8&304&uracil&0.26&0.68&0.23&0.18&0.78&0.66&0.54&0.55&0.67&0.56&0.70\\
97&57-13-6&180&urea&2.20&2.49&2.25&2.38&2.53&2.50&2.47&2.48&2.49&2.45&2.33\\
99&12190-70-4&128&copper dimer&-0.84&-0.17&-0.54&-0.63&-0.17&-0.32&-0.49&-0.42&-0.35&-0.54&-0.34\\
100&544-92-3&126&copper cyanide&-1.58&-0.79&-1.15&-1.21&-0.89&-0.97&-1.07&-1.02&-1.00&-1.11&-0.98\\
\hline
\hline
\end{tabular}
}
\end{table}
\end{landscape}
\restoregeometry

\begin{figure}[htbp]
\centering
\includegraphics[width=\textwidth]{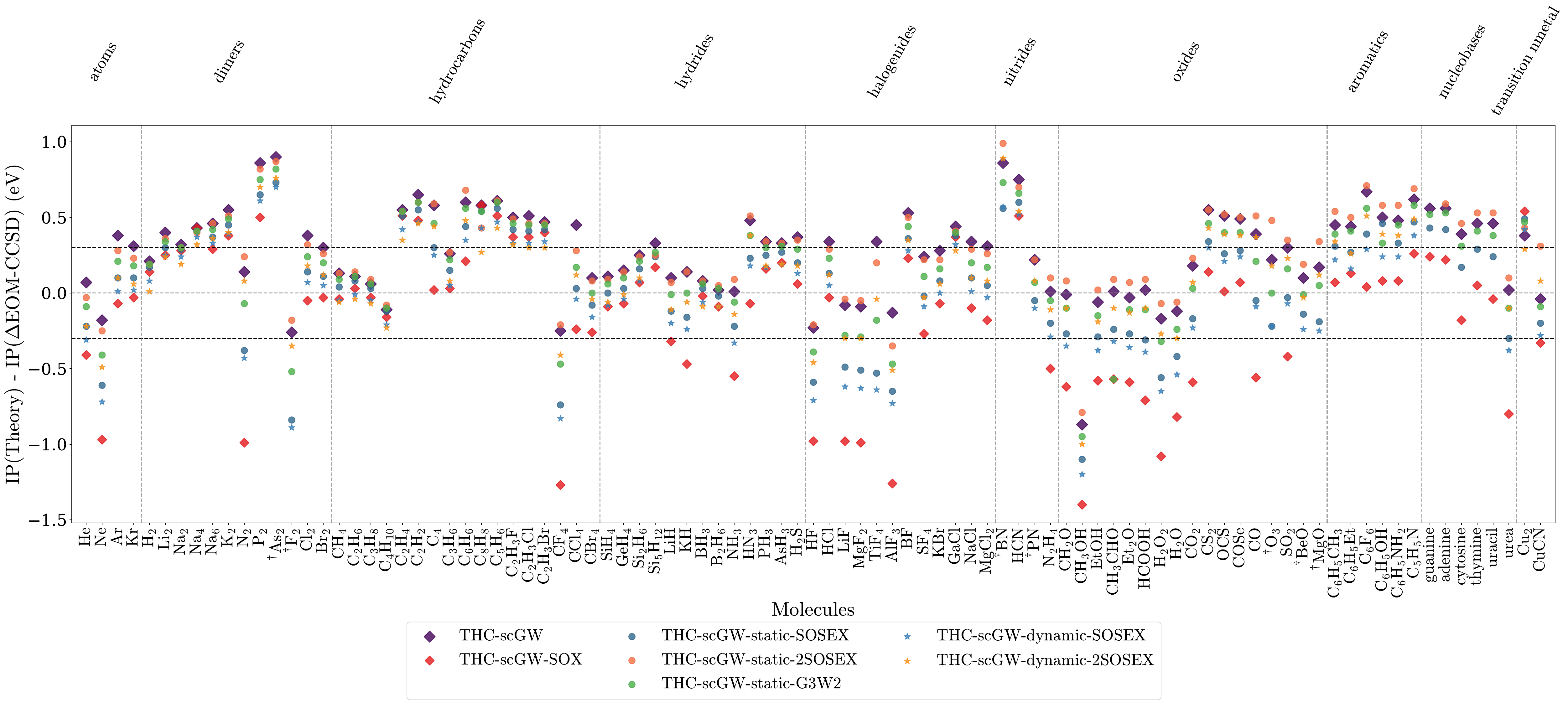}
\caption{Molecule-resolved signed errors of the $GW$100 first ionization potentials relative to EOM-CCSD~\cite{LangeBerkelbach:EOMGW:2018}. Dashed horizontal lines indicate $\pm 0.3$~eV.}
\label{fig:GW100_IP_EOM}
\end{figure}

\section{Intermediate representation grid choice}
We selected the Intermediate Representation (IR) grid size $\Lambda$ so that the grid spans the full range of molecular-orbital energies. More specifically, we chose a positive integer $n$ such that $\Lambda = 10^n > \beta\Delta E$, where $\beta$ is the inverse temperature and $\Delta E = E^{\mathrm{MO}}_{\mathrm{Max}} - E^{\mathrm{MO}}_{\mathrm{Min}}$ is evaluated at the Hartree--Fock level.

For a few cases in $GW$100 subset, THC-sc$GW\Gamma_\Sigma$ calculations exhibited large and progressively increasing IR-grid leakage. For these systems, we increased $\Lambda$ manually and repeated the calculation. This behavior likely indicates that the original grid was not sufficiently dense for those molecules.

For THC-sc$GW$ calculations, CDIIS was activated from iteration 5 with a maximum subspace size of 10, and a simple damping factor of 0.5 was used. For THC-sc$GW\Gamma_\Sigma$ calculations, CDIIS was activated from iteration 10 with a maximum subspace size of 12, and a smaller damping factor of 0.2 was used.

\subsection{29-molecule grid choices}
For the 29-molecule set, the IR-grid choices used in the THC-sc$GW\Gamma_\Sigma$ calculations are summarized in Table~\ref{tab:29-grid}.

\begingroup
\small
\setlength{\tabcolsep}{3pt}
\renewcommand{\arraystretch}{0.7}
\begin{longtable}{r l l c c c}
\caption{IR-grid choices used for the 29-molecule THC-sc$GW\Gamma_\Sigma$ calculations.}
\label{tab:29-grid}\\
\hline
\textbf{G0W0$\Gamma$29 \#} & \textbf{Molecule Name} & \textbf{Molecule} & \textbf{$N_{\mathrm{AO}}$} & \textbf{Grid $\Lambda$} & \textbf{Threshold (a.u.)} \\
\hline
\endfirsthead
\multicolumn{4}{c}{\tablename\ \thetable\ -- continued from previous page} \\
\hline
\textbf{G0W0$\Gamma$29 \#} & \textbf{Molecule Name} & \textbf{Molecule} & \textbf{$N_{\mathrm{AO}}$} & \textbf{Grid $\Lambda$} & \textbf{Threshold (a.u.)} \\
\hline
\endhead
\hline
\multicolumn{6}{r}{Continued on next page} \\
\endfoot
\hline \hline
\endlastfoot
1 & hydrogen & $\mathrm{H_2}$ & 60 &$10^{5}$ & $10^{-9}$ \\
2 & lithium dimer & $\mathrm{Li_2}$ & 110 &$10^{5}$ & $10^{-9}$ \\
3 & nitrogen & $\mathrm{N_2}$ & 110 & $10^{5}$ & $10^{-9}$ \\
4 & phosphorus dimer & $\mathrm{P_2}$ & 118 & $10^{6}$ & $10^{-9}$ \\
5 & chlorine & $\mathrm{Cl_2}$ & 118 & $10^{6}$ & $10^{-9}$ \\
6 & methane & $\mathrm{CH_4}$ & 175 & $10^{5}$ & $10^{-9}$ \\
7 & ethylene & $\mathrm{C_2H_4}$ & 230& $10^{5}$ & $10^{-9}$ \\
8 & ethyne & $\mathrm{C_2H_2}$ & 170 & $10^{6}$ & $10^{-9}$ \\
9 & silane & $\mathrm{SiH_4}$ & 179 & $10^{6}$ & $10^{-9}$ \\
10 & lithium hydride & $\mathrm{LiH}$ & 85 & $10^{5}$ & $10^{-9}$ \\
11 & ammonia & $\mathrm{NH_3}$ & 145 & $10^{5}$ & $10^{-9}$ \\
12 & phosphine & $\mathrm{PH_3}$ & 149 & $10^{5}$ & $10^{-9}$ \\
13 & hydrogen sulfide & $\mathrm{H_2S}$ & 119 & $10^{5}$ & $10^{-9}$ \\
14 & hydrogen fluoride & $\mathrm{HF}$ & 85 & $10^{5}$ & $10^{-9}$ \\
15 & sodium chloride & $\mathrm{NaCl}$ & 118 & $10^{6}$ & $10^{-9}$ \\
16 & hydrogen cyanide & $\mathrm{HCN}$ & 140 & $10^{5}$ & $10^{-9}$ \\
17 & hydrazine & $\mathrm{N_2H_4}$ & 230 & $10^{5}$ & $10^{-9}$ \\
18 & methanol & $\mathrm{CH_3OH}$ & 230 & $10^{5}$ & $10^{-9}$ \\
19 & hydrogen peroxide & $\mathrm{H_2O_2}$ & 170 & $10^{5}$ & $10^{-9}$ \\
20 & water & $\mathrm{H_2O}$ & 115 & $10^{5}$ & $10^{-9}$ \\
21 & carbon dioxide & $\mathrm{CO_2}$ & 165 & $10^{5}$ & $10^{-9}$ \\
22 & carbon monoxide & $\mathrm{CO}$ & 110 & $10^{5}$ & $10^{-9}$ \\
23 & sulfur dioxide & $\mathrm{SO_2}$ & 169 & $10^{6}$ & $10^{-9}$ \\
24 & chlorine fluoride & $\mathrm{ClF}$ & 114 & $10^{6}$ & $10^{-9}$ \\
25 & chloromethane & $\mathrm{CH_3Cl}$ & 204 & $10^{6}$ & $10^{-9}$ \\
26 & methanethiol & $\mathrm{CH_3SH}$ & 234 & $10^{6}$ & $10^{-9}$ \\
27 & silicon monoxide & $\mathrm{SiO}$ & 114 & $10^{6}$ & $10^{-9}$ \\
28 & carbon monosulfide & $\mathrm{CS}$ & 114 & $10^{6}$ & $10^{-9}$ \\
29 & hypochlorous acid & $\mathrm{HClO}$ & 144 & $10^{6}$ & $10^{-9}$ \\
\end{longtable}
\renewcommand{\arraystretch}{1.0}
\endgroup

\subsection{$GW$100 grid choices and convergence thresholds}
For the larger $GW$100 benchmark, both the IR grid and the convergence threshold had to be adjusted for several systems to balance numerical stability and computational cost. The settings used in the final calculations are listed in Table~\ref{tab:gw100-grid-threshold}.

\begingroup
\small
\setlength{\tabcolsep}{3pt}
\renewcommand{\arraystretch}{0.5}
\begin{longtable}{r l l c c c}
\caption{Grid choice and convergence thresholds used for the $GW$100 database.}
\label{tab:gw100-grid-threshold}\\
\hline
\textbf{$GW$100 \#} & \textbf{CAS \#} & \textbf{Molecule} & \textbf{$N_{\mathrm{AO}}$} & \textbf{Grid $\Lambda$} & \textbf{Threshold (a.u.)} \\
\hline
\endfirsthead
\multicolumn{6}{c}{\tablename\ \thetable\ -- continued from previous page} \\
\hline
\textbf{$GW$100 \#} & \textbf{CAS \#} & \textbf{Molecule} & \textbf{$N_{\mathrm{AO}}$} & \textbf{Grid $\Lambda$} & \textbf{Threshold (a.u.)} \\
\hline
\endhead
\hline
\multicolumn{6}{r}{Continued on next page} \\
\endfoot
\hline \hline
\endlastfoot
 1 & 7440-59-7 & helium & 14 & $10^{4}$ & $10^{-7}$ \\
 2 & 7440-01-9 & neon & 31 & $10^{6}$ & $10^{-7}$ \\
 3 & 7440-37-1 & argon & 42 & $10^{6}$ & $10^{-7}$ \\
 4 & 7439-90-9 & krypton & 48 & $10^{6}$ & $10^{-7}$ \\
 6 & 1333-74-0 & hydrogen & 28 & $10^{4}$ & $10^{-7}$ \\
 7 & 14452-59-6 & lithium dimer & 38 & $10^{4}$ & $10^{-7}$ \\
 8 & 25681-79-2 & sodium dimer & 64 & $10^{5}$ & $10^{-7}$ \\
 9 & 39297-86-4 & sodium tetramer & 128 & $10^{5}$ & $10^{-7}$ \\
 10 & 39297-88-6 & sodium hexamer & 192 & $10^{5}$ & $10^{-7}$ \\
 11 & 25681-80-5 & potassium dimer & 66 & $10^{6}$ & $10^{-7}$ \\
 13 & 7727-37-9 & nitrogen & 62 & $10^{5}$ & $10^{-7}$ \\
 14 & 12185-09-0 & phosphorus dimer & 84 & $10^{5}$ & $10^{-7}$ \\
 15 & 23878-46-8 & arsenic dimer & 96 & $10^{6}$ & $10^{-7}$ \\
 16 & 7782-41-4 & fluorine & 62 & $10^{5}$ & $10^{-7}$ \\
 17 & 7782-50-5 & chlorine & 84 & $10^{5}$ & $10^{-7}$ \\
 18 & 7726-95-6 & bromine & 96 & $10^{6}$ & $10^{-7}$ \\
 20 & 74-82-8 & methane & 87 & $10^{5}$ & $10^{-7}$ \\
 21 & 74-84-0 & ethane & 146 & $10^{5}$ & $10^{-7}$ \\
 22 & 74-98-6 & propane & 205 & $10^{5}$ & $10^{-7}$ \\
 23 & 106-97-8 & butane & 264 & $10^{5}$ & $10^{-7}$ \\
 24 & 74-85-1 & ethylene & 118 & $10^{5}$ & $10^{-7}$ \\
 25 & 74-86-2 & ethyne/acetylene & 90 & $10^{5}$ & $10^{-7}$ \\
 26 & 12184-80-4 & tetracarbon & 124 & $10^{5}$ & $10^{-7}$ \\
 27 & 75-19-4 & cyclopropane & 177 & $10^{5}$ & $10^{-7}$ \\
 28 & 71-43-2 & benzene & 270 & $10^{5}$ & $10^{-6}$ \\
 29 & 629-20-9 & cyclooctatetraene & 360 & $10^{5}$ & $10^{-4}$ \\
 30 & 542-92-7 & cyclopentadiene & 239 & $10^{5}$ & $10^{-6}$ \\
 31 & 75-02-5 & vinyl fluoride & 135 & $10^{5}$ & $10^{-7}$ \\
 32 & 75-01-4 & vinyl chloride & 146 & $10^{5}$ & $10^{-7}$ \\
 33 & 593-60-2 & vinyl bromide & 152 & $10^{6}$ & $10^{-7}$ \\
 35 & 75-73-0 & tetrafluoromethane & 155 & $10^{5}$ & $10^{-7}$ \\
 36 & 56-23-5 & tetrachloromethane & 199 & $10^{5}$ & $10^{-7}$ \\
 37 & 558-13-4 & tetrabromomethane & 223 & $10^{6}$ & $10^{-6}$ \\
 39 & 7803-62-5 & silane & 98 & $10^{5}$ & $10^{-7}$ \\
 40 & 7782-65-2 & germane & 104 & $10^{6}$ & $10^{-7}$ \\
 41 & 1590-87-0 & disilane & 168 & $10^{5}$ & $10^{-7}$ \\
 42 & 14868-53-2 & pentasilane & 378 & $10^{5}$ & $10^{-4}$ \\
 43 & 7580-67-8 & lithium hydride & 33 & $10^{4}$ & $10^{-7}$ \\
 44 & 7693-26-7 & potassium hydride & 47 & $10^{5}$ & $10^{-7}$ \\
 45 & 13283-31-3 & borane & 73 & $10^{4}$ & $10^{-7}$ \\
 46 & 19287-45-7 & diborane & 146 & $10^{4}$ & $10^{-7}$ \\
 47 & 7664-41-7 & ammonia & 73 & $10^{5}$ & $10^{-7}$ \\
 48 & 7782-79-8 & hydrogen azide & 107 & $10^{5}$ & $10^{-7}$ \\
 49 & 7803-51-2 & phosphine & 84 & $10^{5}$ & $10^{-7}$ \\
 50 & 7784-42-1 & arsine & 90 & $10^{6}$ & $10^{-7}$ \\
 51 & 7783-06-4 & hydrogen sulfide & 70 & $10^{5}$ & $10^{-7}$ \\
 52 & 7647-01-0 & hydrogen fluoride & 45 & $10^{5}$ & $10^{-7}$ \\
 53 & 7664-39-3 & hydrogen chloride & 56 & $10^{5}$ & $10^{-7}$ \\
 54 & 7789-24-4 & lithium fluoride & 50 & $10^{5}$ & $10^{-7}$ \\
 55 & 7783-40-6 & magnesium fluoride & 94 & $10^{5}$ & $10^{-7}$ \\
 56 & 7783-63-3 & titanium fluoride & 188 & $10^{5}$ & $10^{-6}$ \\
 57 & 7784-18-1 & aluminium fluoride & 135 & $10^{5}$ & $10^{-7}$ \\
 58 & 13768-60-0 & boron monofluoride & 62 & $10^{5}$ & $10^{-7}$ \\
 59 & 7783-60-0 & sulfur tetrafluoride & 166 & $10^{5}$ & $10^{-6}$ \\
 60 & 7758-02-3 & potassium bromide & 81 & $10^{6}$ & $10^{-7}$ \\
 61 & 17108-85-9 & gallium monochloride & 90 & $10^{6}$ & $10^{-6}$ \\
 62 & 7647-14-5 & sodium chloride & 74 & $10^{5}$ & $10^{-7}$ \\
 63 & 7786-30-3 & magnesium chloride & 116 & $10^{5}$ & $10^{-7}$ \\
 65 & 10043-11-5 & boron nitride & 62 & $10^{5}$ & $10^{-7}$ \\
 66 & 74-90-8 & hydrogen cyanide & 76 & $10^{5}$ & $10^{-7}$ \\
 67 & 17739-47-8 & phosphorus mononitride & 73 & $10^{5}$ & $10^{-7}$ \\
 68 & 302-01-2 & hydrazine & 118 & $10^{5}$ & $10^{-7}$ \\
 69 & 50-00-0 & formaldehyde & 90 & $10^{5}$ & $10^{-7}$ \\
 70 & 67-56-1 & methanol & 118 & $10^{5}$ & $10^{-7}$ \\
 71 & 64-17-5 & ethanol & 177 & $10^{5}$ & $10^{-7}$ \\
 72 & 75-07-0 & acetaldehyde & 149 & $10^{5}$ & $10^{-7}$ \\
 73 & 60-29-7 & ethyl ether & 295 & $10^{5}$ & $10^{-6}$ \\
 74 & 64-18-6 & formic acid & 121 & $10^{5}$ & $10^{-7}$ \\
 75 & 7722-84-1 & hydrogen peroxide & 90 & $10^{5}$ & $10^{-7}$ \\
 76 & 7732-18-5 & water & 59 & $10^{5}$ & $10^{-7}$ \\
 77 & 124-38-9 & carbon dioxide & 93 & $10^{5}$ & $10^{-7}$ \\
 78 & 75-15-0 & carbon disulfide & 115 & $10^{5}$ & $10^{-7}$ \\
 79 & 463-58-1 & carbon oxysulfide & 104 & $10^{5}$ & $10^{-7}$ \\
 80 & 1603-84-5 & carbon oxyselenide & 110 & $10^{6}$ & $10^{-7}$ \\
 81 & 630-08-0 & carbon monoxide & 62 & $10^{5}$ & $10^{-7}$ \\
 82 & 10028-15-6 & ozone & 93 & $10^{5}$ & $10^{-7}$ \\
 83 & 7446-09-5 & sulfur dioxide & 104 & $10^{5}$ & $10^{-7}$ \\
 84 & 1304-56-9 & beryllium monoxide & 50 & $10^{5}$ & $10^{-7}$ \\
 85 & 1309-48-4 & magnesium monoxide & 63 & $10^{6}$ & $10^{-7}$ \\
 86 & 108-88-3 & toluene & 329 & $10^{5}$ & $10^{-7}$ \\
 87 & 100-41-4 & ethylbenzene & 388 & $10^{5}$ & $10^{-5}$ \\
 88 & 392-56-3 & hexafluorobenzene & 372 & $10^{5}$ & $10^{-5}$ \\
 89 & 108-95-2 & phenol & 301 & $10^{5}$ & $10^{-7}$ \\
 90 & 62-53-3 & aniline & 315 & $10^{5}$ & $10^{-7}$ \\
 91 & 110-86-1 & pyridine & 256 & $10^{5}$ & $10^{-7}$ \\
 92 & 73-40-5 & guanine & 411 & $10^{5}$ & $10^{-4}$ \\
 93 & 73-24-5 & adenine & 380 & $10^{5}$ & $10^{-5}$ \\
 94 & 71-30-7 & cytosine & 318 & $10^{5}$ & $10^{-5}$ \\
 95 & 65-71-4 & thymine & 363 & $10^{5}$ & $10^{-5}$ \\
 96 & 66-22-8 & uracil & 304 & $10^{5}$ & $10^{-5}$ \\
 97 & 57-13-6 & urea & 180 & $10^{5}$ & $10^{-7}$ \\
 99 & 12190-70-4 & copper dimer & 128 & $10^{6}$ & $10^{-7}$ \\
 100 & 544-92-3 & copper cyanide & 126 & $10^{6}$ & $10^{-7}$ \\
\end{longtable}
\renewcommand{\arraystretch}{1.0}
\endgroup

\newpage
\bibliography{refs/ab_initio, refs/abbr, refs/alg, refs/other_ref, refs/analysis, refs/basis, refs/cc, refs/crit_temp, refs/dft, refs/eff_ham, refs/finite_size, refs/gf2, refs/grid, refs/group, refs/gw, refs/nuclear, refs/programs, refs/rdm, refs/ri_cholesky, refs/so, refs/textbooks, refs/thc, refs/tt_tci,refs/theorems, refs/intermol_int, refs/mol_magnets}